\documentclass[aps,prd,onecolumn,groupedaddress]{revtex4}
\usepackage{graphicx}
\usepackage{graphpap}

\usepackage{dcolumn}
\usepackage{bm}
\usepackage{amssymb}

\newcommand\bmat{\left( \begin{array}{cc}}
\newcommand\emat{\end{array}\right)}

\def\msbar{\ifmmode{\overline{\rm MS}} \else{$\overline{\rm MS}$} \fi}
\def\drbar{\ifmmode{\overline{\rm DR}} \else{$\overline{\rm DR}$} \fi}

\def\ti              {\tilde}

\def\a               {\alpha}
\def\b               {\beta}

\def\e               {\epsilon}
\def\g               {\gamma}

\def\x               {\chi}

\def\sf              {{\ti f}}

\def\st              {{\ti t}}
\def\sb              {{\ti b}}

\def\nt              {\ti \x^0}

\def\kslash          {\not{\!k}}
\def\pslash          {\not{\!p}}

\def\non             {\nonumber}

\begin{document}

\eprint{LPSC 09-082}
\title{SUSY-QCD effects on neutralino dark matter annihilation \\
       beyond scalar or gaugino mass unification}
\author{Bj\"orn Herrmann}
\affiliation{Institut f\"ur Theoretische Physik und Astrophysik,
 Universit\"at W\"urzburg,
 Am Hubland, D-97074 W\"urzburg, Germany}
\author{Michael Klasen}
\author{Karol Kova\v{r}\'{\i}k}
\email[]{kovarik@lpsc.in2p3.fr}
\affiliation{Laboratoire de Physique Subatomique et de Cosmologie,
 Universit\'e Joseph Fourier/CNRS-IN2P3/INPG,
 53 Avenue des Martyrs, F-38026 Grenoble, France}
\date{\today}

\begin{abstract}
 We describe in detail our calculation of the full supersymmetric (SUSY) QCD corrections
 to neutralino annihilation into heavy quarks and extend our numerical analysis of
 the resulting dark matter relic density to scenarios without scalar or gaugino mass
 unification. In these scenarios, the final state is often composed of top
 quarks and the annihilation proceeds through $Z^0$-boson or scalar top-quark
 exchanges. The impact of the corrections is again shown to be sizable, so that
 they must be taken into account systematically in global analyses of the
 supersymmetry parameter space. 
\end{abstract}

\maketitle

\section{Introduction \label{sec1}} 
The search for physics beyond the Standard Model (SM) is no longer restricted to
colliders only. In fact, the most compelling evidence for new physics comes today
from cosmological observations such as the mission of the Wilkinson Microwave
Anisotropy Probe (WMAP), which have determined the matter and energy decomposition
of our universe with unprecedented precision. These observations indicate the
existence of Cold Dark Matter (CDM) in the universe, which cannot be accounted for
by the SM and likely consists of Weakly Interacting Massive Particles (WIMPs) with
non-relativistic velocities. A combination of the five-year measurement of the
cosmic microwave background by the WMAP mission with supernova and baryonic
acoustic oscillation data yields the narrow $2\sigma$-interval for the relic
density of dark matter \cite{WMAP}
\begin{equation}\label{cWMAP}
  0.1097 < \Omega_{\rm CDM}h^2 < 0.1165\,,
\end{equation}
where $h$ denotes the present Hubble expansion rate $H_0$ in units of 100 km
s$^{-1}$ Mpc$^{-1}$ .

The measured relic density of dark matter can be used to constrain extensions of
the Standard Model, which provide a viable WIMP candidate. In the Minimal
Supersymmetric Standard Model (MSSM) with $R$-parity conservation, this could be
the Lightest Supersymmetric Particle (LSP) $\tilde{\chi}$, if it is neutral and a
color singlet. One can then calculate its relic density, compare it with the
experimental limits in Eq.\ (\ref{cWMAP}), and identify
the favored regions of the MSSM parameter space. The relic density
\begin{equation}
  \Omega_{\tilde{\chi}}h^2 ~=~ \frac{n_0\,m_{\tilde{\chi}}}{\rho_c}
\end{equation}
is proportional to the present number density $n_0$ and the mass
$m_{\tilde{\chi}}$ of the LSP. $\rho_c=3H_0^2/(8\pi G_{\rm N})$ is the critical
density of our universe, and $G_{\rm N}$ is the gravitational constant. The
present number density $n_0$ is obtained by solving the Boltzmann equation
describing the time evolution of the number density
\begin{equation}
 \frac{dn_{\tilde{\chi}}}{dt} ~=~ -3 H n_{\tilde{\chi}} - \langle
 \sigma_{\rm ann}v \rangle \big( n_{\tilde{\chi}}^2 - n_{\rm eq}^2 \big) \,.
\end{equation}
The first term on the right-hand side corresponds to a dilution due to the
expansion of the universe, and the second term corresponds to a decrease due to
annihilations and co-annihilations of the relic particle into SM particles
\cite{Gondolo1991}. Here, $H$ denotes the time-dependent Hubble expansion
parameter, and $n_{\rm eq}$ the density of the relic particle in thermal
equilibrium. Details of the dark matter interactions enter the Boltzmann equation
through the thermally averaged cross section $\langle \sigma_{\rm ann}v \rangle$.
The cross section takes into account the thermal velocity distribution of the
relic particle and is calculated for a given temperature $T$ by
\begin{equation}
 \langle \sigma_{\rm ann}v \rangle ~=~ \frac{4}{m_{\tilde{\chi}}^4 T K_2^2\!
 \left( \frac{m_{\tilde{\chi}}}{T} \right)}\int {\rm d}p_{\rm cm} p_{\rm cm}^3
 \sqrt{s} K_1\!\left(       \frac{\sqrt{s}}{T} \right) \sigma_{\rm ann}(s)\,,
\end{equation}
where $K_1$ and $K_2$ are the modified Bessel functions of the first and second
kind, respectively. The center-of-momentum energy $\sqrt{s}$ is related to the
particle mass $m_{\tilde{\chi}}$ and the relative momentum $p_{\rm cm}$ of the
annihilating pair through $s=4 \big( m_{\tilde{\chi}}^2 + p_{\rm cm}^2 \big)$
\cite{Gondolo1991}.

In order to keep up with current and future experimental improvements, one has to
understand and reduce the different uncertainties involved in the analysis, both
for the prediction of the dark matter relic density and the extraction of new
mass parameters from cosmological data. These uncertainties include, e.g., a
modification of the Hubble expansion rate due to quintessence or an effective dark
energy density \cite{Arbey2009}, differences in the new physical particle masses
obtained with different spectrum codes \cite{Belanger2005}, or a lack of precision
in the annihilation cross-section of dark matter particles \cite{Baro2007}. In
this paper, we focus on the impact of next-to-leading order QCD and SUSY-QCD
corrections on the latter, but other possible uncertainties will also be briefly
discussed.

In many scenarios of the MSSM, the lightest neutralino is the LSP and therefore a
suitable dark matter candidate. The thermally averaged cross section is then
obtained by computing all relevant neutralino annihilation cross sections into SM
particles. Most prominent are the processes with two-particle final states such as
a fermion-antifermion pair or a combination of gauge ($W^\pm,Z^0$) and Higgs
bosons ($h^0,H^0,A^0,H^\pm$) \cite{Jungman,DreesNojiri}. In this paper, we focus
on the annihilation into a massive quark-antiquark pair, since the leading order
cross section with a fermion-antifermion final state is proportional to the mass
of the fermion, which disfavors the light quarks. Moreover, the annihilation into
heavy quarks is important in the regions of parameter space allowed by Eq.\
(\ref{cWMAP}). We now present the full details of our calculation and investigate
scenarios with dominant top-quark final state contributions, extending our
analysis of Refs.~\cite{Letter1,Letter2} beyond minimal supergravity (mSUGRA)
models. In mSUGRA models, one is constrained by having only five universal
high-scale parameters [$m_0$, $m_{1/2}$, $A_0$, $\tan\beta$, and sgn($\mu$)], and
in regions of parameter space, where quark final states are important, the cross
section is dominated by an exchange of Higgs bosons. Here, we relax the
unification of either the scalar masses  or the gaugino masses, one at a time.
This allows for scenarios different from mSUGRA, where the annihilation cross
section is not necessarily dominated by Higgs-boson exchanges. Apart from a
Higgs-boson dominated scenario, we thus analyze scenarios, where $Z^0$-boson
exchanges in the $s$-channel or squark exchanges in the $t$- and $u$-channels play
an important role. The full QCD and SUSY-QCD corrections in these scenarios
turn out to be significant, and we have therefore included them into the public
code {\tt micrOMEGAs} \cite{micromegas}.

This paper is organized as follows: In Sec.\ \ref{sec2} and appendices
\ref{asec1} and \ref{asec2}, we give all necessary details of the calculation
related to the virtual loop corrections, the renormalization procedure, and the
calculation of real gluon emission, in particular the subtraction of the induced
soft and collinear singularities. We then continue in Sec.\ \ref{sec3} with a
discussion of the MSSM models beyond scalar or gaugino mass unification. In Sec.\
\ref{sec4}, we analyze the impact of the radiative corrections on the relic
density in these models. Finally, our results are summarized in \ref{sec5}.

\section{Calculation details \label{sec2}}

\begin{figure}
  \begin{center}
    \includegraphics[scale=1.0]{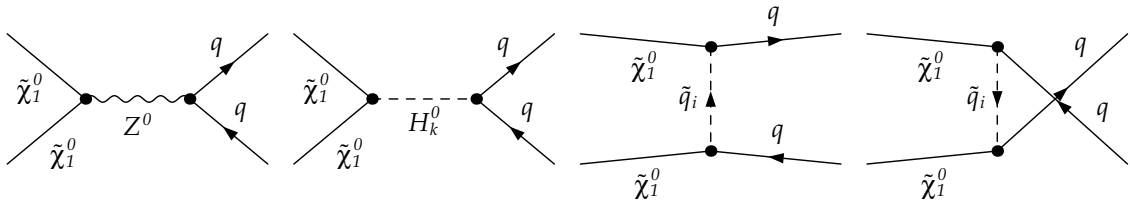}
  \end{center}
  \caption{Tree-level Feynman diagrams for the annihilation of a neutralino pair
  into a quark-antiquark pair through the exchange of a $Z^0$-boson, a neutral
  Higgs-boson $H_k^0=(h^0,H^0,A^0)$, or a squark $\tilde{q}_i$ ($i=1,2$).}
  \label{figborn}
\end{figure}

The annihilation of neutralinos into quarks proceeds at tree-level through an
exchange of a $Z^0$-boson and Higgs-bosons in the $s$-channel and through the
exchanges of scalar quarks in the $t$- and $u$-channels (see Fig.~\ref{figborn}).
By definition, a WIMP can only have electroweak interactions, and in order to
reach a sufficient annihilation rate to explain the dark matter relic density, the
cross section has to be enhanced, e.g., by a resonance. One often finds that there
is one contribution which dominates the whole cross section. This allows us, by
choosing different scenarios, to study contributions from various channels.
Moreover, it allows us also to isolate effects of radiative corrections coming
from different sources.

We have computed the full QCD and SUSY-QCD corrections to neutralino annihilation
into quarks. The next-to-leading order cross section contains virtual
contributions stemming from loop diagrams and real contributions, which are due to
the radiation of an additional gluon. Symbolically, the next-to-leading order
(NLO) cross section can be written as
\begin{figure}[t!]
\begin{picture}(450,475)
 \put(0,400){\includegraphics[scale=1.0]{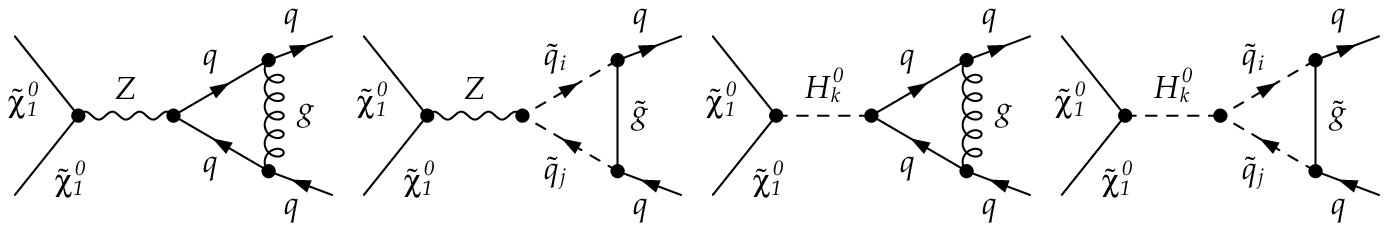}}
 \put(0,320){\includegraphics[scale=1.0]{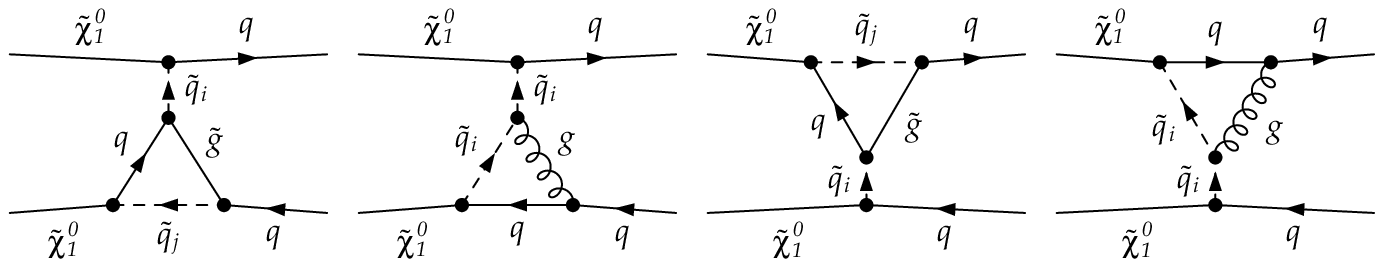}}
 \put(0,240){\includegraphics[scale=1.0]{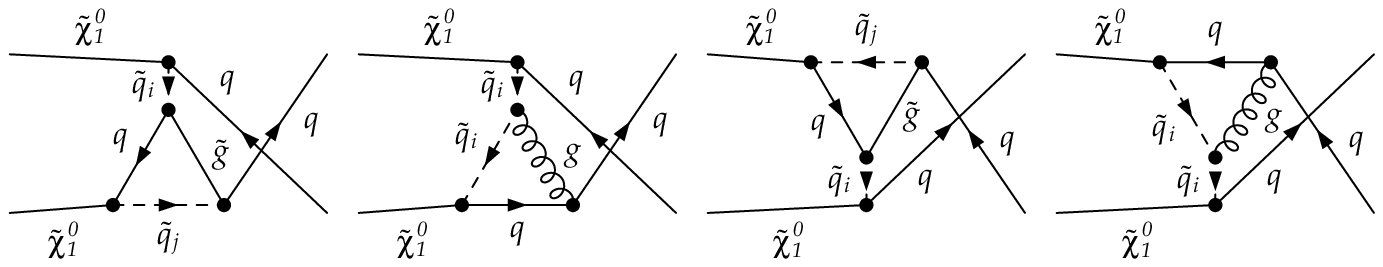}}
 \put(0,160){\includegraphics[scale=1.0]{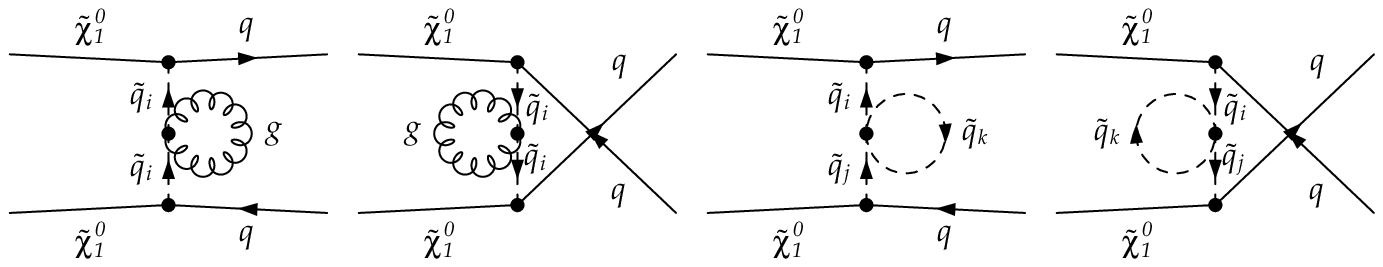}}
 \put(0,80){\includegraphics[scale=1.0]{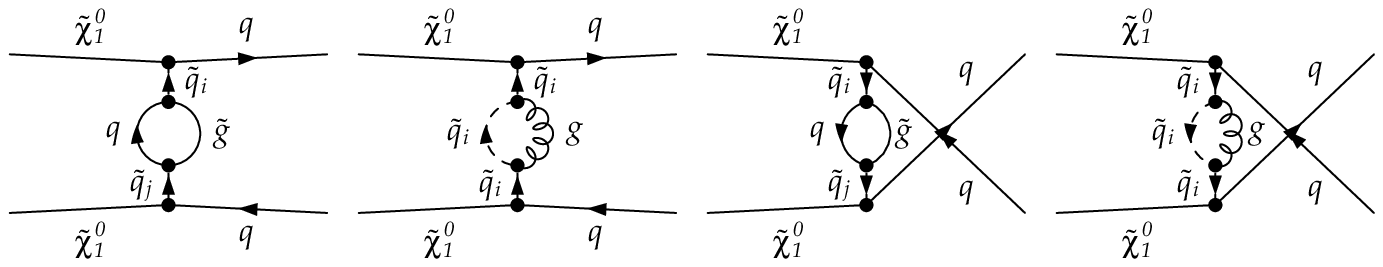}}
 \put(0,0){\includegraphics[scale=1.0]{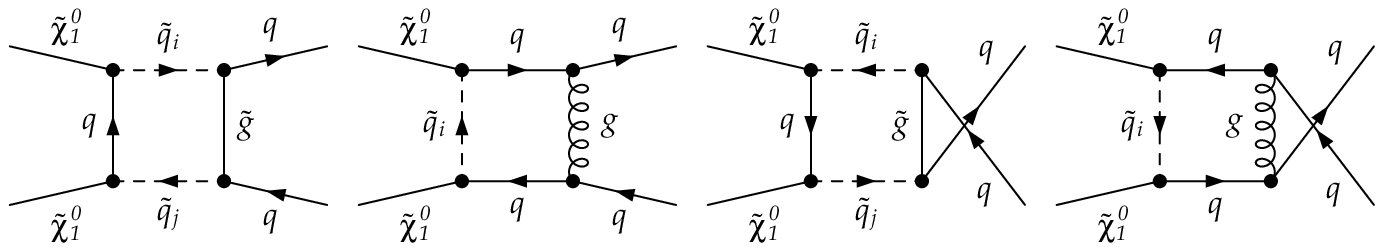}}
\end{picture}
	\caption{SUSY-QCD loop diagrams contributing to the annihilation of neutralinos into quarks.}\label{FDloop}
\end{figure}
\begin{figure}[t!]
\begin{picture}(450,160)
 \put(0,80){\includegraphics[scale=1.0]{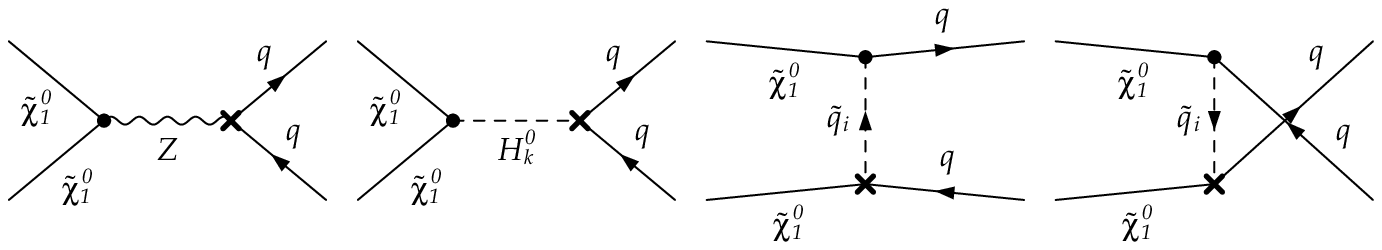}}
 \put(0,0){\includegraphics[scale=1.0]{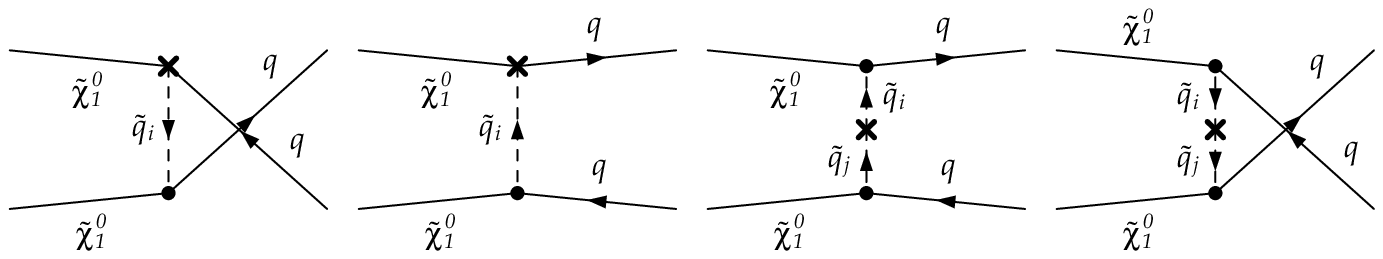}}
\end{picture}
\caption{SUSY-QCD counterterm diagrams contributing to the annihilation of neutralinos into quarks.}\label{FDcount}
\end{figure}
\begin{equation}\label{NLOs}
 \sigma^{\rm NLO} = \int_{2\rightarrow 2}{\rm d}\sigma^{\rm V} + 
 \int_{2\rightarrow 3}{\rm d}\sigma^{\rm R}\,,
\end{equation}
where $\sigma^{\rm V,R}$ denote the virtual and real emission parts integrated
over the two- and three-particle phase space, respectively.
The cross section $\sigma^{\rm V}$, which includes the tree-level and the one-loop
virtual corrections, is given by
\begin{equation}
\sigma_{\rm ann}^{\rm V}(s) = \int\frac{1}{16\pi^2}	\frac{1}{2\kappa
	(s,m_{\tilde{\chi}}^2,m_{\tilde{\chi}}^2)}\frac{\kappa
(s,m_q^2,m_q^2)}{2 s}\Big[|\mathcal{M}_{\rm tree}|^2 + 2\Re\,\big( \mathcal{M}_{\rm 1-loop}\,\mathcal{M}^*_{\rm tree}\big)\Big]\;{\rm d}\Omega\,,
\end{equation}
where $\kappa (x, y, z) = \sqrt{(x-y-z)^2 - 4 yz}$. For our notation and
conventions, we refer the reader to App.\ \ref{asec1}. There are two types of
contributions to the one-loop amplitude $\mathcal{M}_{\rm 1-loop}$, those coming
from the loop diagrams and those coming from the counterterms. The loop diagrams,
depicted in Fig.~\ref{FDloop}, contain ultraviolet (UV) and infrared (IR)
divergent loop integrals. We regulate both types of divergences dimensionally
($d=4-2\e$) and
evaluate the loop integrals in the dimensional reduction scheme
($\overline{\rm DR}$). The full analytic results for the loop diagrams are given
in App.\ \ref{asec2}.

The UV divergences are compensated by counterterms (see Fig.~\ref{FDcount})
related to quarks and their scalar super-partners, the squarks. All counterterm
vertices with quark or squark legs contain wave-function renormalization factors
$\delta Z_L$, $\delta Z_R$ and $\delta Z_{ij}$, that follow from replacing the
(s)quark fields by 
\begin{eqnarray}
\left(\begin{array}{c}
 q_L \\ q_R
      \end{array}\right) \rightarrow \left(\begin{array}{cc}
1+\frac{1}{2}\delta Z_L & 0 \\ 0 & 1+\frac{1}{2}\delta Z_R
     \end{array}\right)\left(
\begin{array}{c}
 q_L \\ q_R
     \end{array}\right)\,,\qquad
\tilde{q}_i \rightarrow (\delta_{ij}+\frac{1}{2}\delta Z_{ij})\tilde{q}_j
\,.
\end{eqnarray}
Although in principle only the wave-function renormalization constants of the external quarks have to be included, we also include the squark renormalization constants, as they allow us to perform simpler UV-convergence checks. In the full calculation, the squark wave-function renormalization constants cancel out. The wave-function renormalization constants are determined by requiring the residues of the propagators to remain at unity even at one-loop order. This condition gives
\begin{eqnarray}\label{offdiag}
\delta Z_L &=& \Re\,\Big[ -\Pi_{L} (m_q^2)- m_q^2\big(\dot\Pi_{L}
(m_q^2)+ \dot\Pi_{R} (m_q^2)\big)+\frac{1}{2 m_q}\big(\Pi_{SL}
(m_q^2)-\Pi_{SR} (m_q^2)\big)\nonumber
\\
&&  - m_q\big(\dot\Pi_{SL} (m_q^2)+\dot\Pi_{SR}
(m_q^2)\big)\Big]\,,
\\ 
\delta Z_R &=& \delta Z_L(L \leftrightarrow R)\,,
\\
\delta Z_{ii} &=& - \Re\,\Big[\dot\Pi_{ii}^{\tilde{q}} (m_{\tilde{q}_i}^2)\Big]\,,\qquad\quad \delta
Z_{ij}~=~\frac{2\,\Re\,\big[\Pi_{ij}^{\tilde{q}}(m_{\tilde{q}_j}^2)\big]}{m_{\tilde{q}_{i}}^2-m_{\tilde{q}_{j}}^2}\qquad {\rm for}\quad i\neq j\,,
\end{eqnarray}
where $\Pi_{L,R}(k^2)$ and $\Pi_{SL,SR}(k^2)$ stand for the vector and the scalar parts of the two-point Green's function as defined in Ref.~\cite{Kovarik2005} and $\dot\Pi (m^2)=\left[\frac{\partial}{\partial k^2}\Pi
(k^2)\right]_{k^2=m^2 }$. After performing the wave-function renormalization, the remaining divergences are canceled by renormalizing the coupling constants. In our case, the coupling counterterms that receive contributions proportional to the strong coupling constant $\alpha_s$ are the quark Yukawa couplings through the masses of the quarks, the squarks masses and the squark mixing angle $\theta_{\tilde{q}}$.
A very important contribution, in particular in scenarios with a dominant Higgs-boson exchange, comes from renormalizing the Yukawa couplings of the quarks  \cite{Letter2}. In our calculation, we use the $\overline{\rm DR}$ Yukawa couplings for both the top and the bottom quarks. As the quark masses that serve as inputs are defined in different schemes, we take two distinct approaches for top and bottom quarks. For top quarks, the input is the on-shell mass $m_t=172.4\,{\rm GeV}$ measured at the Tevatron \cite{CDFD02008}, and the $\overline{\rm DR}$-mass of the top quark (and hence the $\overline{\rm DR}$ Yukawa coupling) is obtained by subtracting the finite on-shell counterterm
\begin{eqnarray}
	\delta m_{q} & = & \frac{1}{2}\, \Re \Big[ m_{q}
	\big( \Pi_L (m_{q}) + \Pi_R (m_{q}) \big) +
	\Pi_{SL}(m_{q}) + \Pi_{SR}(m_{q}) \Big] \,.
\end{eqnarray}
On the other hand, the input mass $m_b(m_b)$ for bottom quarks is extracted in the $\overline{\rm MS}$ renormalization scheme from the Standard Model analysis of $\Upsilon$ sum rules \cite{MBmass}. In order to obtain the appropriate bottom Yukawa coupling in the $\overline{\rm DR}$ renormalization scheme within the MSSM, we first use the Standard Model next-to-next-to-leading order (NNLO) renormalization group evolution to obtain the mass of the bottom quark at the scale $Q=2m_{\tilde{\chi}}$ \cite{Baer2002}. Still in the SM, we then convert $m_b^{\overline{\rm MS}}(Q)$ to $m_b^{\overline{\rm DR}}(Q)$ \cite{Baer2002}, and finally we apply the threshold corrections including also contributions from SUSY particles in the loop. For the last step, we take into account the fact that the sbottom-gluino and stop-chargino one-loop contributions are considerably enhanced for large $\tan\beta$ or large $A_b$ and can be resummed to all orders in perturbation theory \cite{Carena2000,Spira2003}. Denoting the resummable part by $\Delta_b$ and the finite one-loop remainder by $\Delta m_b$, the bottom quark mass is then given by
\begin{equation}
        m_{b}^{\overline{\rm DR},\,{\rm MSSM}}(Q) ~=~ \frac{m_{b}^{\overline{\rm DR},\,{\rm SM}}(Q)}{1+\Delta_b} - \Delta m_b\, .
\end{equation}
In the squark sector, we pick five independent quantities, $m_{{\ti t}_1}$, $m_{{\ti t}_2}$, $m_{{\ti b}_1}$, $\theta_{\ti t}$, and $\theta_{\ti b}$, in order to respect the SU(2) symmetry. We renormalize the masses of three squarks in the on-shell scheme, which leads to the counterterm
\begin{equation}
	\delta m_{\tilde{q}_i}^2 = \Re\, \Big[\Pi_{ii}^{\tilde{q}}(m_{\tilde{q}_i}^2)\Big] \,.
\end{equation}
The remaining mass of the heavier scalar bottom quark $m_{\tilde{b}_2}$ is treated as dependent,
\begin{equation}
\delta m_{\tilde{b}_2}^2 = \frac{1}{\sin^2\theta_{\ti b}}\Big[\delta m_{\tilde{t}_1}^2 \cos^2\theta_{\ti t} + \delta m_{\tilde{t}_2}^2 \sin^2\theta_{\ti t} - \delta m_{\tilde{b}_1}^2 \cos^2\theta_{\ti b} + (m_{\tilde{t}_2}^2 - m_{\tilde{t}_1}^2 )\sin 2\theta_{\ti t}\,\delta \theta_{\ti t} - (m_{\tilde{b}_2}^2 - m_{\tilde{b}_1}^2 )\sin 2\theta_{\ti b}\,\delta \theta_{\ti b} \Big] \,.	
\end{equation} 
The squark mixing angle is renormalized in the $\overline{\rm DR}$ scheme and so the corresponding counterterms of the squark mixing matrices $R^{\tilde q}_{ij}$ contain only the divergent parts. The counterterms can be determined as
\begin{equation}
\delta R^{\tilde{q}}_{ij}=\sum_{k=1}^2\frac{1}{4}(\delta
Z^{{\rm div}}_{ik}-\delta Z^{{\rm div}}_{ki})R^{\tilde q}_{kj}\,,
\end{equation}
using only the divergent parts of the wave-function renormalization constants. This is equivalent to fixing the mixing angle as
\begin{eqnarray}\label{dthetasf}
   \delta \theta_{\tilde{q}} & = & \frac{1}{4}\, \left(
   \delta Z_{12}^{{\rm div}} - \delta Z_{21}^{{\rm div}}\right)
   \,=\, \frac{1}{2\big(m_{\tilde{q}_1}^2 \!-\! m_{\tilde{q}_2}^2\big)}\, \Re\!
   \left[ \Pi_{12}^{\tilde{q}\,{\rm div}}(m_{\tilde{q}_{2}}^2) + \Pi_{21}^{\tilde{q}\,{\rm div}}
   (m_{\tilde{q}_{1}}^2)
    \right] \,.
\end{eqnarray}
 
\begin{figure}[t!]
\begin{picture}(450,240)
 \put(0,160){\includegraphics[scale=1.0]{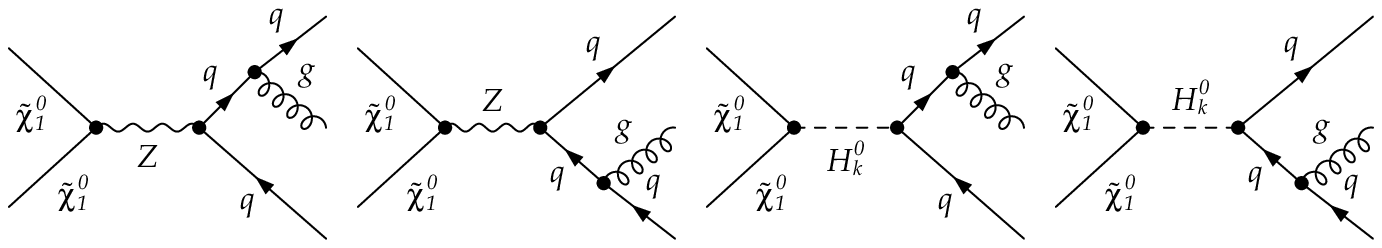}}
 \put(0,80){\includegraphics[scale=1.0]{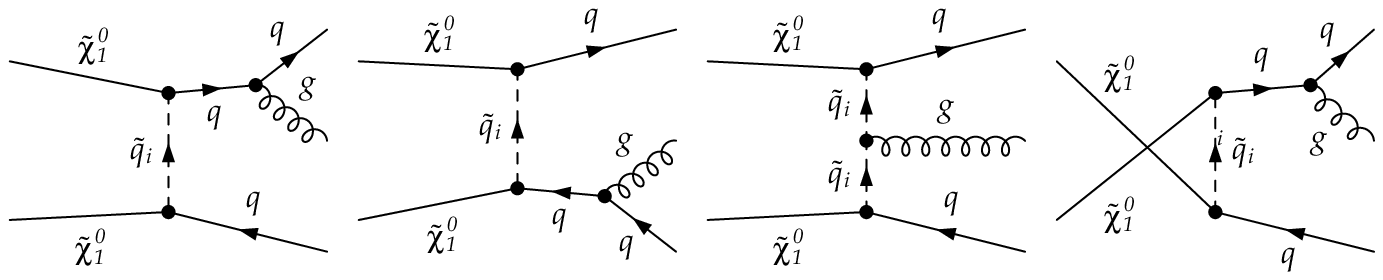}}
 \put(0,0){\includegraphics[scale=1.0]{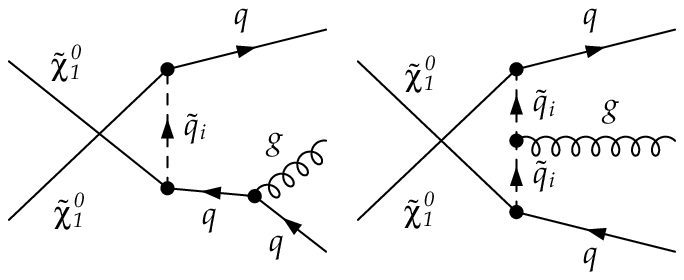}}
\end{picture}
	\caption{Bremsstrahlung diagrams contributing to the annihilation of neutralinos into quarks.}\label{FDbrem}
\end{figure}

After renormalization, the virtual cross section $\sigma^{\rm V}$ still contains IR divergences, that come from an exchange of a gluon in the loops. In order to compensate them, one has to include the real cross section $\sigma^{\rm R}$ coming from diagrams with an additional gluon in the final state (see Fig.~\ref{FDbrem}). The symbolic formula of Eq.~(\ref{NLOs}) cannot be directly applied, since the divergences appear as $1/\e$ and $1/\e^2$ poles in the number of dimensions in the one-loop amplitude and as a divergent matrix element in certain regions of the $2\rightarrow 3$ parameter space. A convenient way to combine these two cross sections and to cancel the divergences is the dipole subtraction method \cite{Catani2002}. Using this method, we can write the total next-to-leading order cross section $\sigma^{\rm NLO}$ as
\begin{equation}
\sigma^{\rm NLO} = 
\int_{2\rightarrow 2}
	  \left[{\rm d}\sigma^{\rm V} + \int_1 {\rm d}\sigma^{\rm A} \right]_{\epsilon=0} + \int_{2\rightarrow 3}\Big[ \left({\rm d}\sigma^{\rm R}\right)_{\epsilon =0} - \left({\rm d}\sigma^{\rm A}\right)_{\epsilon=0} \;\Big]\,,
\end{equation}
where we introduced an auxiliary cross section $\sigma^{\rm A}$. The auxiliary cross section does not contribute to the total cross section and serves only as a tool to cancel the IR divergences. It has the same divergence structure as the real cross section and at the same time its structure allows for a partial integration of the gluon phase space, so that it can also be added to the virtual cross section, canceling the divergences in each part. The $2\rightarrow 3$ matrix element leading to the auxiliary cross section for the real part is constructed from two dipole contributions  ${\cal D}_{31,2}$ and ${\cal D}_{32,1}$ as 
\begin{equation}
	|{\cal M}_{\rm aux}^{2\rightarrow 3}|^2 = {\cal D}_{31,2}(k_1,k_2,k_3) + {\cal D}_{32,1}(k_1,k_2,k_3)\,,
\end{equation}
where the $k_i$ are the four-momenta of the final state quarks and of the gluon. The dipole contributions ${\cal D}_{31,2}$ and ${\cal D}_{32,1}$ are related by a simple interchange $k_1\leftrightarrow k_2$ with ${\cal D}_{31,2}$ given by
\begin{eqnarray}\nonumber
	  {\cal D}_{31,2}(k_1,k_2,k_3) &=& C_F\frac{8\pi\alpha_s}s
	  \;|{\cal M}_{\rm tree}|^2\;\times\\ &&
	  \frac1{1-x_2}
	  \left\{
	    \frac{2(1-2\mu_q^2)}{2-x_1-x_2}
	    -\sqrt{\frac{1-4\mu_q^2}{x_2^2-4\mu_q^2}}
	    \frac{x_2-2\mu_q^2}{1-2\mu_q^2}
	    \left[2 + \frac{x_1 - 1}{x_2-2\mu_q^2}
	      +\frac{2\mu_q^2}{1-x_2}
	    \right]\right\}\,,\label{dip312}	
\end{eqnarray}
where $x_i = 2k_i.q/q^2$, $\mu_q = m_q/\sqrt{s}$, and $q^2=(k_1+k_2+k_3)^2 = s$ \cite{Catani2002}. The leading order matrix element $|{\cal M}_{\rm tree}|^2$ appearing in Eq.~(\ref{dip312}) is calculated using different kinematics with redefined 4-momenta  
\begin{equation}
	k_1^\mu \rightarrow \widetilde{k}_{31}^\mu = \frac12 q^\mu-
	  \frac{\sqrt{1-4\mu_q^2}}{\sqrt{x_2^2-4\mu_q^2}}
	  \left(k_2^\mu-\frac12 x_2 q^\mu\right),\qquad
  	k_2^\mu \rightarrow \widetilde{k}_2^\mu = \frac12 q^\mu+
  \frac{\sqrt{1-4\mu_q^2}}{\sqrt{x_2^2-4\mu_q^2}}
  \left(k_2^\mu-\frac12 x_2 q^\mu\right)\,.
\end{equation}
The auxiliary matrix element that cancels the infrared divergences of the virtual matrix element is
\begin{eqnarray}\nonumber
	  |{\cal M}_{\rm aux}^{2\rightarrow 2}|^2 &=& 
	  2\,C_F\frac{\alpha_s}{2\pi}\,\frac{(4\pi)^\epsilon}{\Gamma(1-\epsilon)}\,
	  |{\cal M}_{\rm tree}|^2\times\\ &&
	   \left[\left(\frac{\mu^2}{s_{12}}\right)^\epsilon
	    \left({\cal V}_q(s_{12},m_q,m_q;\epsilon)-\frac{\pi^2}3\right)
	    +\frac1{C_F}\Gamma_q(m_q;\epsilon)
	    +\frac32\ln\frac{\mu^2}{s_{12}}
	    +5-\frac{\pi^2}6\right]\,,
\end{eqnarray}
where $s_{12} = s-2m_q^2$. The function ${\cal V}_q$ is composed of a singular part, 
\begin{equation}
  {\cal V}^{(\rm S)}(s_{12},m_q,m_q;\epsilon)=\frac{1+\beta^2}{2\beta}
  \left[\frac1\epsilon\ln\frac{1-\beta}{1+\beta}-\frac12\ln^2\frac{1-\beta}{1+\beta}
    -\frac{\pi^2}6
  +\ln\frac{1-\beta}{1+\beta}\ln\frac2{1+\beta^2}\right]
\end{equation}
with $\beta = \sqrt{1-4\mu_q^2}$, and a non-singular part given by
\begin{eqnarray}
{\cal V}_q^{(\rm NS)}(s_{12},m_q,m_q) &=&
  \frac32\ln\frac{1+\beta^2}2+\frac{1+\beta^2}{2\beta}\left[2\ln\frac{1-\beta}{1+\beta}\ln\frac{2(1+\beta^2)}{(1+\beta)^2}
    +2{\rm Li}_2\left(\frac{1-\beta}{1+\beta}\right)^2
    -2{\rm Li}_2\left(\frac{2\beta}{1+\beta}\right)-\frac{\pi^2}6\right]
\nonumber\\&&
  +\ln(1-\frac12\sqrt{1-\beta^2})-2\ln(1-\sqrt{1-\beta^2})
  -\frac{1-\beta^2}{1+\beta^2}\ln\frac{\sqrt{1-\beta^2}}{2-\sqrt{1-\beta^2}}
\nonumber\\&&
  -\frac{\sqrt{1-\beta^2}}{2-\sqrt{1-\beta^2}}+2\frac{1-\beta^2-\sqrt{1-\beta^2}}{1+\beta^2}
  +\frac{\pi^2}2\,.
\end{eqnarray}
The function $\Gamma_q$ is defined as
\begin{equation}
  \Gamma_q(m_q;\epsilon) =
  C_F\left[\frac1\epsilon+\frac12\ln\frac{m_q^2}{Q^2}-2\right]\,,
\end{equation}
where $Q$ is the renormalization scale. Apart from convergence checks of our calculation, we also performed checks of the finite part of the one-loop diagrams against the results obtained by automatic computer packages {\tt FeynArts} and {\tt FormCalc} \cite{FeynArts} and parts of the calculation also against existing results, e.g. in Ref.~\cite{PhDs}. The UV and IR divergent parts of our loop integrals were checked and agree with the results given in Ref.~\cite{Ellis2007}.
%
\section{SUSY models beyond minimal supergravity \label{sec3}}
In our previous publications, we studied the impact of the SUSY-QCD corrections to
neutralino annihilation within minimal supergravity (mSUGRA) models, defined at
the grand unification scale $M_{\rm GUT}$ in terms of a universal scalar mass
$m_0$, a universal gaugino mass $m_{1/2}$, a common trilinear coupling $A_0$, the
ratio $\tan\beta$ of the Higgs doublet vacuum expectation values, and the sign of
the higgsino mass parameter $\mu$ \cite{Letter1, Letter2}. In mSUGRA models, the
lightest neutralino is often a $B$-ino and annihilates preferably through resonant
Higgs-boson exchanges. For example, in the focus-point region at high scalar
masses $m_0$ the cross section is dominated by heavy $CP$-even Higgs bosons
decaying into top quarks. Regions with important bottom quark final states include
those with small gaugino masses $m_{1/2}$, where light $CP$-even Higgs-boson
exchanges dominate, and the $A^0$-funnel region at high values of $\tan\beta$,
where the bottom Yukawa coupling is enhanced and the annihilation proceeds through
pseudo-scalar Higgs-boson exchanges. At large $\tan\beta$, the bottom-quark
contribution can also become sizable in the focus-point region.

In this work, we study in detail numerically the annihilation of neutralinos into
top quark-antiquark pairs through the exchange of $Z^0$-bosons in the $s$-channel
and of top squarks in the $t$- and $u$-channels (see Fig.\ \ref{figborn}). These
channels can be enhanced in models, where some of the unification conditions have
been relaxed, which is well motivated theoretically \cite{Anderson:1996bg,%
Anderson:1999uia,Olechowski:1994gm}. To be concrete, we focus on two sets of
models based on the SO(10) Grand Unified Theory (GUT): models with non-universal
Higgs masses (NUHM), and models without gaugino mass unification. SO(10) theories
are particularly promising, as they involve complete {\bf 16}-dimensional matter
multiplets with a right-handed neutrino and can be embedded in string theories
involving larger groups like E$_8$ or SO(32) \cite{Ratz2007,Nilles2008}.

\subsection{SUSY models with non-universal Higgs masses \label{sec3A}}
In SO(10) SUSY GUTs, the matter superfields of one generation belonging to a
{\bf 16}-dimensional representation are completely mass degenerate, if the
SUSY-breaking masses are acquired above the SO(10)-breaking scale.
Flavor-blind mechanisms can furthermore lead to universal masses of all matter
scalars. However, if the Higgs doublets $H_u$ and $H_d$ belong to different,
e.g.\ {\bf 10}-dimensional, representations, the corresponding
SUSY-breaking masses need not be the same, i.e.\ the Higgs masses need not be
universal \cite{Baer:2005bu,Ellis2002,Ellis2002b,Ellis2002c}.
In the scalar part of the general MSSM Lagrangian,
\begin{eqnarray}\nonumber
{\cal L}_{\rm soft} ~\subset~ 	&& -\left ( \widetilde {\overline u} \,{\bf a_u}\, \widetilde Q H_u
	- \widetilde {\overline d} \,{\bf a_d}\, \widetilde Q H_d
	- \widetilde {\overline e} \,{\bf a_e}\, \widetilde L H_d
	+ {\rm h.c.}\right ) 
\\ \nonumber	
&&  -\widetilde Q^\dagger \, {\bf m^2_{Q}}\, \widetilde Q
	-\widetilde L^\dagger \,{\bf m^2_{L}}\,\widetilde L
	-\widetilde {\overline u} \,{\bf m^2_{{\overline u}}}\, {\widetilde {\overline u}}^\dagger
	-\widetilde {\overline d} \,{\bf m^2_{{\overline d}}} \, {\widetilde {\overline d}}^\dagger
	-\widetilde {\overline e} \,{\bf m^2_{{\overline e}}}\, {\widetilde {\overline e}}^\dagger
\\
&&
- \, m_{H_u}^2 H_u^* H_u - m_{H_d}^2 H_d^* H_d
- \big( b H_u H_d + {\rm h.c.} \big)\,,
\end{eqnarray}
the trilinear scalar couplings ${\bf a_i}$ and the SUSY-breaking scalar masses
${\bf m^2_i}$ are still unified to $A_0$ and $m_0$ at the GUT scale,
but the SUSY-breaking Higgs mass parameters $m_{H_u}$ and $m_{H_d}$ are in
general different. Models with non-universal Higgs masses (NUHM) are therefore
defined by the parameters $m_0$, $m_{1/2}$, $A_0$, $\tan\beta$,
${\rm sgn}\,(\mu)$, $m_{H_u}$, and $m_{H_d}$. Note that in our NUHM models the
gaugino masses remain unified to $m_{1/2}$ and electroweak symmetry-breaking
(EWSB) is still achieved radiatively, albeit through modified renormalization
group equations (RGEs). This leads to a more constrained parameter space in the
$m_0-m_{1/2}$ plane as compared to mSUGRA models and to the fact that the minimum
conditions of the tree-level Higgs potential
\begin{eqnarray}
	\sin\beta &=& \frac{-2b}{m_{H_u}^2+m_{H_d}^2+2\mu^2}\,,\\ \label{murel}
	\frac{m_Z^2}{2} &=& \frac{m_{H_d}^2-m_{H_u}^2\tan^2\beta}{\tan^2\beta-1}-\mu^2\,, \ {\rm and}\\
	m_A^2 &=& m_{H_u}^2 + m_{H_d}^2 + 2\mu^2
\end{eqnarray}
allow to replace the parameter $b$ by $\tan\beta$, but not to determine the
superpotential parameter $\mu$ and the pseudo-scalar Higgs boson mass $m_A$ as
functions of $m_0$ and $m_{1/2}$, as was the case in mSUGRA. It is possible to
replace the free parameters $m_{H_u}$ and $m_{H_d}$ by the low-scale parameters
$\mu$ and $m_A$. Although we don't make use of this fact in our analysis, it can
be useful to consider $\mu$ and $m_A$ as free parameters of the model, since
they strongly influence the annihilation of neutralinos into
quarks and especially the relative weight of each channel. Having $m_A$ as a free
parameter means that one can find scenarios, where $m_A^2\simeq 2m_{\chi}^2$
and the Higgs-boson exchange dominates the cross section even for smaller values
of $\tan\beta$. In mSUGRA, such a scenario was only allowed for values of $\tan
\beta\gtrsim 40$. We study this case by choosing the benchmark point I given in
Tab.\ \ref{tab1}. Furthermore, the higgsino parameter $\mu$ influences the
higgsino component of the neutralino. By making it
larger, one can enhance the contributions from the $Z^0$-boson
exchange. This can be achieved, as discussed in Ref.\ \cite{Baer:2005bu}, by
starting with large positive values of $m_{H_u}$ and $m_{H_d}$ at the GUT scale.
By virtue of the RGE evolution, $m_{H_u}^2$ is driven to small negative values,
while $m_{H_d}^2$ remains positive, which results in a small value of $\mu$ as
given by Eq.\ (\ref{murel}). This in turn gives rise to a larger higgsino fraction
of the neutralino and enhances the coupling to $Z^0$-bosons. Such a scenario
corresponds to our point II in Tab.\ \ref{tab1}. Finally, one can make use of the
NUHM RGEs to reduce the values of ${\bf m^2_{\overline u}}$  and
${\bf m^2_{\overline d}}$. This can be obtained by choosing a large and negative
difference $m_{H_u}^2 - m_{H_d}^2$. On top of that, by choosing $A_0$ to be large
and negative, one induces a larger splitting in the third-generation sfermion
masses, leading to a scenario (point III in Tab.\ \ref{tab1}) with a small top
squark mass and enhanced $t$- and $u$-channel exchanges of top squarks.

\begin{table}
  \caption{High-scale parameters together with the corresponding neutralino relic density, the contribution from top and bottom quark-antiquark final states to the annihilation cross section obtained with {\tt micrOMEGAs 2.1}, and the mass eigenvalues of the lightest neutralino and the lightest stop for our three selected NUHM scenarios.}
  \label{tab1}
  \begin{tabular}{c|ccccc|cc|ccc|cc|}
    & $m_0$ [GeV] & $M_2$ [GeV] & $A_0$ [GeV] & $\tan\beta$ & $\rm{sgn}(\mu)$ & $m_{H_{u}}$ [GeV] & $m_{H_{d}}$ [GeV] & $\Omega_{\rm CDM}h^2$ & $t\bar{t}$ & $b\bar{b}$ & $m_{\tilde{\chi}_1^0}$ (GeV) & $m_{\tilde{t}_1}$ (GeV) \\
    \hline
    I & 500 & 500 & 0 & 10 & + & 1500 & 1000 & 0.118 & 21.0\% & 64.0\% & 207.2 & 606.4 \\
    II  & 620 & 580 & 0 & 10 & + & 1020 & 1020 & 0.118 & 51.0\% & -- & 223.7 & 923.8 \\
    III   & 500 & 500 & -1200 & 10 & + & 1250 & 2290 & 0.113 & 93.4\% & -- & 200.7 & 259.3 \\
    \hline
  \end{tabular}
\end{table}
\subsection{SUSY models without gaugino mass unification \label{sec3B}}
In the gaugino sector, the situation is somewhat similar to the one in Sec.\
\ref{sec3A}. The breaking of the SO(10) symmetry can proceed via a step
involving its SU(5) subgroup. The SU(5) later breaks into the Standard Model
gauge groups SU(3)$\times$SU(2)$\times$U(1) \cite{Ellis1985} and also
determines the properties of the SUSY-breaking mechanism. As pointed out
in Refs.\ \cite{Anderson:1996bg,Anderson:1999uia}, the breaking of supersymmetry itself
is induced by an $F$-term, while the gaugino masses are generated through a chiral
superfield $\Phi$, whose auxiliary component $F_{\Phi}$ acquires a vacuum
expectation value. Assuming that the gauginos belong to the adjoint representation
{\bf 24} of SU(5), the fields $\Phi$ and $F_{\Phi}$ can in principle belong to
any of the irreducible representations appearing in the symmetric product
\begin{equation}
  \big( {\bf 24}\otimes{\bf 24} \big)_{\rm sym} ~=~ {\bf 1} \oplus {\bf 24} \oplus {\bf 75} \oplus {\bf 200},
\end{equation}
or any linear combination thereof. The relations between the gaugino masses $M_i$ at the
unification scale are given by the embedding coefficients of the Standard Model
groups in SU(5). Note that these possibilities are all compatible with gauge
coupling unification, but only the case where the SUSY-breaking field $F_{\Phi}$
is taken to be a pure singlet ({\bf 1}) leads to the gaugino mass universality
featured by the mSUGRA model.

The soft SUSY-breaking Lagrangian contains mass terms for the $B$-ino, $W$-ino, and gluino,
\begin{equation}
  {\cal L}_{\rm soft} ~\subset~ -\frac{1}{2} \Big( M_1 \tilde{B}\tilde{B} + M_2 \tilde{W}\tilde{W} + M_3 \tilde{g}\tilde{g} + {\rm h.c.} \Big) .
\end{equation}
As argued above, the values of $M_1$, $M_2$, and $M_3$ at the unification scale can be considered as independent parameters. 
Here, we adopt a commonly used parametrization and introduce the dimensionless parameters
\begin{equation}
  x_1 ~=~ \frac{M_1}{M_2} \qquad {\rm and}\qquad x_3 ~=~ \frac{M_3}{M_2},
\end{equation}
which will be used together with the $W$-ino mass parameter $M_2$ to describe the gaugino sector. The case $x_1=x_3=1$ reproduces the mSUGRA model with the five parameters $m_0$, $m_{1/2}$, $A_0$, $\tan\beta$, and $\rm{sgn}(\mu)$ at the unification scale.

\begin{table}
  \caption{High-scale parameters together with the corresponding neutralino relic density, the contribution from top quark-antiquark final states to the annihilation cross section obtained with {\tt micrOMEGAs 2.1}, and the mass eigenvalues of the lightest neutralino and the lightest stop for our two selected non-universal gaugino mass scenarios.}
  \label{tab2}
  \begin{tabular}{c|ccccc|cc|cc|cc|}
    & $m_0$ [GeV] & $M_2$ [GeV] & $A_0$ [GeV] & $\tan\beta$ & $\rm{sgn}(\mu)$ & $x_1$ & $x_3$ & $\Omega_{\rm CDM}h^2$ & $t\bar{t}$ & $m_{\tilde{\chi}_1^0}$ (GeV) & $m_{\tilde{t}_1}$ (GeV) \\
    \hline
    IV & 320 & 700 & -350 & 10 & + & 2/3 & 1/3 & 0.114 & 79.2\% & 183.4 & 281.9 \\
    V & 1500 & 600 & 0 & 10 & + & 1 & 4/9 & 0.104 & 50.4\% & 235.6 & 939.0 \\
    \hline
  \end{tabular}
\end{table}

Models with non-universal gaugino masses have been shown to favor annihilation processes where neutralino annihilation into quarks is mediated by a $Z^0$-boson exchange and squark exchanges \cite{Orloff2002, Martin2007, Baer2007}. It has also been shown that the key parameter in this context is the gluino mass $M_3$, since it influences practically all sectors of the low-energy mass spectrum through the renormalization group evolution. A decrease in $M_3$ induces a decrease of the mass squared $m_{H_u}^2$, which through electroweak symmetry breaking conditions induces a decrease in the higgsino mass parameter $\mu$ as well. This in turn increases the higgsino fraction of the neutralino and lowers the pseudo-scalar Higgs mass $m_A$. Moreover, having $M_3$ independent of the other gaugino mass parameters, it is straightforward to obtain lighter squarks, in particular the scalar tops. This effect can still be enhanced by decreasing the scalar mass parameter $m_0$ and by adjusting the trilinear coupling $A_0$, which influences the squark mass splitting. With scalar tops being light (even becoming eventually the next-to-lightest SUSY particle), the squark exchange dominates the cross section for a $B$-ino-like neutralino, where a low value of $\tan\beta$ suppresses the Higgs-boson exchange (our point IV in Tab.~\ref{tab2}). As the higgsino fraction of the lightest neutralino increases, the squark exchange is enhanced due to the large Yukawa couplings. In the case of a large higgsino fraction, also the $Z^0$-boson exchange becomes important and can take over if the lightest stop is not too close in mass to the lightest neutralino. We analyze the consequences of such a scenario by choosing the point V in Tab.~\ref{tab2}.

\section{Numerical Results and Discussion \label{sec4}}

Starting from the high-scale parameters, we use the public computer program
{\tt SPheno~2.2.3} \cite{Spheno} for the numerical evaluation of the
renormalization group running in order to obtain the SUSY-breaking parameters at
the electroweak scale. The relic density of the neutralino is then evaluated
numerically using the public code {\tt micrOMEGAs~2.1} \cite{micromegas}, where
we have included our calculation of the neutralino annihilation cross section
into third-generation quarks as discussed in Sec.~\ref{sec2}. For the Standard
Model input parameters, such as masses and couplings, we refer the reader to
Ref.~\cite{PDG2008}, except for the value of the top quark pole mass,
$m_{\rm top}=172.4$~GeV, which has been taken from Ref.~\cite{CDFD02008}.

We have chosen five typical parameter points shown in Tabs.~\ref{tab1} and
\ref{tab2}, that have a dominant neutralino annihilation into top or bottom quarks
through Higgs-boson, $Z^0$-boson or squark exchanges and whose relic density lies
reasonably close to the WMAP range of Eq.~(\ref{cWMAP}). In all scenarios we only
consider relatively low values of $\tan\beta=10$. In Sec.\ \ref{sec4A} we analyze
the first three parameter points (Tab.~\ref{tab1}) where we make use of the
possible non-universality of the Higgs-boson masses to construct scenarios with
cross sections dominated by Higgs-boson, $Z^0$-boson, and squark exchanges,
respectively. In Sec.\ \ref{sec4B}, we investigate the point IV, which corresponds
to one of the scenarios without gaugino mass unification discussed in
Ref.~\cite{Orloff2002}, and the point V, which is motivated by one of the
``compressed SUSY'' scenarios proposed in Ref.~\cite{Martin2007}.
Our chosen parameter points also satisfy electroweak precision and low-energy
constraints such as the measurements of the $\rho$-parameter, the anomalous
magnetic moment of the muon, and the branching ratio of the decay $b\rightarrow
s\gamma$. Due to the large experimental error on the measurement of $\Delta\rho$
\cite{PDG2008}, however, only regions featuring very high SUSY masses are excluded
at the $2\sigma$-level, so that this constraint does not affect our analysis. We
have taken into account the anomalous magnetic moment of the muon by taking the
higgsino mass parameter $\mu>0$. Negative values are disfavored, since they
increase the gap between the recent experimental value and the Standard Model 
prediction \cite{Moroi1995}. 
The most stringent constraint is given by the inclusive branching ratio of the
decay $b\to s\gamma$. Recent experimental measurements from BaBar, Belle, and CLEO
lead to the combined value \cite{BSGamma}
\begin{equation}
  {\rm BR}(b\to s\gamma) ~=~ \big(3.52\pm 0.25 \big)\cdot 10^{-4} .
\label{eqBSG}
\end{equation}
The theoretical prediction of the SUSY contribution  is particularly sensitive to
the masses of the chargino, the charged Higgs boson, and the lightest scalar
quark. We have verified that our parameter points lead to values that lie within
$2\sigma$ of the above limit using the public codes {\tt FeynHiggs 2.6.5}
\cite{FeynHiggs} and {\tt SusyBSG 1.3} \cite{SusyBSG}. Note that also the direct
mass limits from collider searches are fulfilled \cite{PDG2008}. 

Although the focus of the present paper is the improvement of the annihilation
cross section through SUSY-QCD corrections, one should keep in mind that there are
also other sources of uncertainties in the calculation of the relic density. From
the particle physics side, it is well known that differences in the low-energy
mass spectrum may occur when using different spectrum generators. This, in
consequence, may induce a sizable difference in the prediction of the dark matter
relic density or other observables \cite{Belanger2005}. Note that, in this
context, also the numerical value of the Standard Model parameters, in particular
the top-quark mass, can influence the favored regions of parameter space,
in particular in the case of dominant annihilation into top-quark final states.
Corrections to the annihilation cross section are also induced by the electroweak
interaction \cite{Baro2007}, but these are generally smaller than the strong
corrections and beyond the scope of this work. Concerning cosmology, it has been
shown that modifications of the standard cosmological model may also affect the
prediction of the dark matter relic density. Including, e.g., an energy content
such as quintessence or an effective dark density due to extra dimensions modifies
the expansion rate or the entropy density of the early universe (see e.g.\
\cite{Arbey2009}) and thus enters into the calculation of the relic density.

In the left panels of Figs.~\ref{figxsec1}, \ref{figxsec23} and \ref{figxsec45},
we show the cross section for the annihilation of a neutralino pair into a bottom
and/or top quark-antiquark pair as a function of the center-of-momentum energy
$p_{\rm cm}$ calculated at the tree-level with $\overline{\rm DR}$ Yukawa
couplings (solid lines) as well as the leading contributions from the individual
annihilation channels (dashed lines) and their interferences (dotted lines). Note
that some of the latter have been multiplied by $(-1)$ in order to fit into the
logarithmic plot. We also show, in arbitrary units, the thermal velocity
distribution function, involved in the calculation of the thermal average
evaluated at the freeze-out temperature (shaded areas). 

In the right panels of Figs.~\ref{figxsec1}, \ref{figxsec23} and \ref{figxsec45},
we show again the total annihilation cross section into bottom and/or top
quark-antiquark pairs at different levels of precision, i.e.\ at leading order
with $\overline{\rm DR}$ couplings (dash-dotted lines), in the approximation
included in the {\tt micrOMEGAs} code (dashed lines), which uses effective
couplings to absorb the leading loop effects, and with our full one-loop QCD and
SUSY-QCD corrections (solid lines). The shaded area indicates again the thermal
velocity distribution of the neutralinos at the freeze-out temperature in
arbitrary units.

In order to generalize our results, the remaining Figs.~\ref{figscan1},
\ref{figscan2}, \ref{figscan3}, \ref{figscan4}, and \ref{figscan5} show scans in
various two parameter planes around our parameter points in Tabs.~\ref{tab1} and
\ref{tab2} and display the contours allowed by WMAP calculated at tree-level
(green), with the approximation implemented in {\tt micrOMEGAs} (red), and with
our full SUSY QCD corrections (blue). We also show dependence of the relic
density on various physical parameters, e.g.\ the mass of the lightest neutralino.
\subsection{Relic density in models with non-universal Higgs masses \label{sec4A}}
\begin{figure}
  \begin{center}
    \includegraphics[scale=0.4]{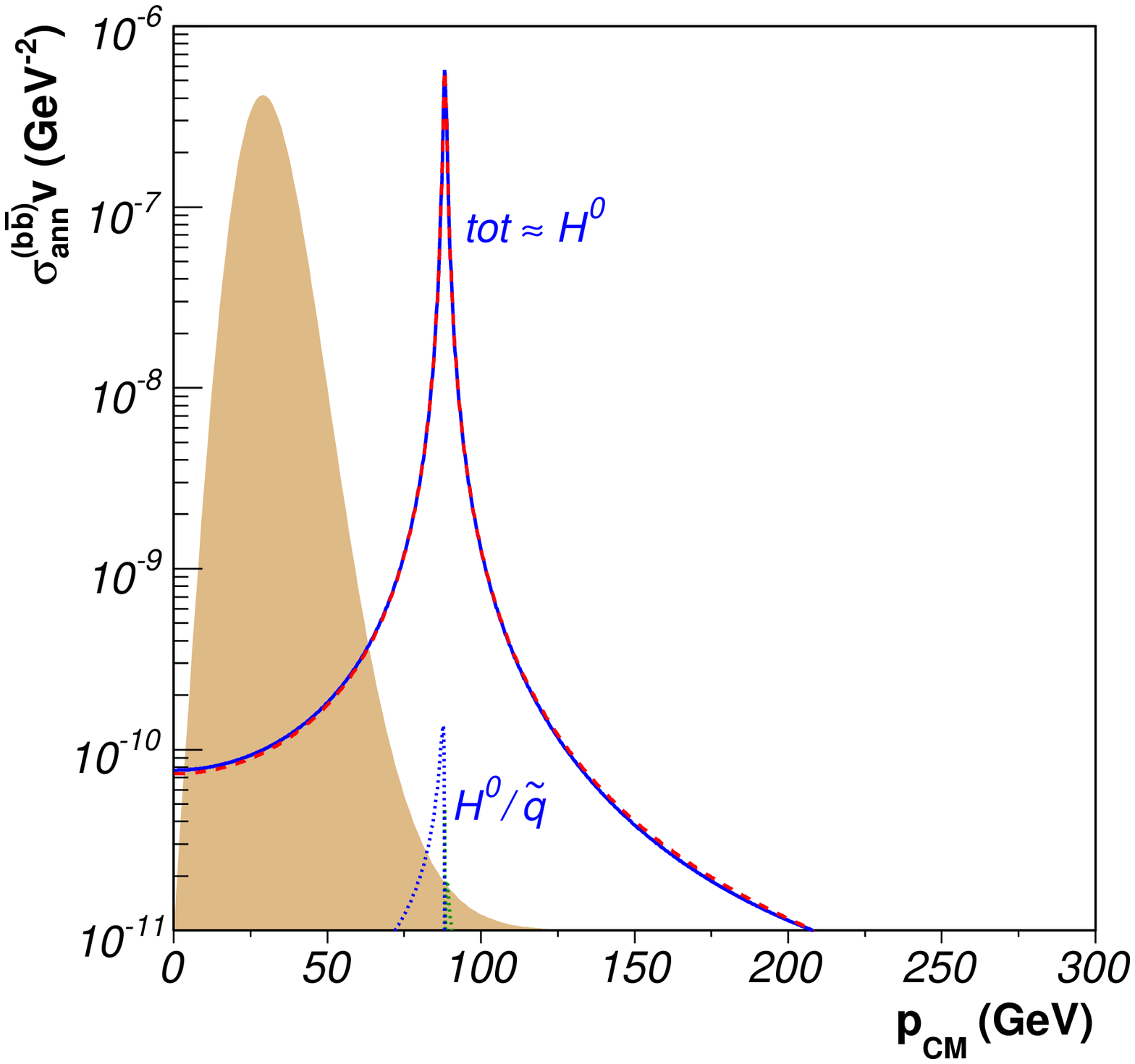}
    \includegraphics[scale=0.4]{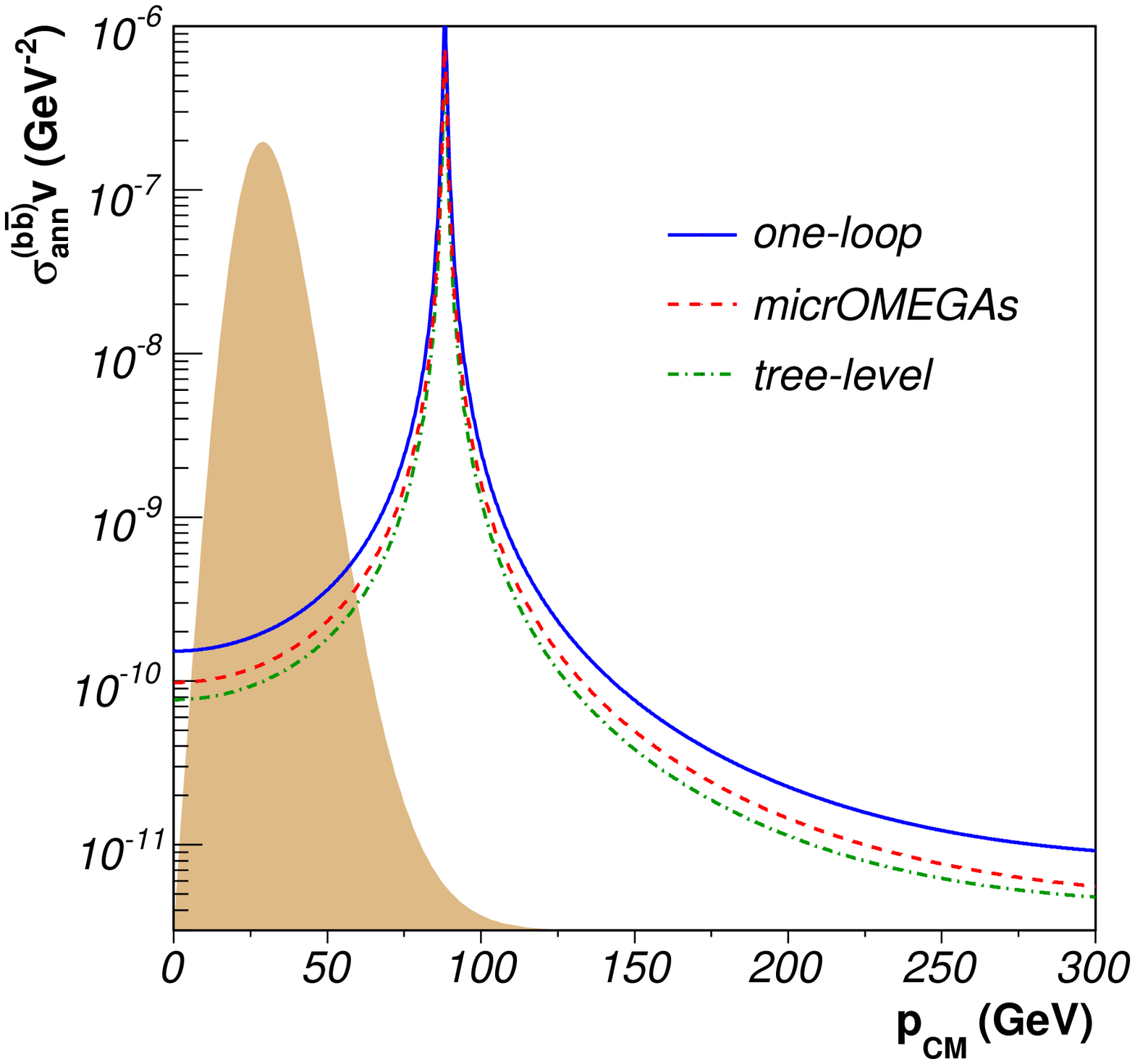}
    \includegraphics[scale=0.4]{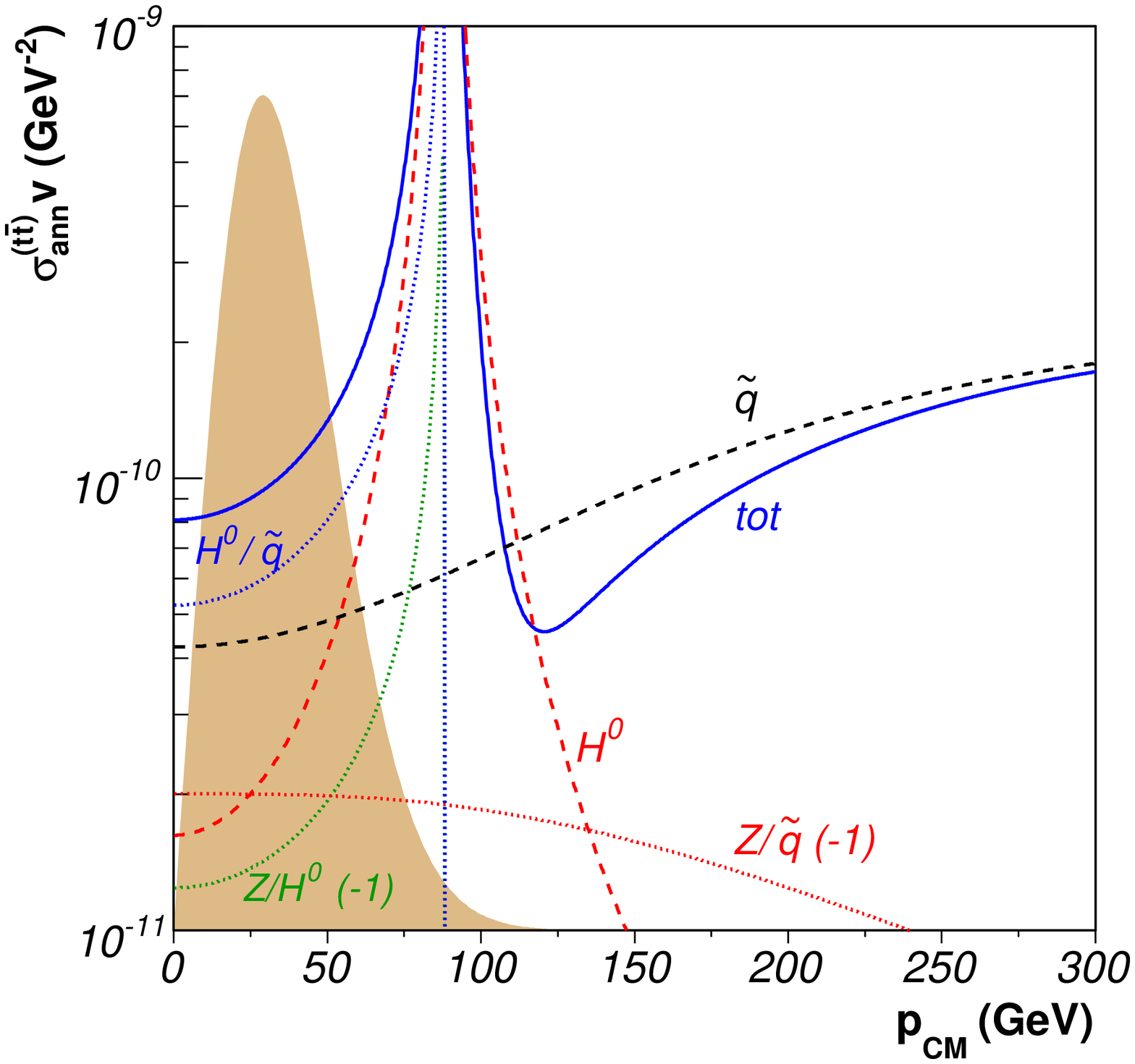}
    \includegraphics[scale=0.4]{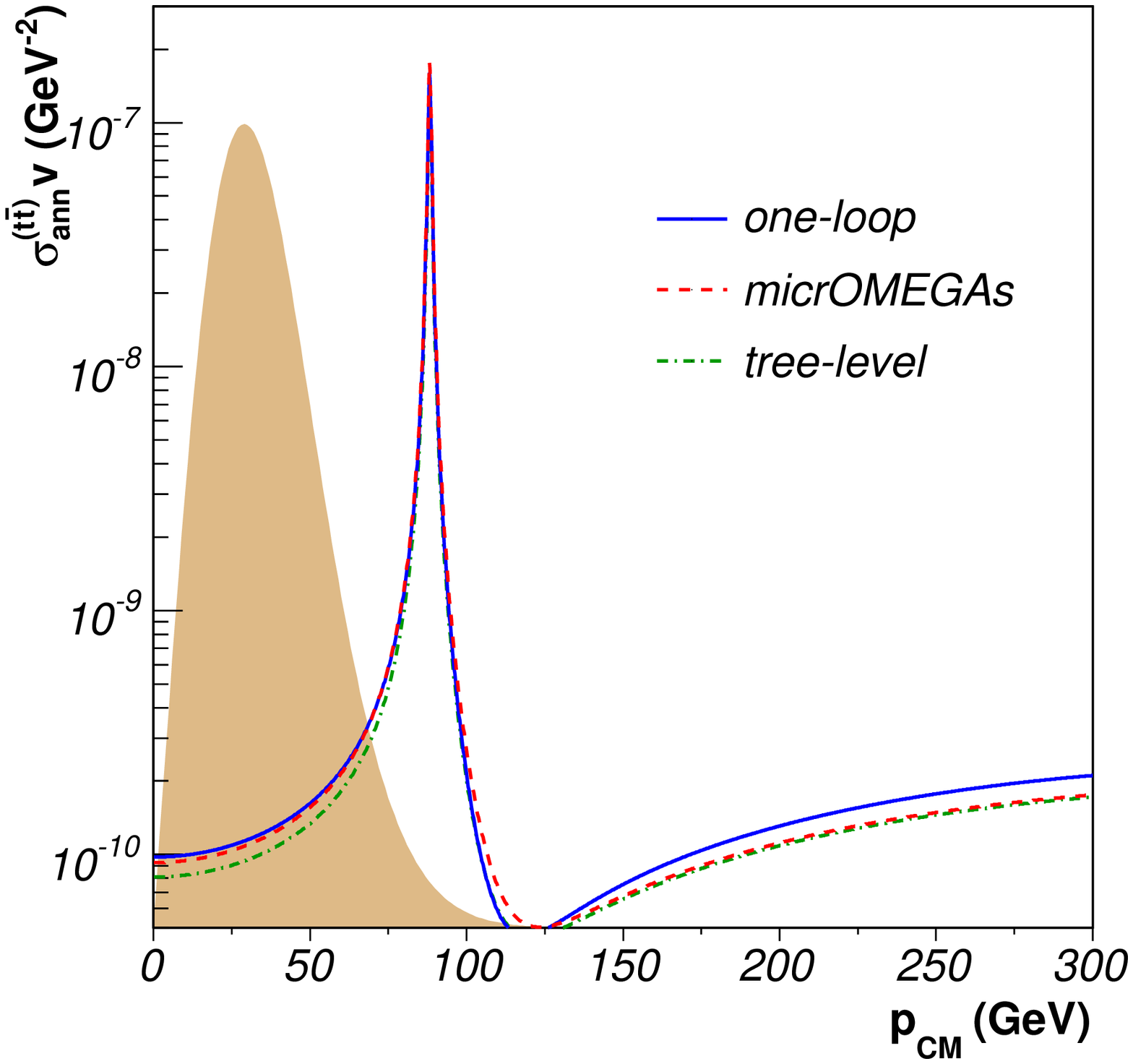}
  \end{center}
  \caption{The contributions of the different diagrams to the annihilation cross section of a neutralino pair into bottom (above) and top (below) quark-antiquark pairs (left) and the effect of the radiative corrections on the annihilation cross sections (right) as a function of the center-of-momentum energy $p_{\rm cm}$ for our parameter point I. The shaded area indicates the velocity distribution of the neutralino at the freeze-out temperature in arbitrary units.}
  \label{figxsec1}
\end{figure}

\begin{figure}
  \begin{center}
    \includegraphics[scale=0.4]{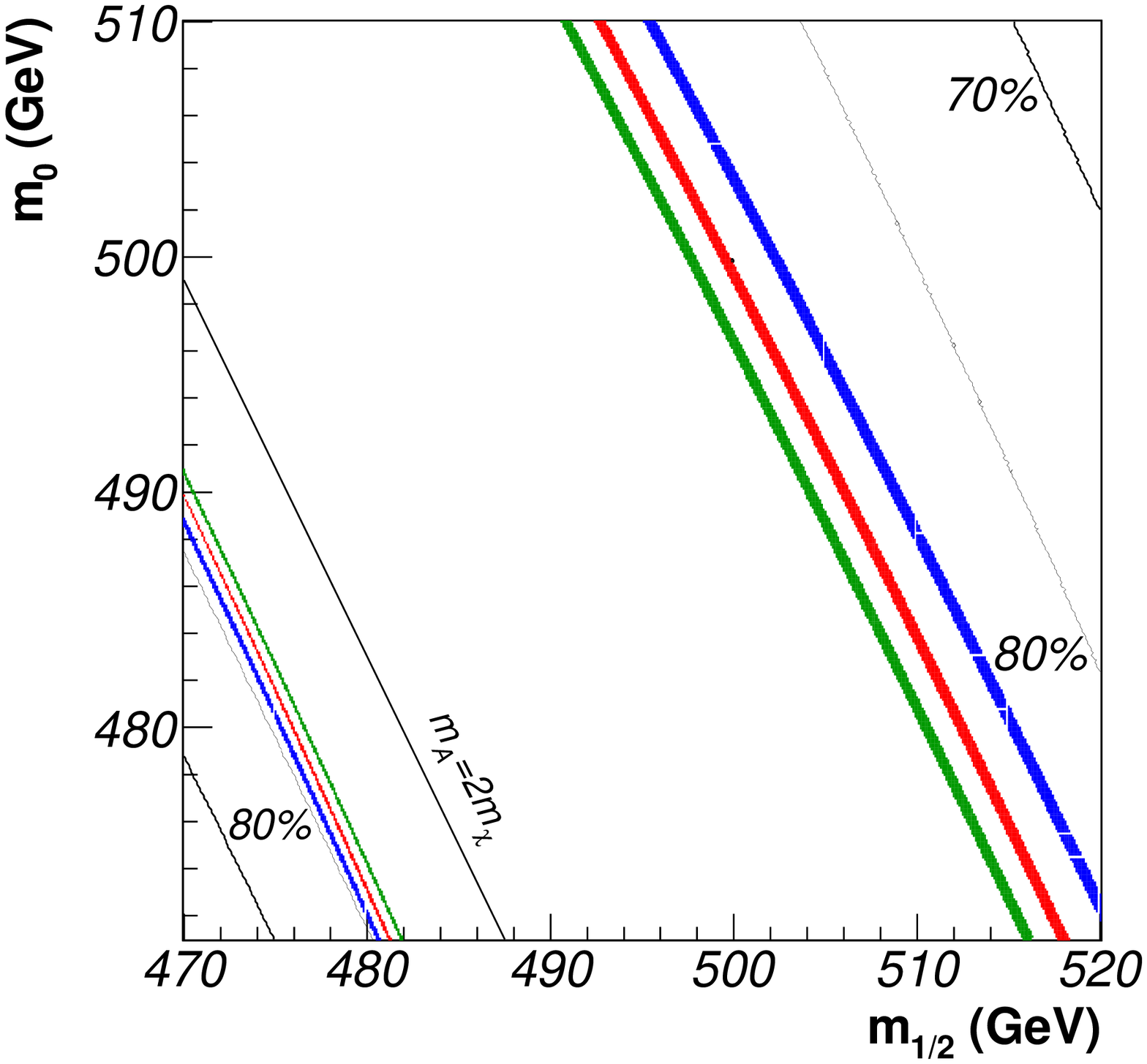}
    \includegraphics[scale=0.4]{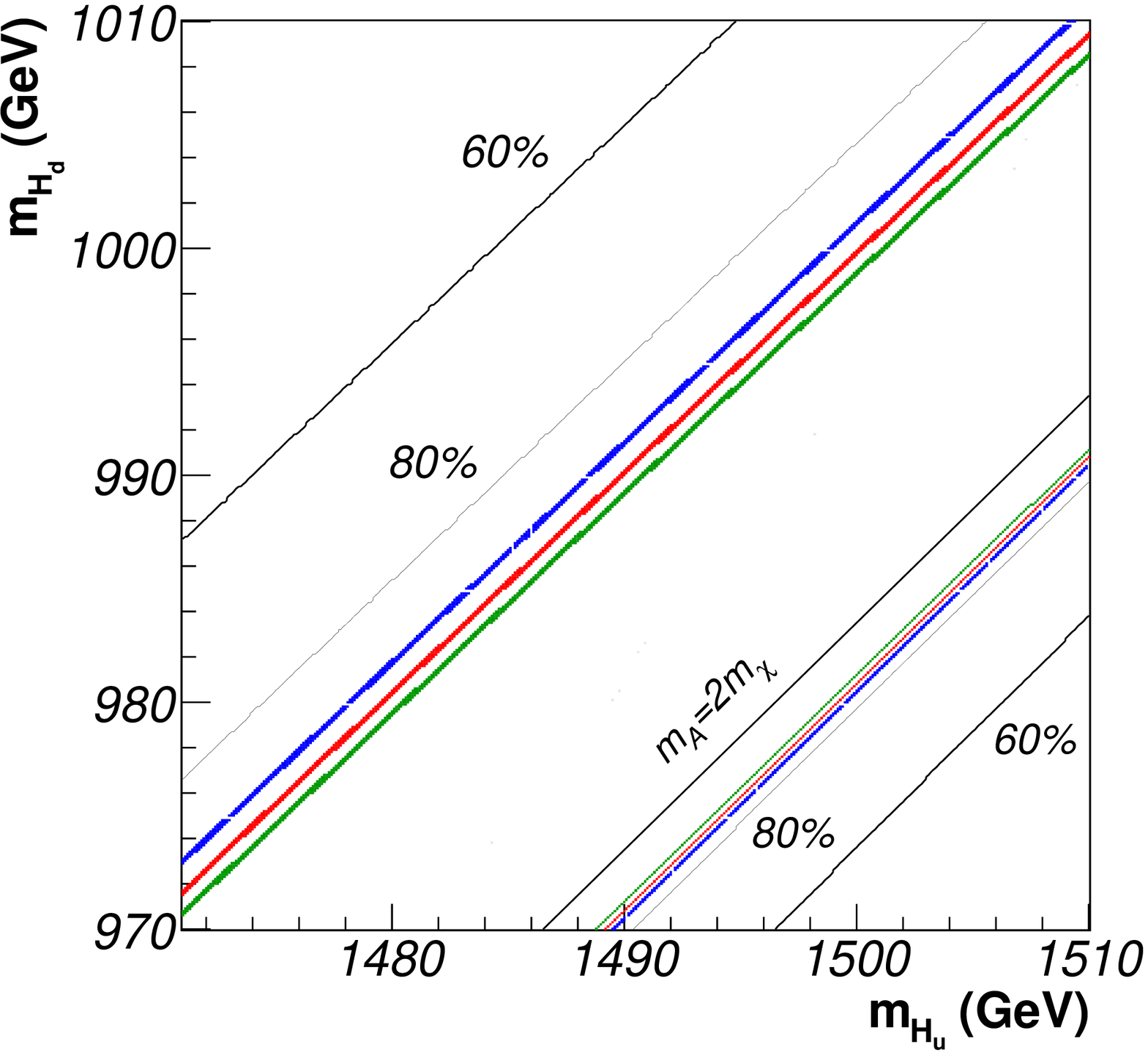}
    \includegraphics[scale=0.4]{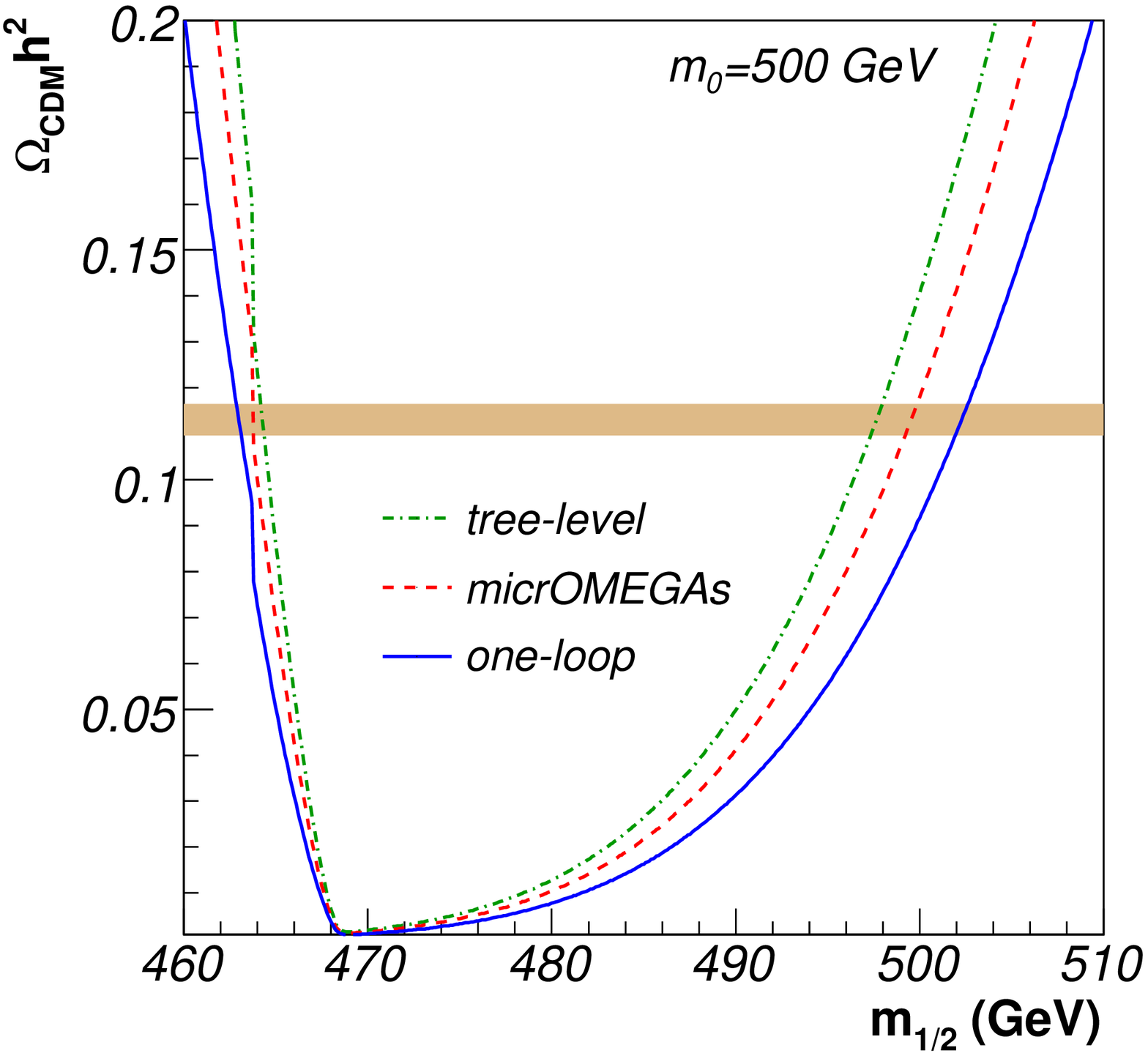}
    \includegraphics[scale=0.4]{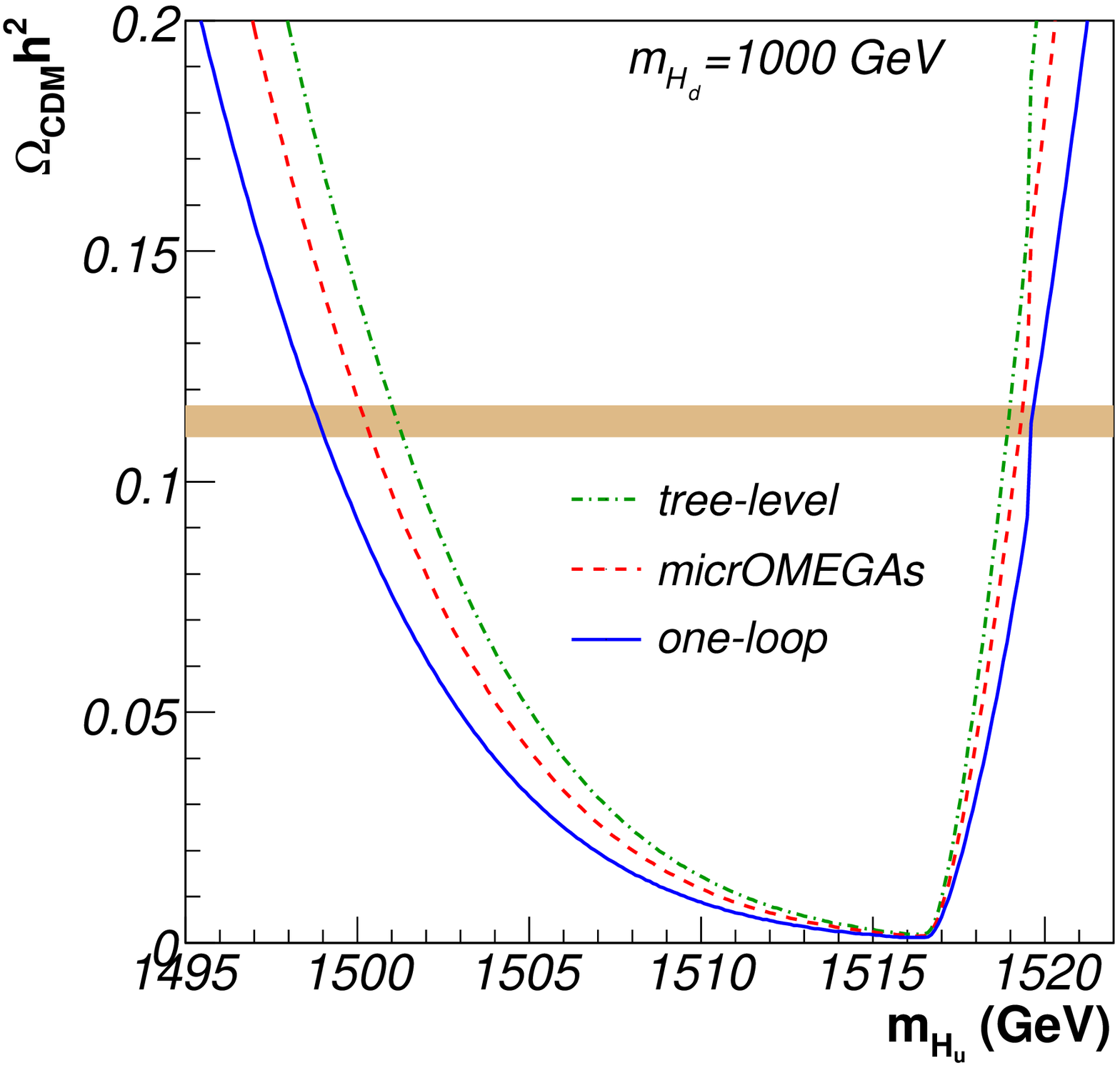}
  \end{center}
  \caption{Top: Cosmologically favored regions in the $m_0$-$m_{1/2}$-plane (top left) and $m_{H_u}$-$m_{H_d}$-plane (top right) for scans around our parameter point I. We show the regions that satisfy the constraint from Eq.~(\ref{cWMAP}) for our tree-level calculation (green), the calculation implemented in {\tt micrOMEGAs} (red), and our calculation including the full SUSY-QCD corrections (blue). We also indicate the contributions from quark-antiquark final states to the total annihilation cross section by isolines. 
Bottom: The prediction of the neutralino relic density $\Omega_{\rm CDM}h^2$ including the tree-level (green dash-dotted) cross section, the approximation included in {\tt micrOMEGAs} (red dashed), and the full one-loop SUSY-QCD corrected cross-section (blue solid) as a function of the gaugino mass parameter $m_{1/2}$ (left) and as a function of $m_{H_u}$ parameter (right). The shaded area indicates the favored region of Eq.~(\ref{cWMAP}).}
  \label{figscan1}
\end{figure}

\begin{figure}
  \begin{center}
    \includegraphics[scale=0.4]{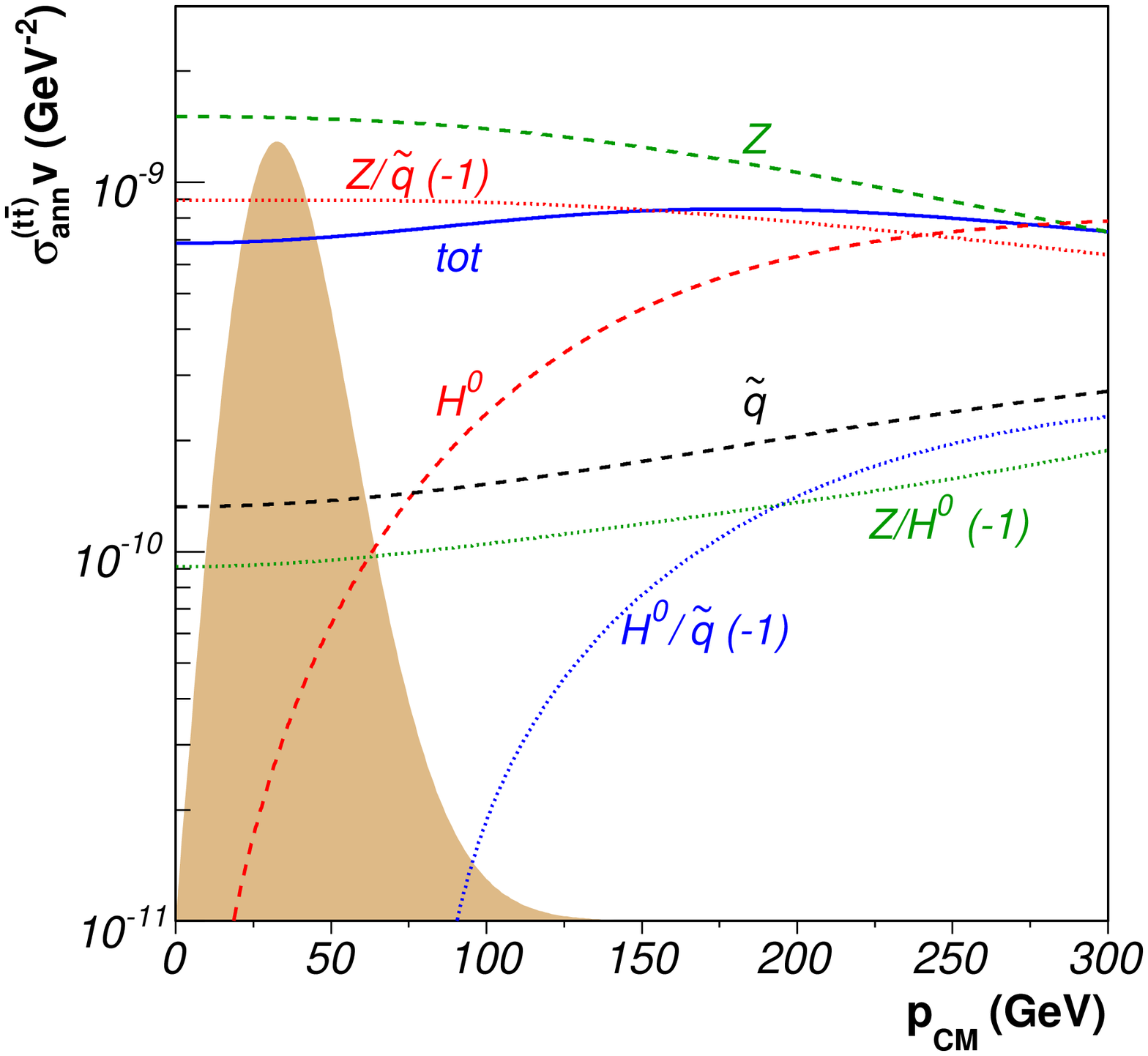}
    \includegraphics[scale=0.4]{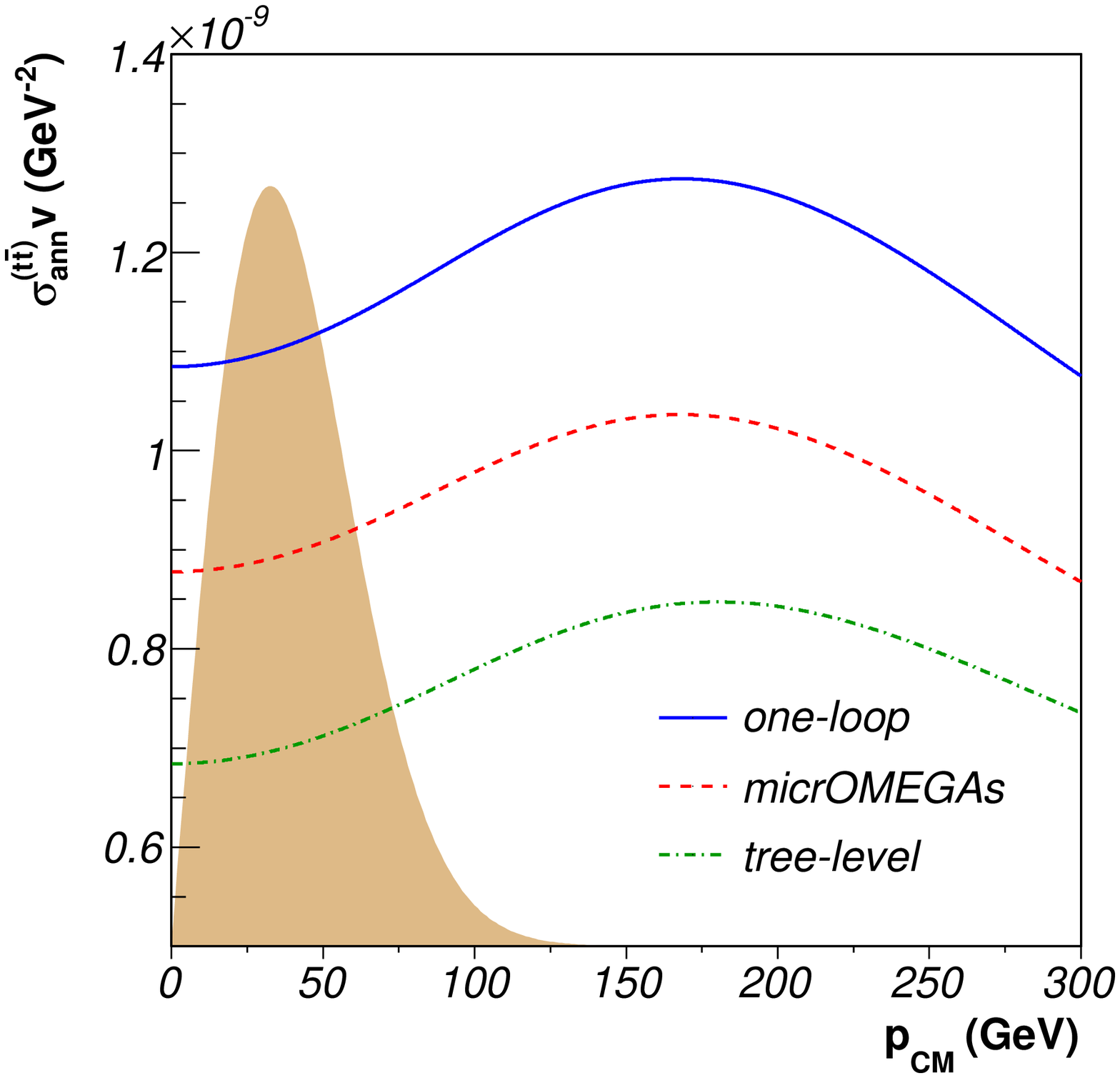}
    \includegraphics[scale=0.4]{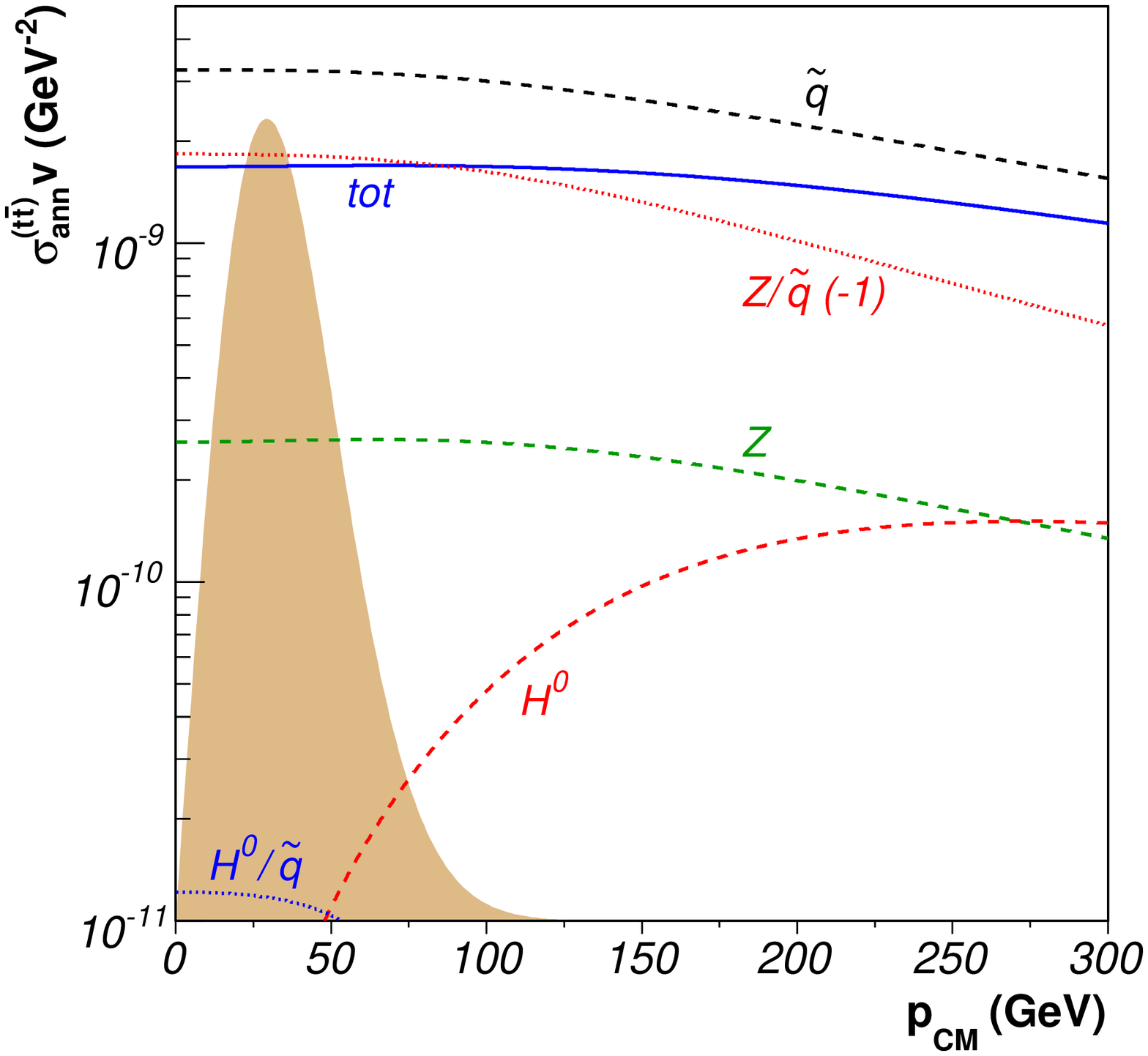}
    \includegraphics[scale=0.4]{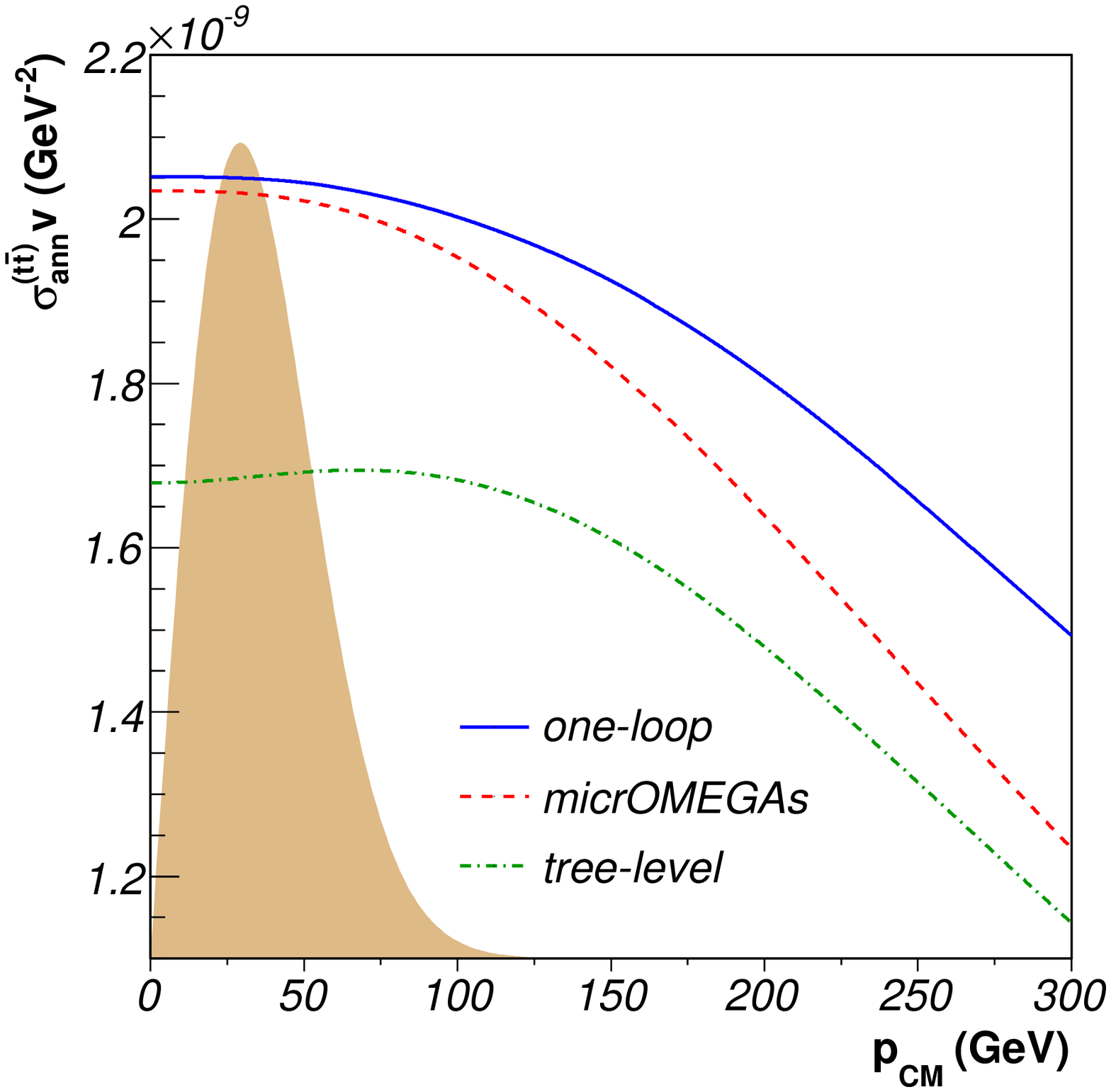}
  \end{center}
  \caption{The contributions of the different diagrams to the annihilation cross section of a neutralino pair into top quark-antiquark pairs (left) and the effect of the radiative corrections on the annihilation cross section (right) as a function of the center-of-momentum energy $p_{\rm cm}$ for our parameter points II (above) and III (below). The shaded area indicates the velocity distribution of the neutralino at the freeze-out temperature in arbitrary units.}
  \label{figxsec23}
\end{figure}

We begin our detailed numerical discussion by analyzing a scenario, where the annihilation cross section into quarks is dominated by an exchange of a Higgs boson. This scenario corresponds to our parameter point I from Tab.~\ref{tab1}. Here the heavy CP-even and the CP-odd Higgs boson resonances coincide with both Higgs boson masses at $450.3$ GeV. Due to the dominance of the Higgs-boson exchange, the neutralino annihilates predominantly into bottom quarks. This is a consequence of the fact that, although the bottom quarks have a much smaller mass compared to the top quarks, their couplings to the Higgs bosons are enhanced by $\tan\beta$ or $\cos\alpha$. As opposed to mSUGRA, the neutralino in this scenario is rather heavy, which allows for an effective annihilation into top quarks as well. In the bottom panels of Fig.~\ref{figscan1}, a small discontinuity indicates the place, where the top quark final state starts to contribute. These circumstances result in a mixture of top and bottom quark final states, which cannot be reached in mSUGRA for such a low value of $\tan\beta$. 
\begin{figure}
  \begin{center}
    \includegraphics[scale=0.4]{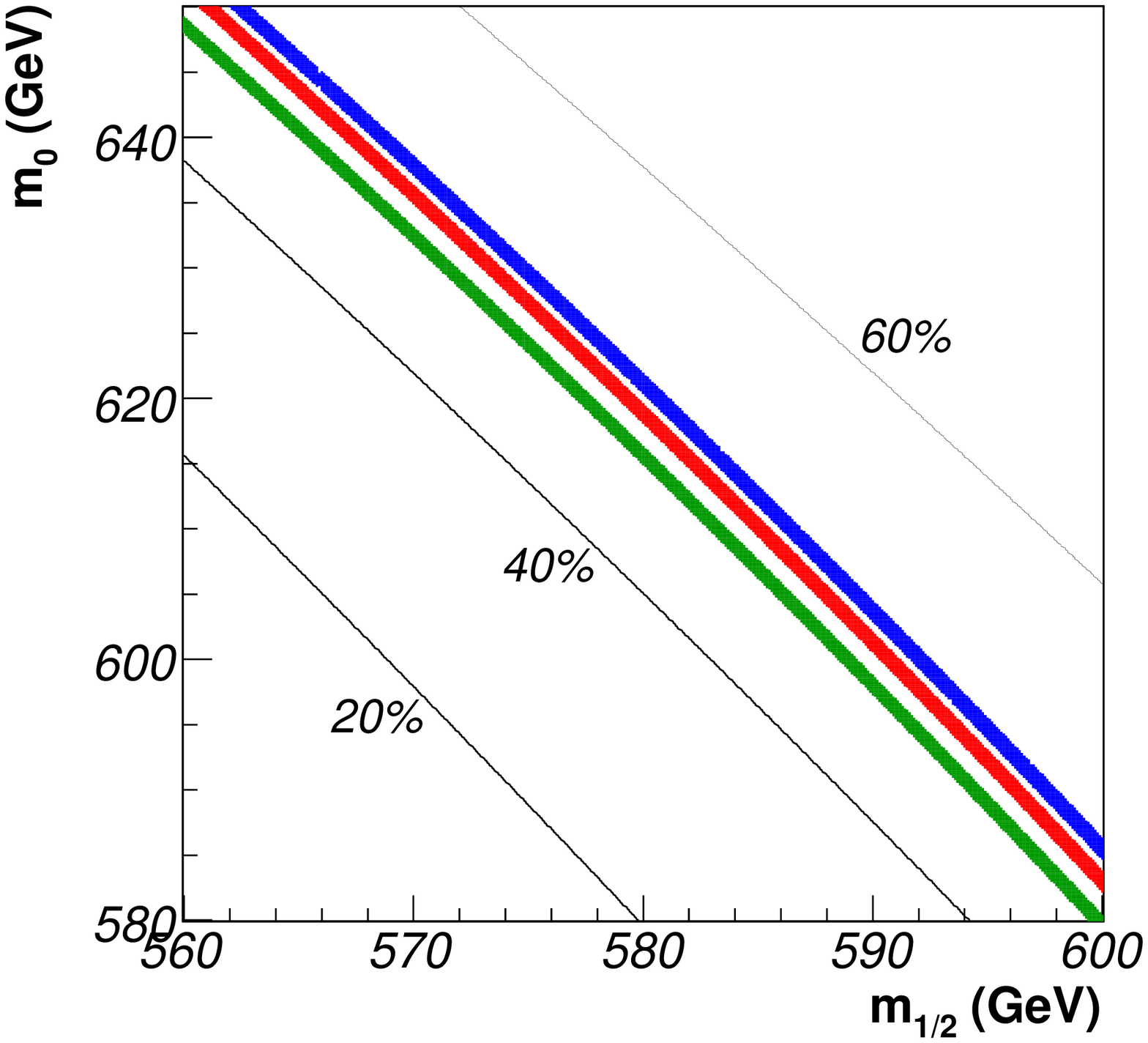}
    \includegraphics[scale=0.4]{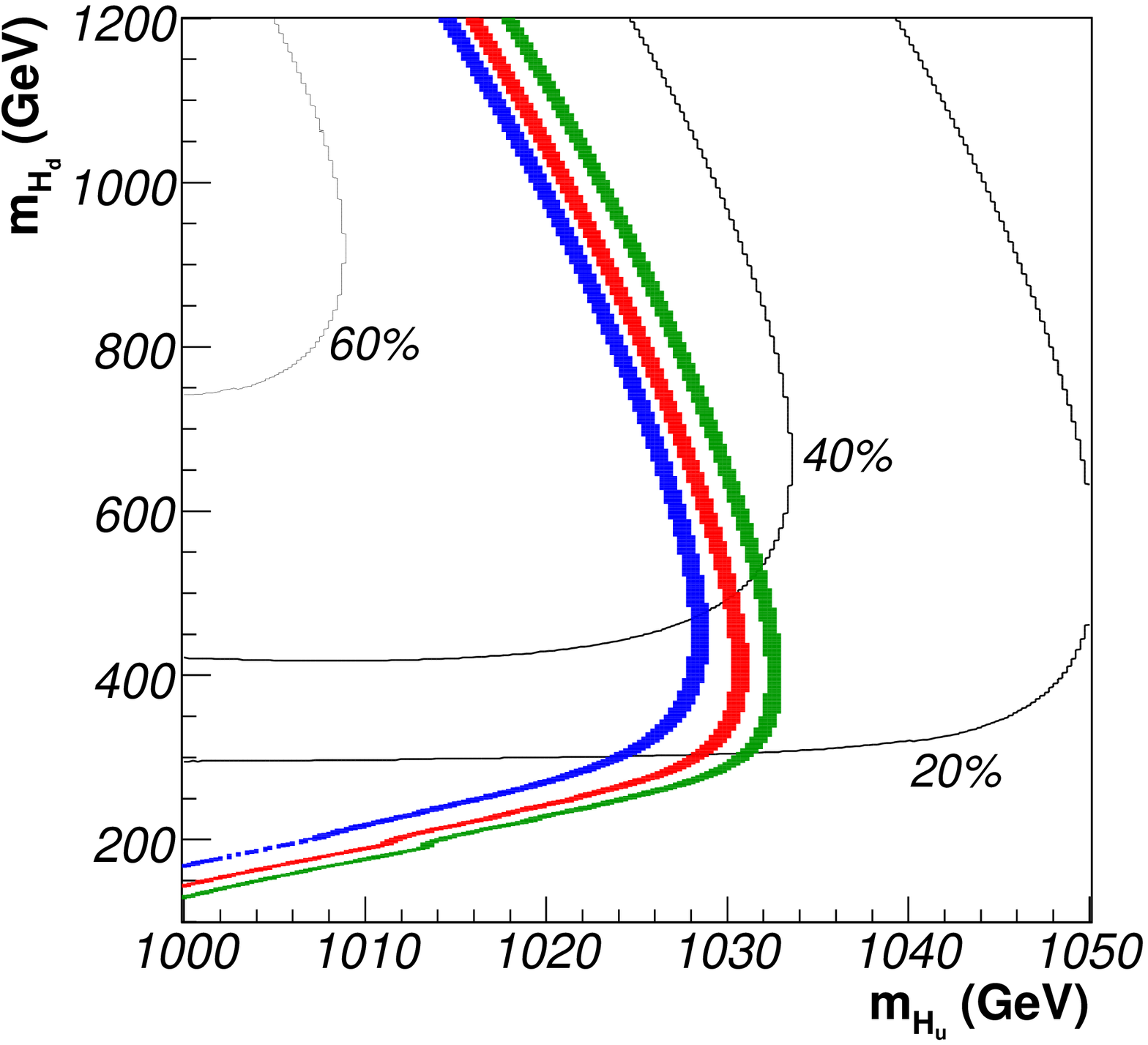}
    \includegraphics[scale=0.4]{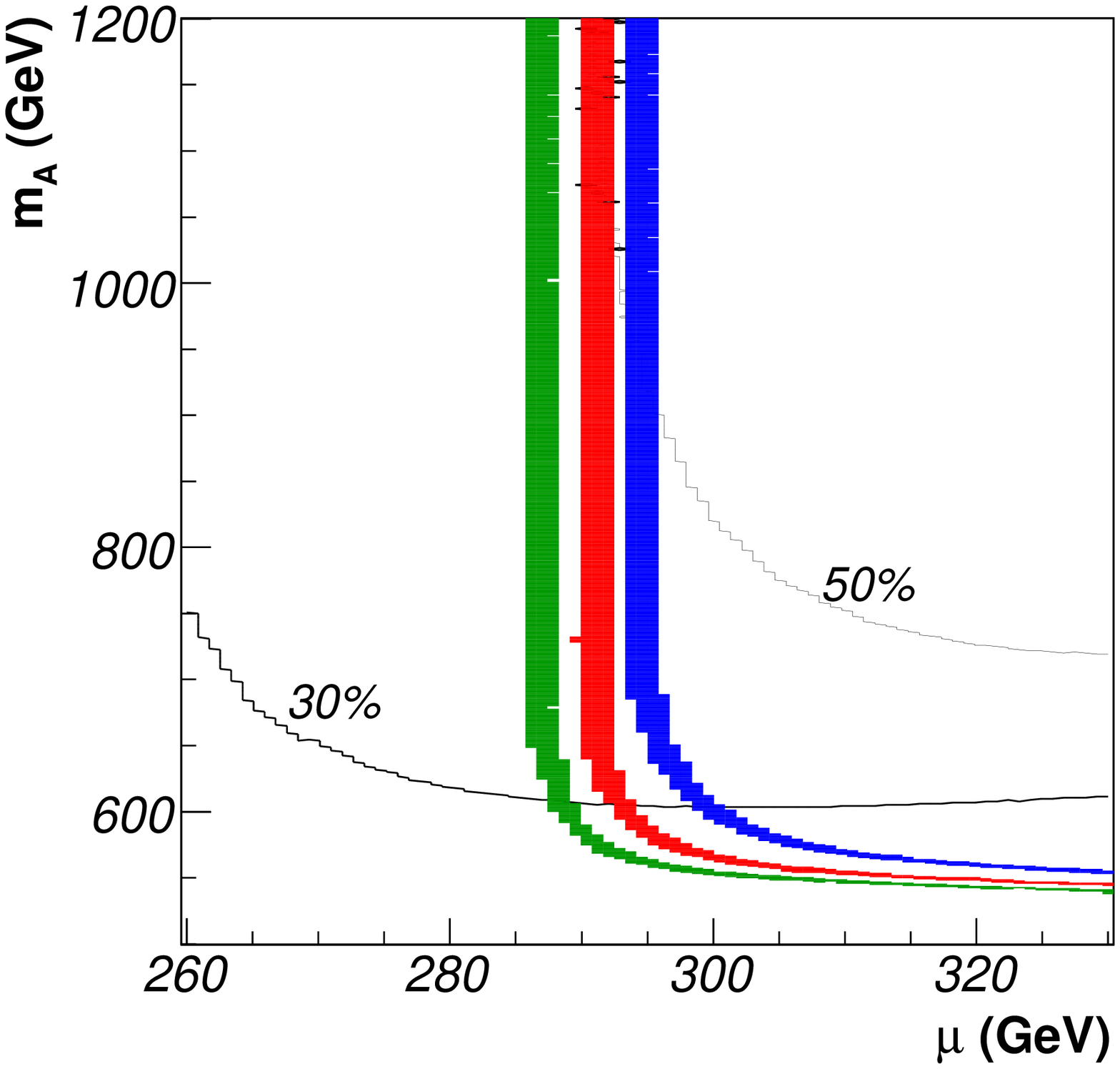}
    \includegraphics[scale=0.4]{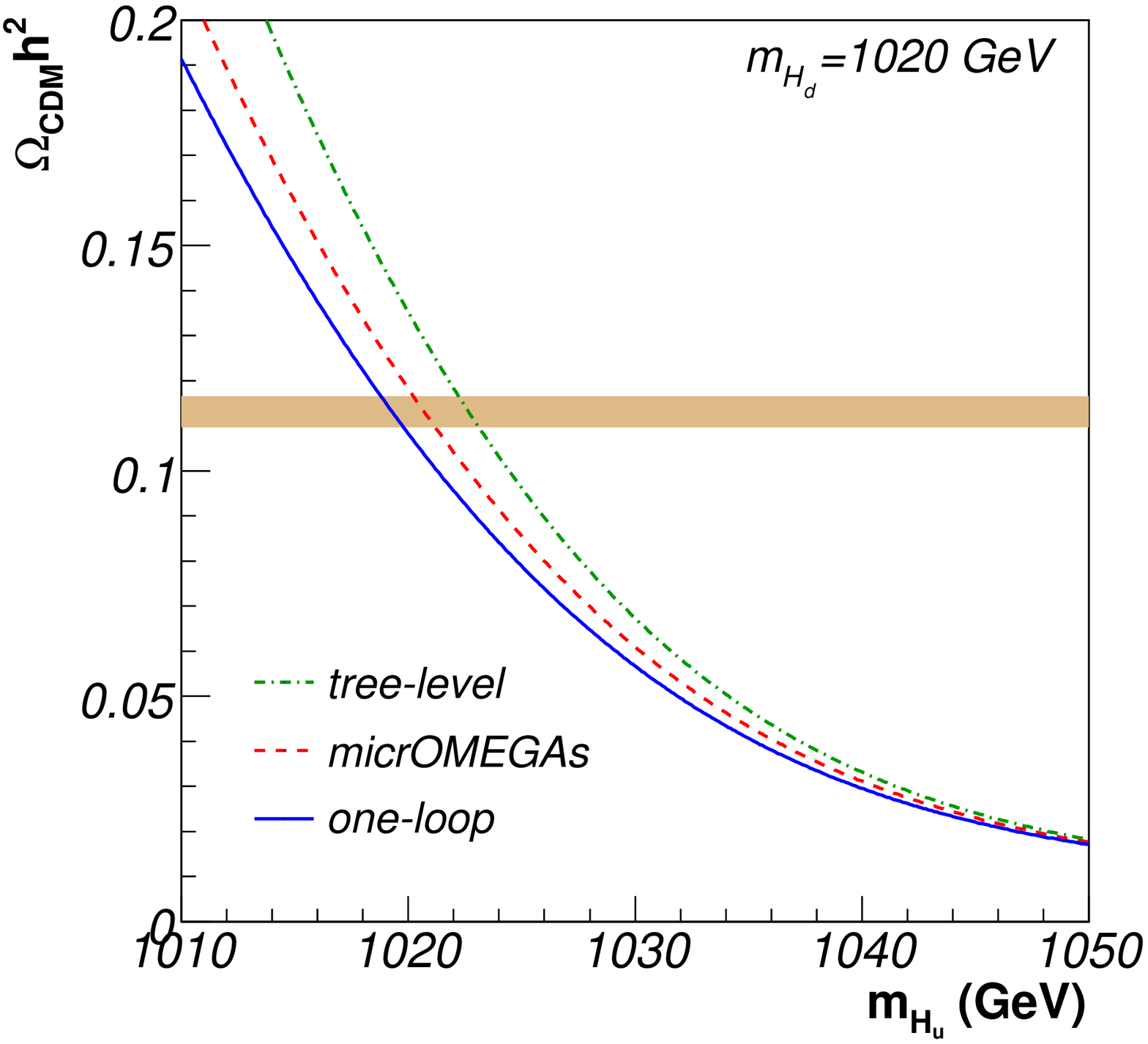}
  \end{center}
  \caption{Top and Bottom left: Scans around our parameter point II in the $m_0$-$m_{1/2}$-plane, $m_{H_u}$-$m_{H_d}$-plane, and $m_{A}$-$\mu$-plane. Bottom right: The prediction of the neutralino relic density $\Omega_{\rm CDM}h^2$ as a function of the high-scale Higgs-mass parameter $m_{H_u}$. The labels are the same as in Fig.\ \ref{figscan1}.}
  \label{figscan2}
\end{figure}
A few general features can be observed in Figs.~\ref{figxsec1} and \ref{figscan1}. First, on-resonance annihilation of neutralinos would reduce the relic density too much, so that regions allowed by Eq.~(\ref{cWMAP}) sit on each side of the resonance peak. In Fig.~\ref{figxsec1} one can clearly see that the resonance is not aligned with the kinematic region where the biggest contribution to the relic density comes from. The top panels in Fig.~\ref{figscan1} display the bands where quark final states dominate the annihilation. In addition, we show a line that corresponds to the position of the Higgs resonance peak. The WMAP allowed regions are situated on each side of the resonance where the width of the these regions reflects the steepness of the peak.
The SUSY-QCD corrections to the relic density calculated here are a combination of corrections to the processes with either bottom or top quarks final states. The corrections to the annihilation into bottom quarks are essentially confined to the Higgs-quark-antiquark vertex. The bulk of the correction comes from the gluon exchange and from the SUSY corrections which become large at large $\tan\beta$ and have to be resummed. As we take a moderate $\tan\beta=10$ even the non-resummable corrections play a role here. These corrections amount to the difference between our full SUSY-QCD result and the effective coupling approximation implemented in {\tt micrOMEGAs} in this scenario.
\begin{figure}
  \begin{center}
    \includegraphics[scale=0.4]{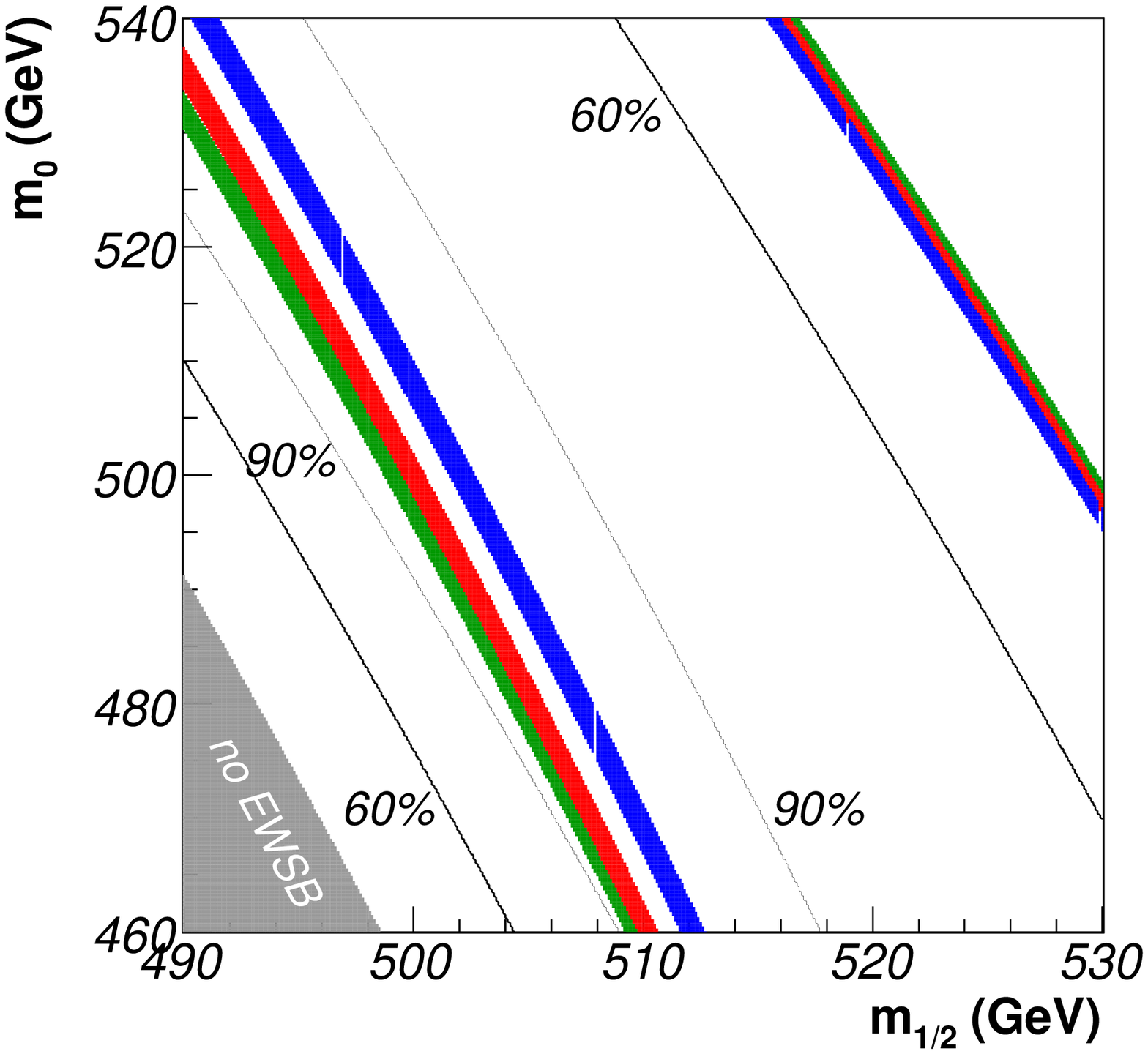}
    \includegraphics[scale=0.4]{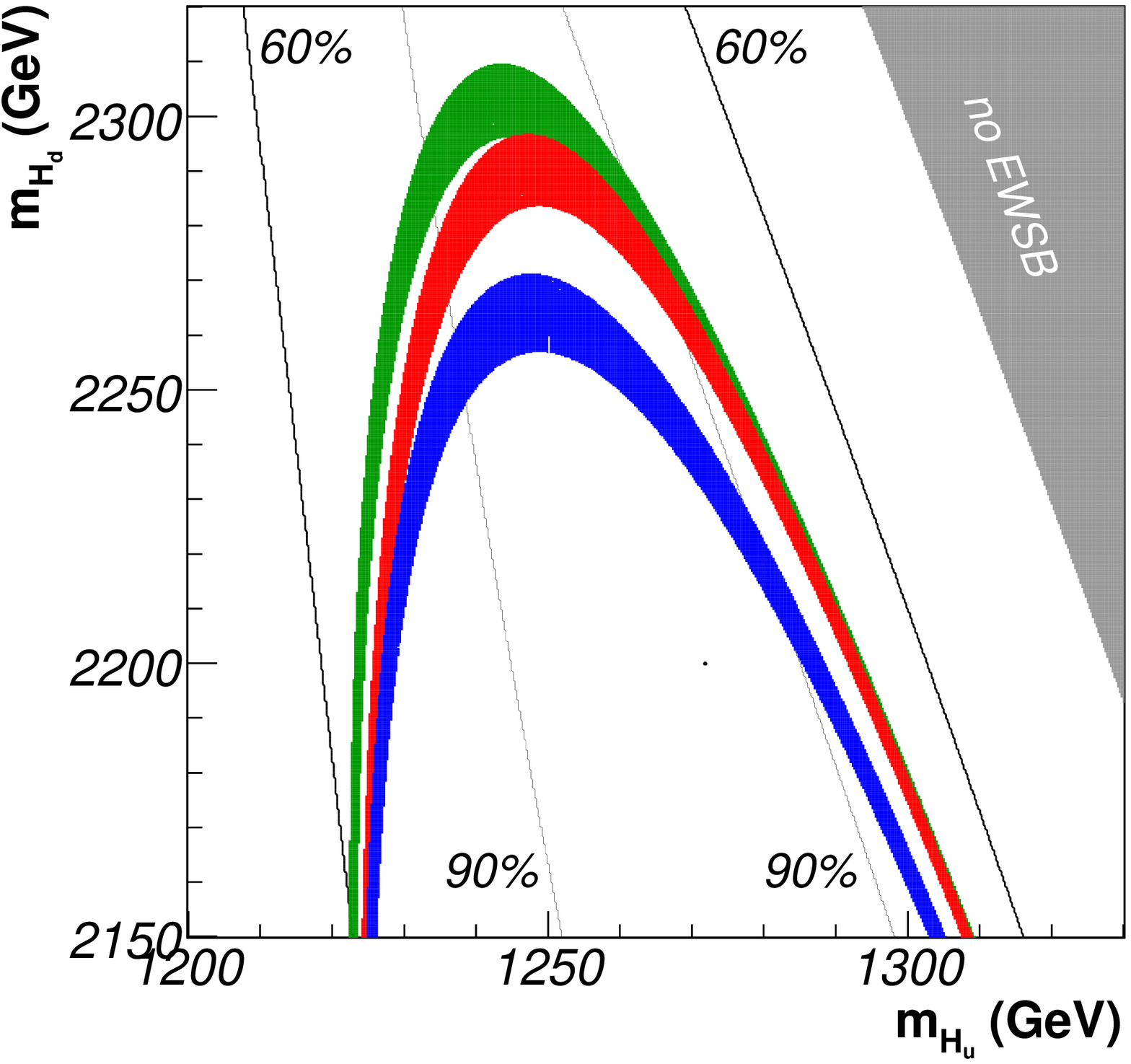}
    \includegraphics[scale=0.4]{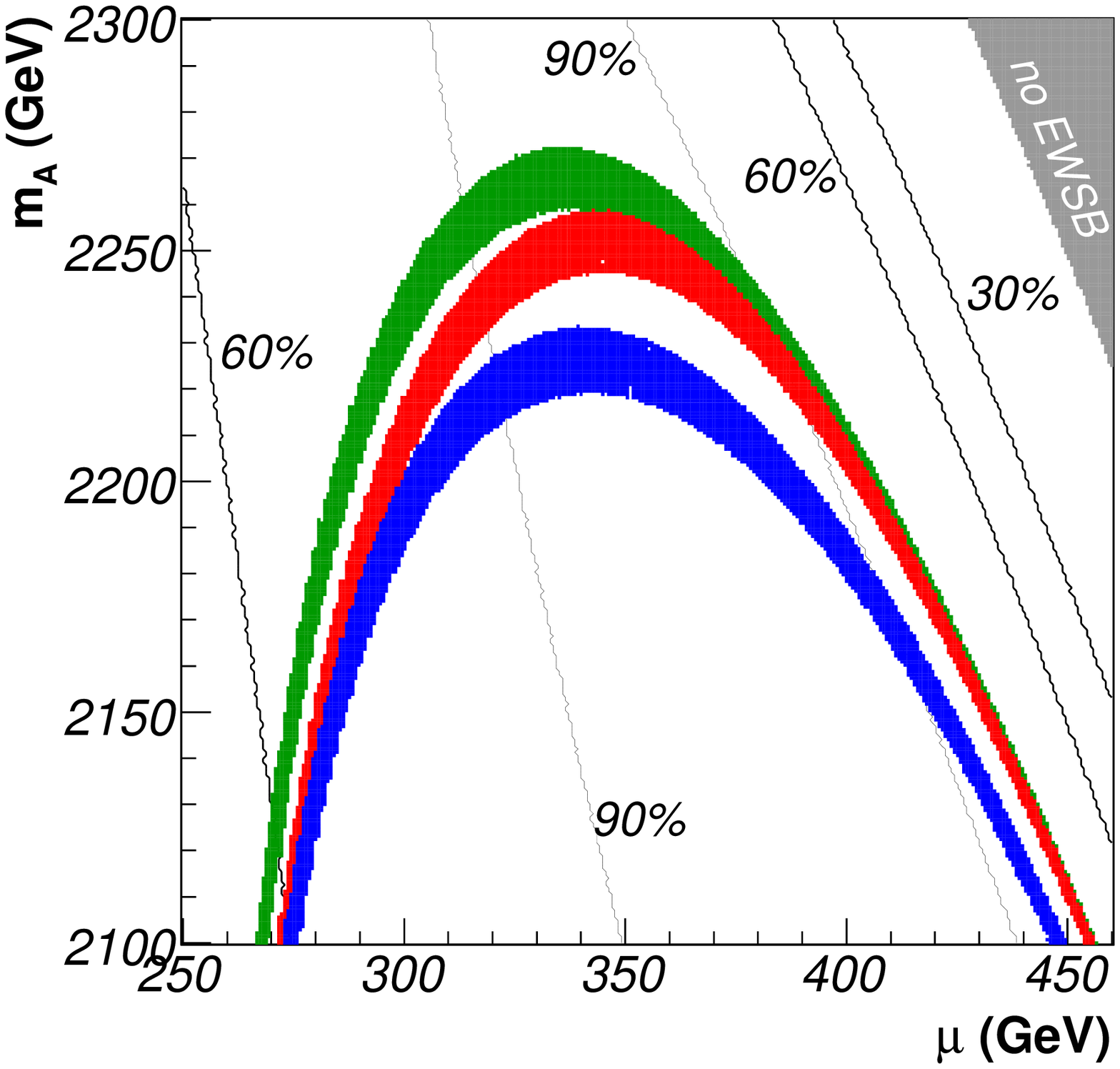}
    \includegraphics[scale=0.4]{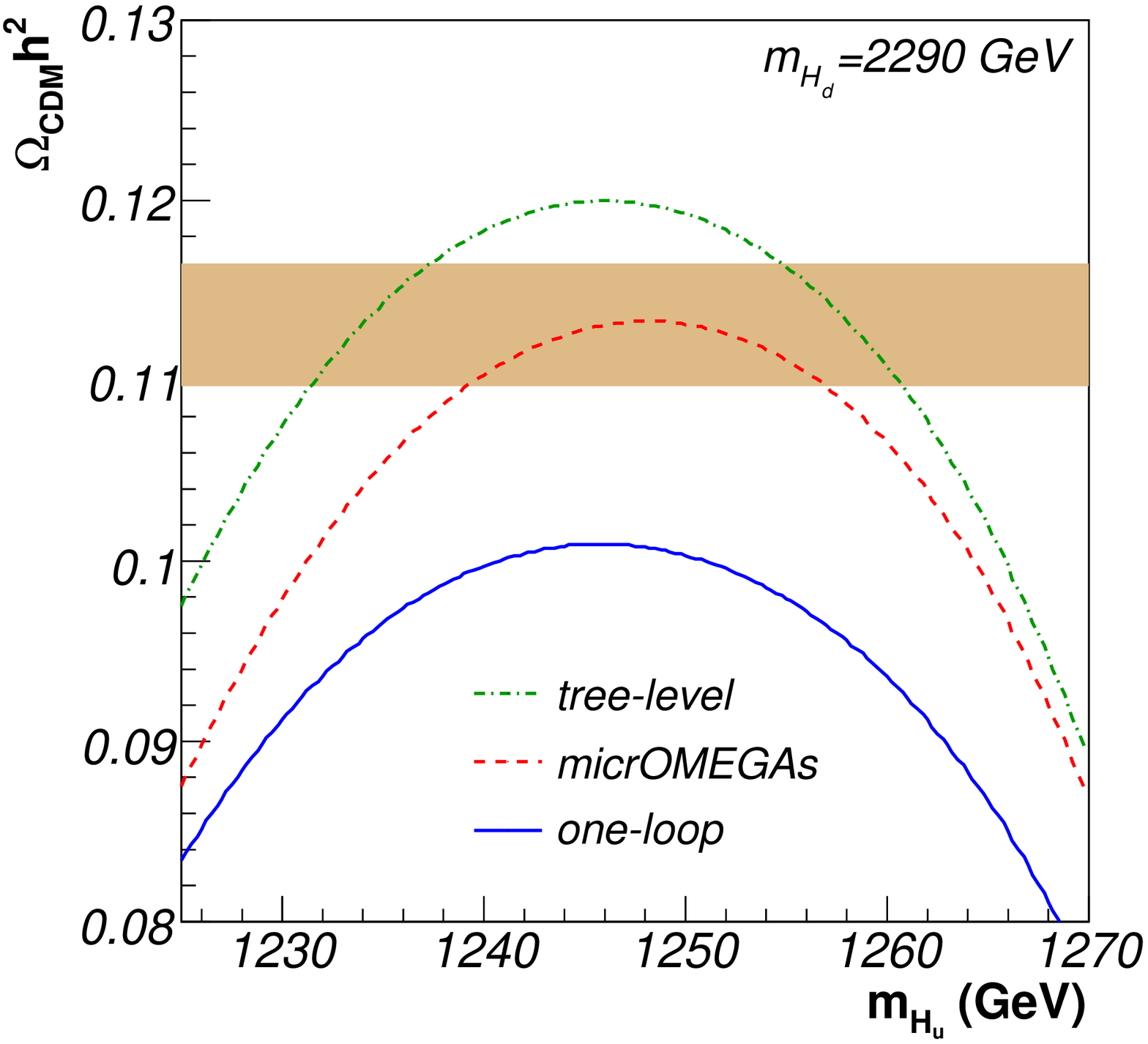}
  \end{center}
  \caption{Top and Bottom left: Scans around our parameter point III in the $m_0$-$m_{1/2}$-plane, $m_{H_u}$-$m_{H_d}$-plane, and $m_{A}$-$\mu$-plane. Bottom right: The prediction of the neutralino relic density $\Omega_{\rm CDM}h^2$ as a function of the high-scale Higgs-mass parameter $m_{H_u}$. The labels are the same as in Figs.\ \ref{figscan1} and \ref{figscan2}.}
  \label{figscan3}
\end{figure}
The situation with the top quarks is more complicated. As can be seen from Fig.~\ref{figxsec1}, the squark exchanges in the $t$- and $u$-channels and their interference with the Higgs-boson exchange become now also sizable, which renders also the corrections to squark-quark-neutralino vertex and to the squark propagator important. Overall for this scenario, the effect of the full SUSY-QCD corrections amounts to a 20\% shift in the prediction of the relic density with respect to the calculation implemented in {\tt micrOMEGAs}, as can be seen in the bottom panels of Fig.\ \ref{figscan1}. In consequence, the cosmologically favored regions of parameter space are shifted away from the position of the Higgs pole in order to compensate the larger annihilation cross-section. These shifts amount to up to 5 GeV for the common scalar and gaugino masses and a few GeV for the masses of the Higgs doublets at the unification scale (see Fig.\ \ref{figscan1}). This is about two times larger than the experimental precision, and therefore distinct bands appear.

The remaining analyzed scenarios were chosen so as to investigate effects not related to Higgs-boson exchanges. The Higgs-boson masses are very heavy in these scenarios ($m_{H^0},m_A>1100$ GeV) and suppress the importance of the Higgs-boson $s$-channel exchange. The parameter point II in Tab.~\ref{tab1} was chosen as described in Sec.~\ref{sec3A} so that the $Z^0$-boson exchange is enhanced. The only viable way to do this is to increase the higgsino fraction of the neutralino, which in turn increases the coupling of neutralinos to the $Z^0$ boson. Another possibility would be to sit on the $Z^0$-boson resonance, which would, however, lead to very small neutralino and chargino masses that are already ruled out by the LEP experiments. The parameter point III in Tab.~\ref{tab1} was picked to use the Higgs potential high-scale parameters to drive down the squark masses via the renormalization group evolution. On top of that, by choosing $A_0=-1200$ GeV we induced a large mixing of the third generation squarks. The mixing is biggest for scalar tops increasing the contribution of their exchange. A feature common to both scenarios is a distinct destructive interference between the squark and the $Z^0$ boson which links these two scenarios (see Fig.~\ref{figxsec23}). This fact makes corrections to both $Z^0$-boson and squark exchange important in each of the two scenarios.
In the case of the parameter point II, the corrections are about 20\% in the cross section as compared to the cross section in {\tt micrOMEGAs}. This is not reflected in all regions of the WMAP allowed regions in Fig.~\ref{figscan2}, since the contribution of the quark-antiquark final state falls to only 60\%-40\%, as we lower the higgsino parameter $\mu$. This decreases the masses of the lightest chargino and of the second-lightest neutralino, which, through their $t$-channel exchange, increase the annihilation cross sections into $W^+W^-$ and $Z^0Z^0$ final states. Nevertheless, the full SUSY QCD corrections shift the contour in the $m_A - \mu$ plane by $5$ GeV in $\mu$ and in all instances shift the contour by more than the current experimental precision.  
The effect of the corrections to the cross section is not screened by other final states in the case of the scenario with a dominant squark exchange (point III in Tab.~\ref{tab1}). The top quark final state accounts in certain regions for more than 90\% of the annihilation cross section, and the mass of the scalar top is not light enough to allow for efficient co-annihilations. The hyperbolic shape of the WMAP allowed regions in the $m_{H_u}-m_{H_d}$ plane is governed by the Higgs-mass parameter combination $m_{H_u}^2-m_{H_d}^2$. The effect of the corrections on the cross section is about 20\% and causes a shift of the preferred value of the pseudo-scalar Higgs boson mass by about $50$ GeV.
\subsection{Relic density in models without gaugino mass unification \label{sec4B}}
\begin{figure}
  \begin{center}
    \includegraphics[scale=0.4]{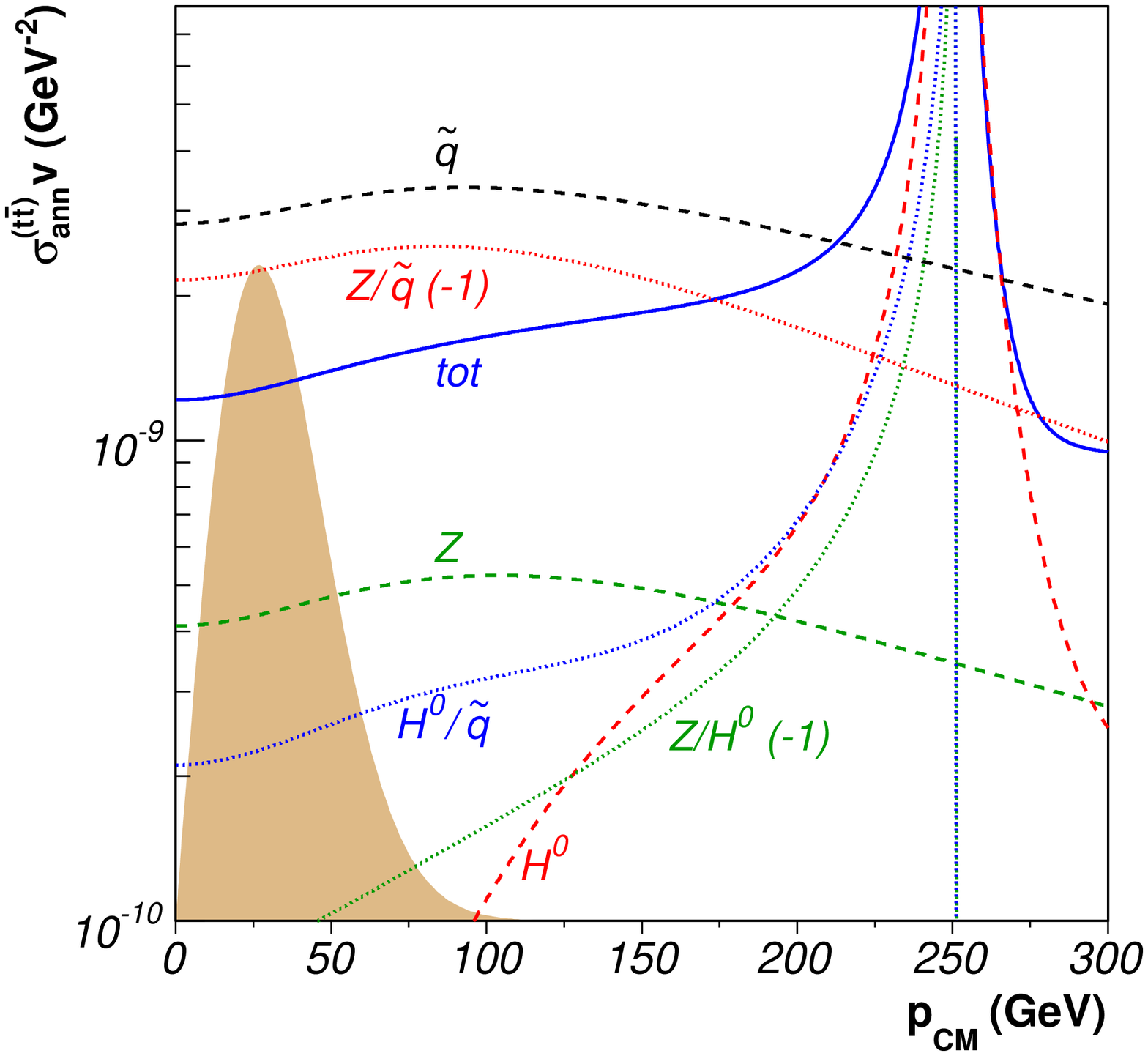}
    \includegraphics[scale=0.4]{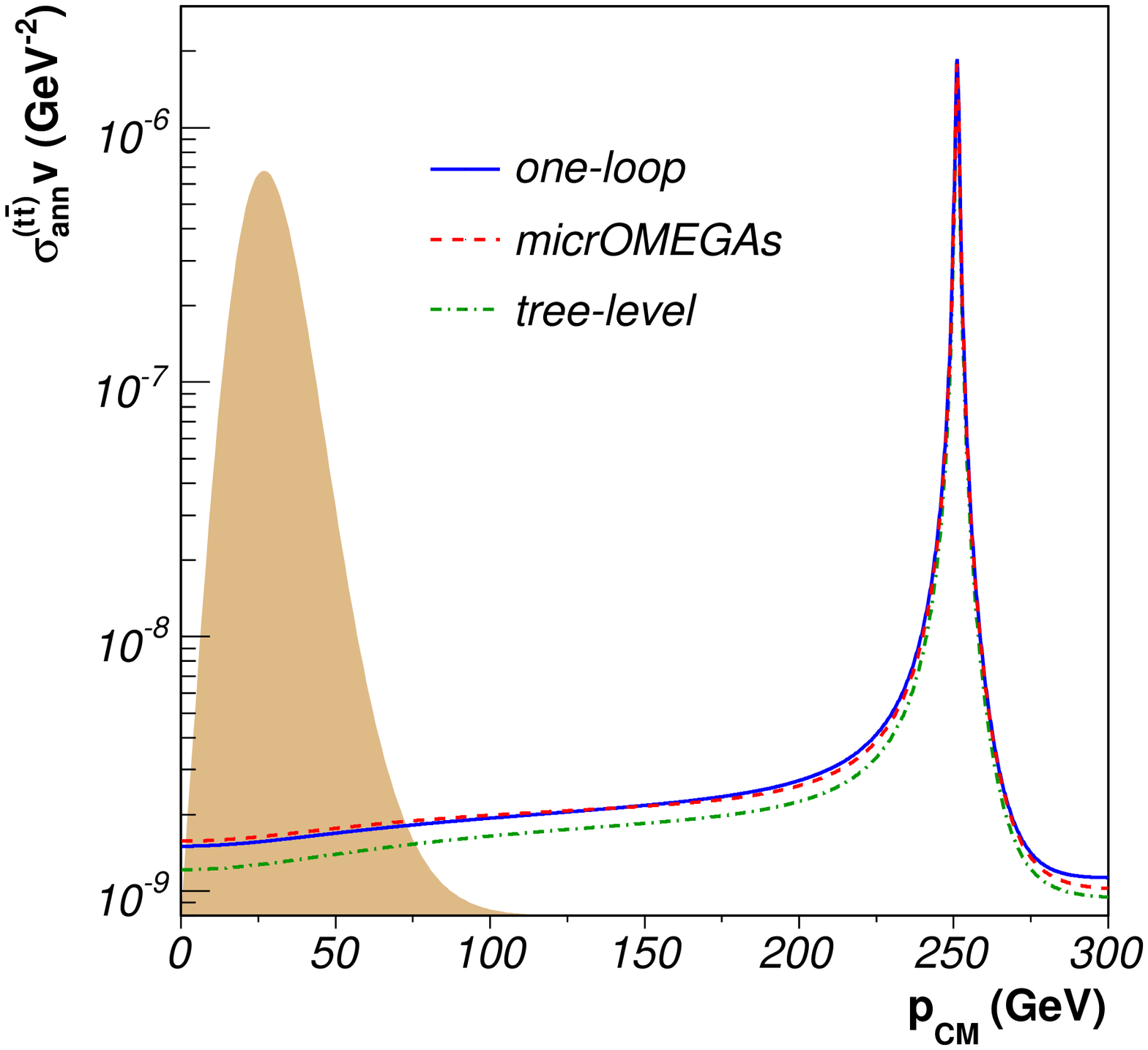}
    \includegraphics[scale=0.4]{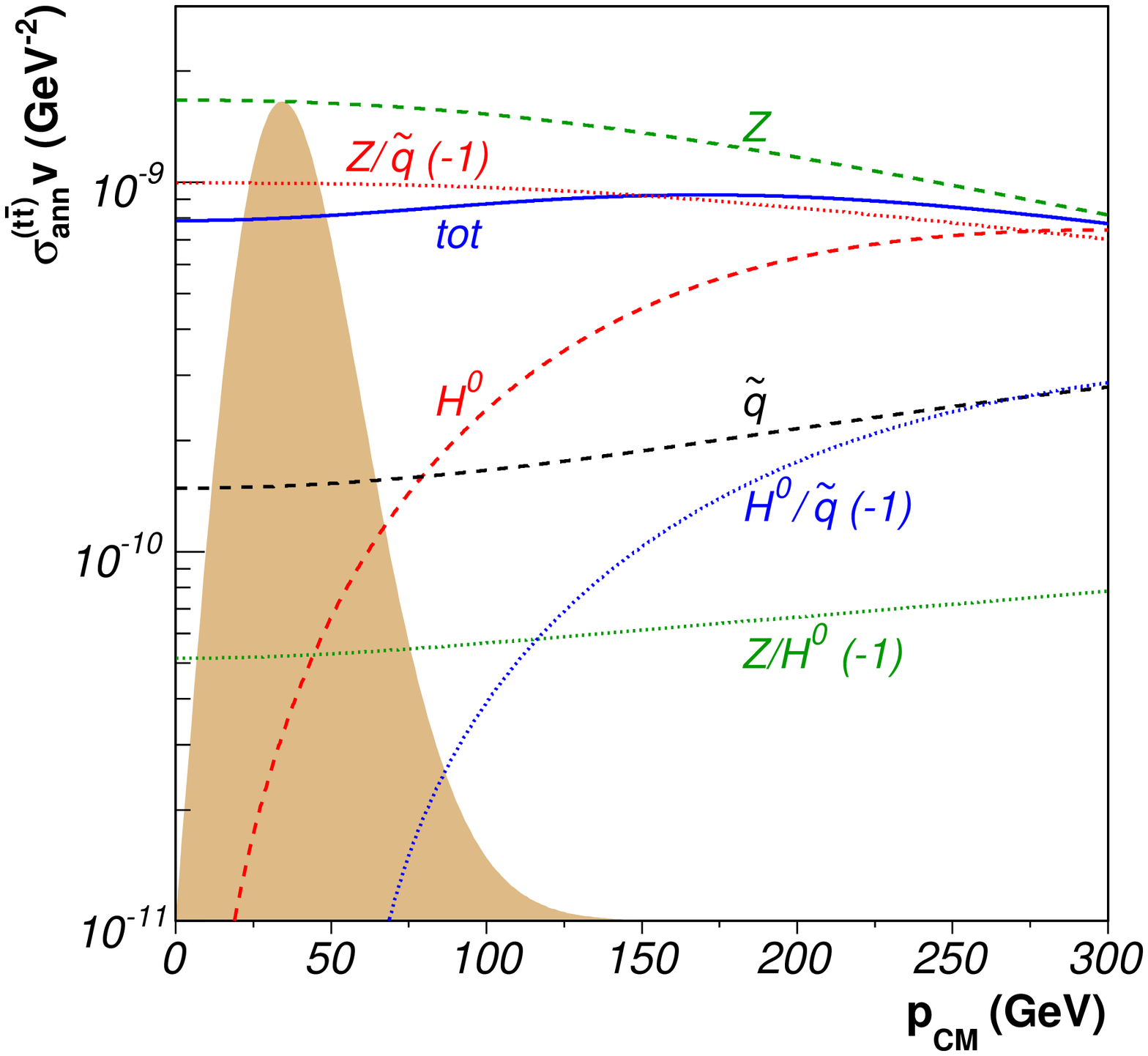}
    \includegraphics[scale=0.4]{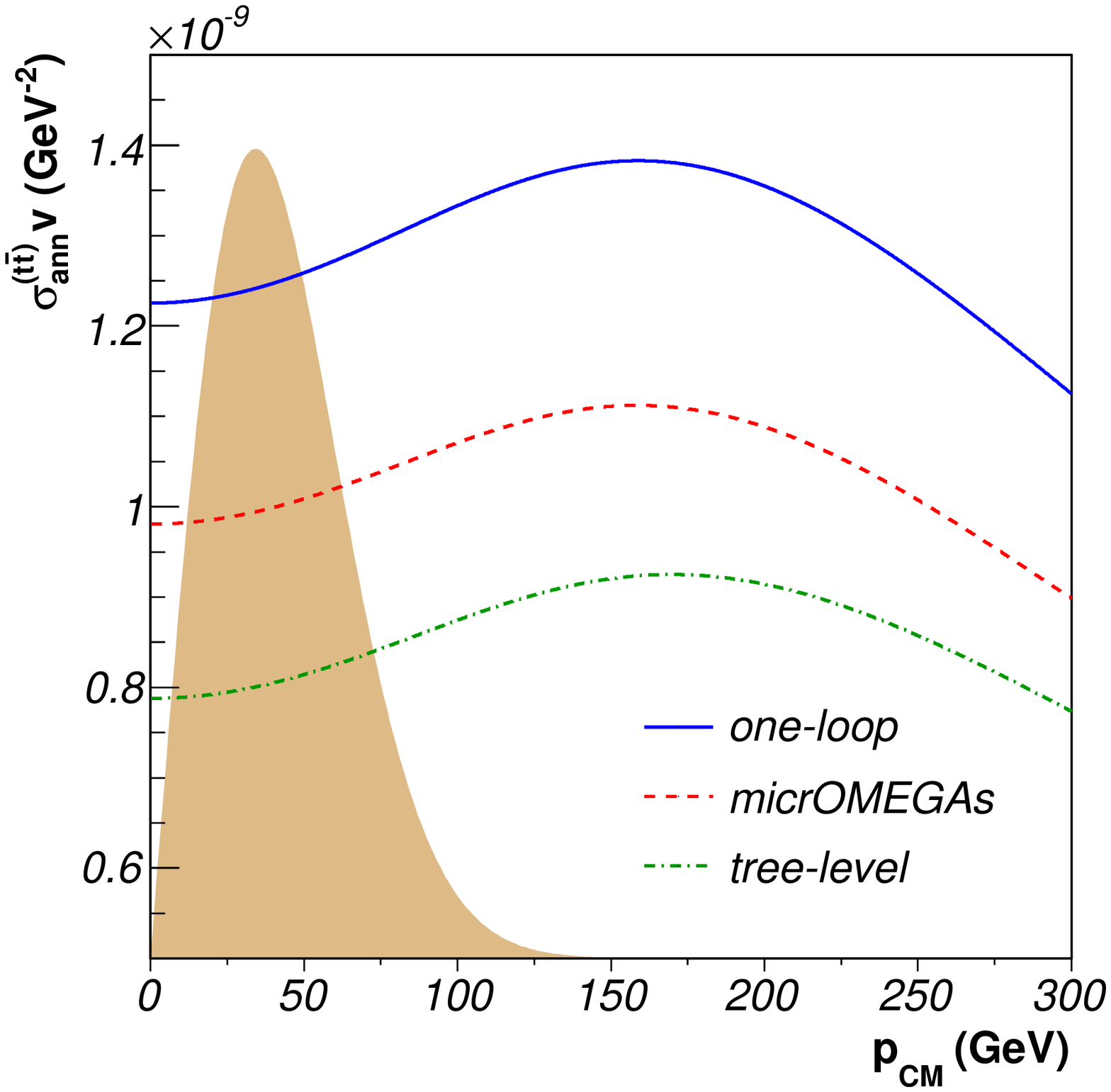}
  \end{center}
  \caption{The contributions of the different diagrams to the annihilation cross section of a neutralino pair into top quark-antiquark pairs (left) and the effect of the radiative corrections on the annihilation cross section (right) as a function of the center-of-momentum energy $p_{\rm cm}$ for our parameter points IV and V. The shaded area indicates the velocity distribution of the neutralino at the freeze-out temperature in arbitrary units.}
  \label{figxsec45}
\end{figure}

\begin{figure}
  \begin{center}
    \includegraphics[scale=0.4]{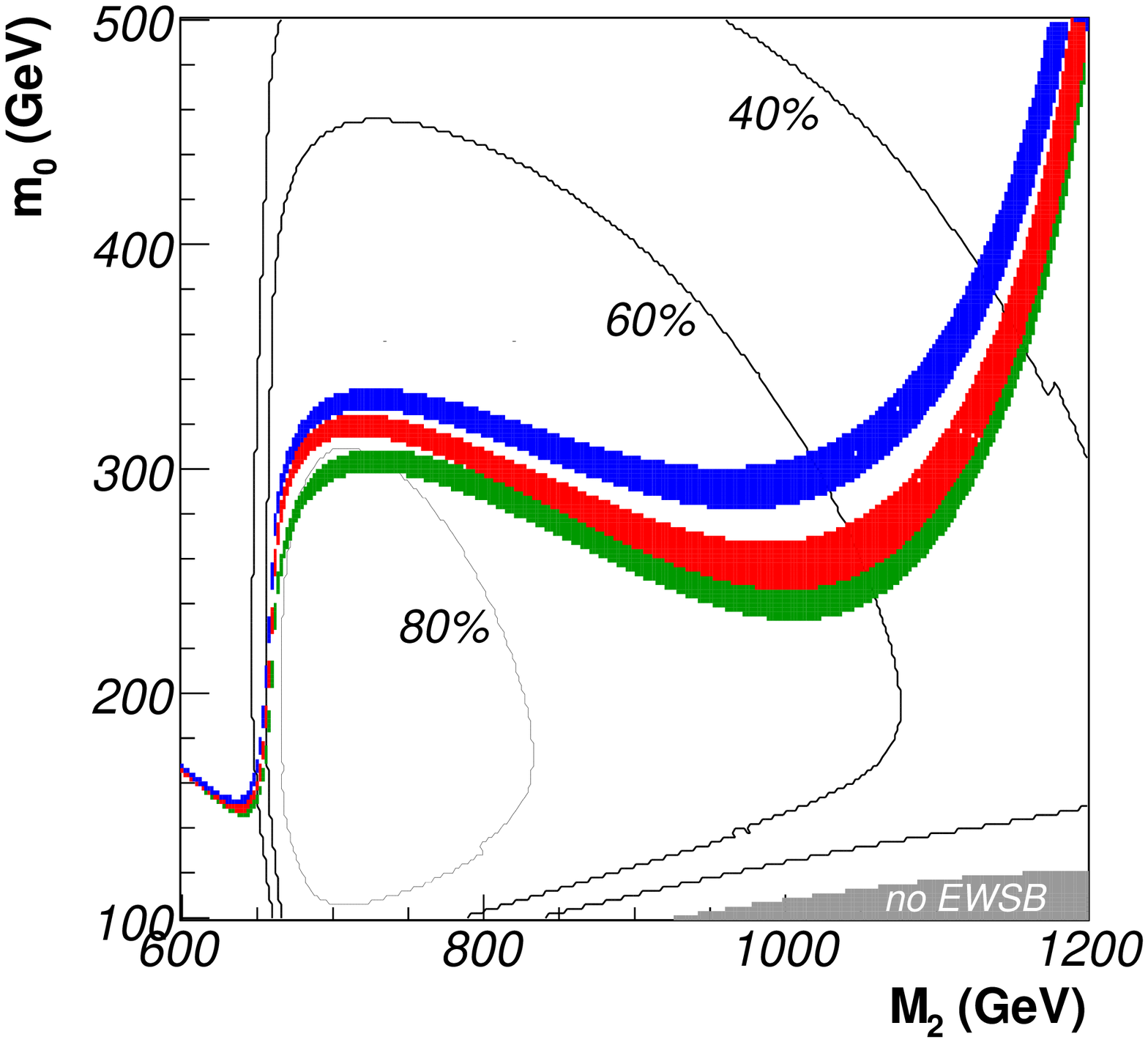}
    \includegraphics[scale=0.4]{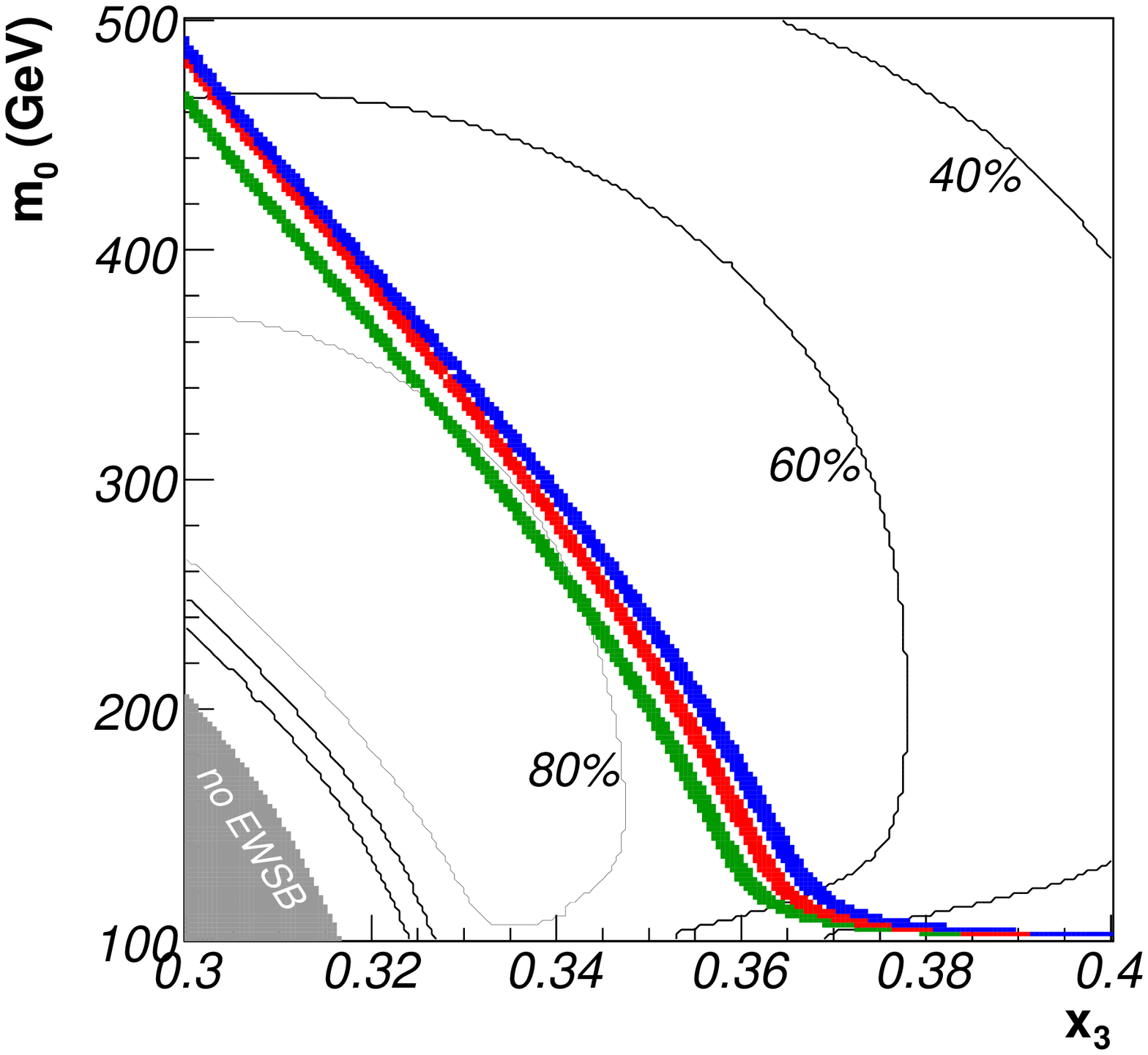}
    \includegraphics[scale=0.4]{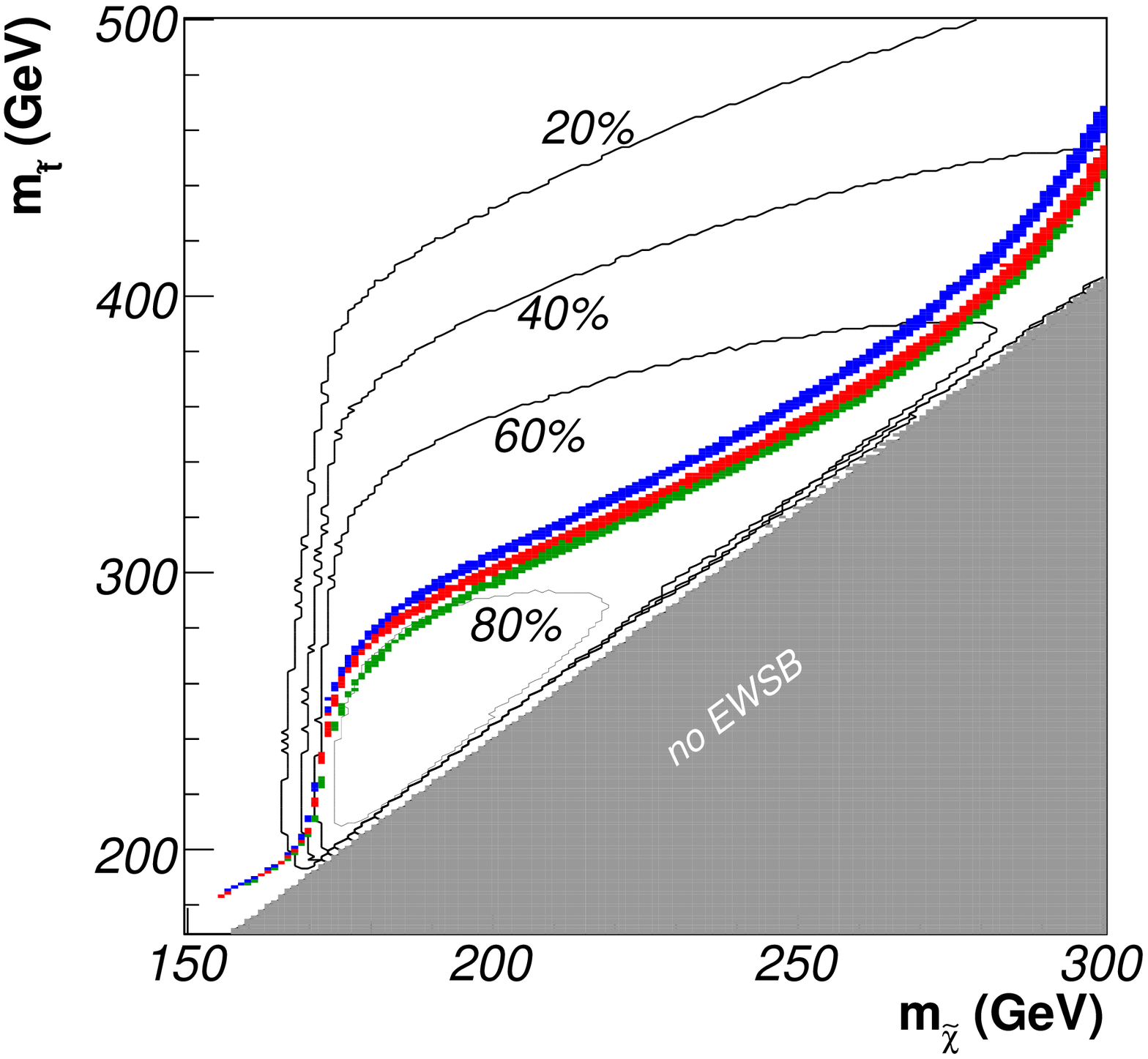}
    \includegraphics[scale=0.4]{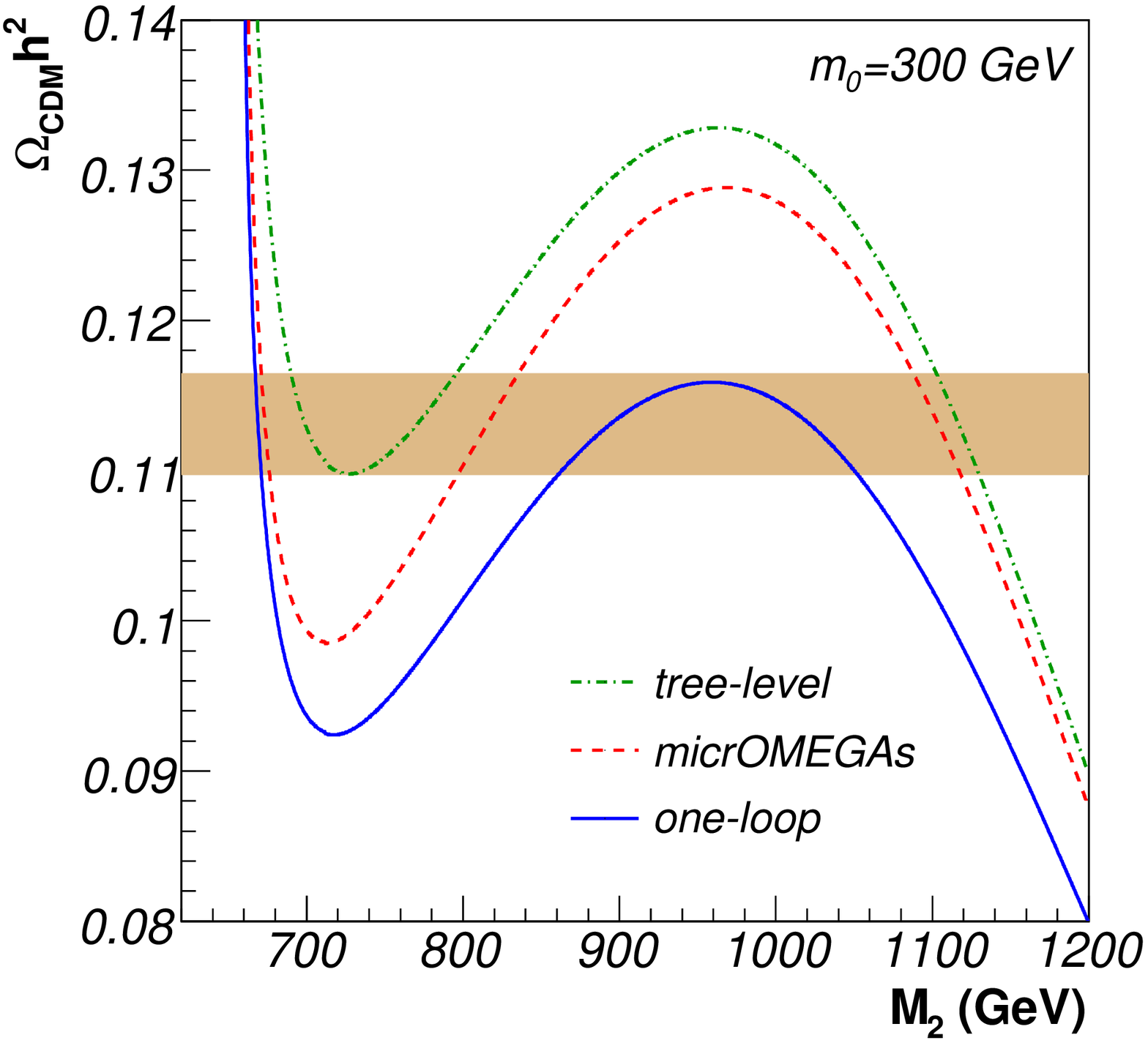}
  \end{center}
  \caption{Top and bottom left: Cosmologically favored regions in the $m_0$-$M_2$-plane (top left), $m_{\tilde{\chi}}$-$m_{\tilde{t}}$-plane (bottom left), and $m_0$-$x_3$-plane (top right) for scans around our parameter point IV. We show the regions that satisfy the constraint from Eq.~(\ref{cWMAP}) for our tree-level calculation (green), the calculation implemented in {\tt micrOMEGAs} (red), and our calculation including the full SUSY-QCD corrections (blue). We also indicate the contributions from top-quark final states to the total annihilation cross section by isolines. Excluded regions due to unphysical solutions of the renormalization group equations are shown in gray. Bottom right: The prediction of the neutralino relic density $\Omega_{\rm CDM}h^2$ including the tree-level (green dash-dotted) cross section, the approximation included in {\tt micrOMEGAs} (red dashed), and the full one-loop SUSY-QCD corrected cross-section (blue solid) as a function of the high-scale gaugino mass parameter $M_2$ for fixed $m_0=300$ GeV. The shaded area indicates the favored region of Eq.~(\ref{cWMAP}).}
\label{figscan4}
\end{figure}

The relevant masses of our points III and IV featuring very light stops are very similar, as can be seen in Tabs.~\ref{tab1} and \ref{tab2}. The same holds for the annihilation cross sections. For the benchmark point IV, the exchange of a stop in the $t$- or $u$-channel is therefore favored, which is well visible in the top left panel of Fig.~\ref{figxsec45}. The contribution from the $Z^0$-boson exchange is here one order of magnitude lower than the one from squark exchange. Again, an important role is played by the squark $Z^0$-boson interference effect, which is here even stronger than in the case of point III. The differences are the Higgs boson masses, which were more than $2000$~GeV for point III, but are $m_{H^0,A^0}\simeq 622$~GeV here (see Fig.~\ref{figxsec45}). The mass difference can be traced back to the mechanism by which we lowered the squark masses. For point III, we had to induce a big difference between the high-scale Higgs potential parameters, which drove the Higgs-boson masses higher, whereas for point IV we changed freely the gluino mass parameter influencing the RGEs for the squark masses, which has not such a big effect on the masses of the Higgs bosons. Due to the large contribution from the exchange of the light squark, the annihilation into top quarks accounts for up to 80\% of the total annihilation cross section (see Fig.\ \ref{figscan4}). The subdominant final states here are again $W^+W^-$ and $Z^0Z^0$, but also pairs of bottom quarks or leptons. Note that also here the mass difference between the lightest neutralino and the lightest stop is not small enough for efficient co-annihilations.
One small, but important difference is that the destructive interference is stronger for point IV than for point III. This leads to negative correction effects when including the full SUSY QCD corrections (when compared to the cross section of {\tt micrOMEGAs}). The reason is that for this particular point the corrections to the squark exchange are effectively taken into account by including $\overline{\rm DR}$ Yukawa couplings and the only difference between our full calculation and the {\tt micrOMEGAs} approximation are the corrections to the interference which push the full result down (as seen in the top right part of Fig.~\ref{figxsec45}).
\begin{figure}
  \begin{center}
    \includegraphics[scale=0.4]{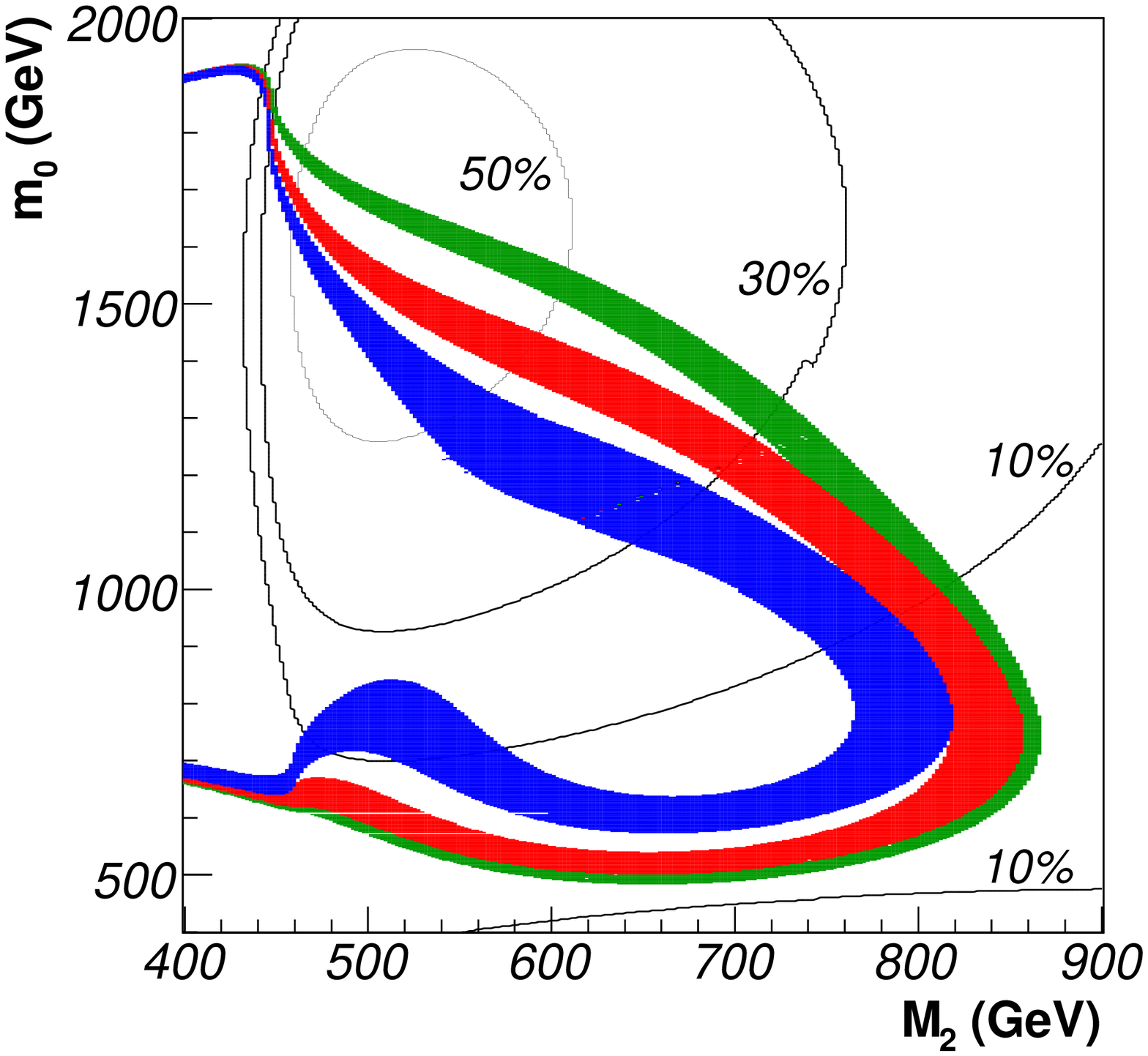}
    \includegraphics[scale=0.4]{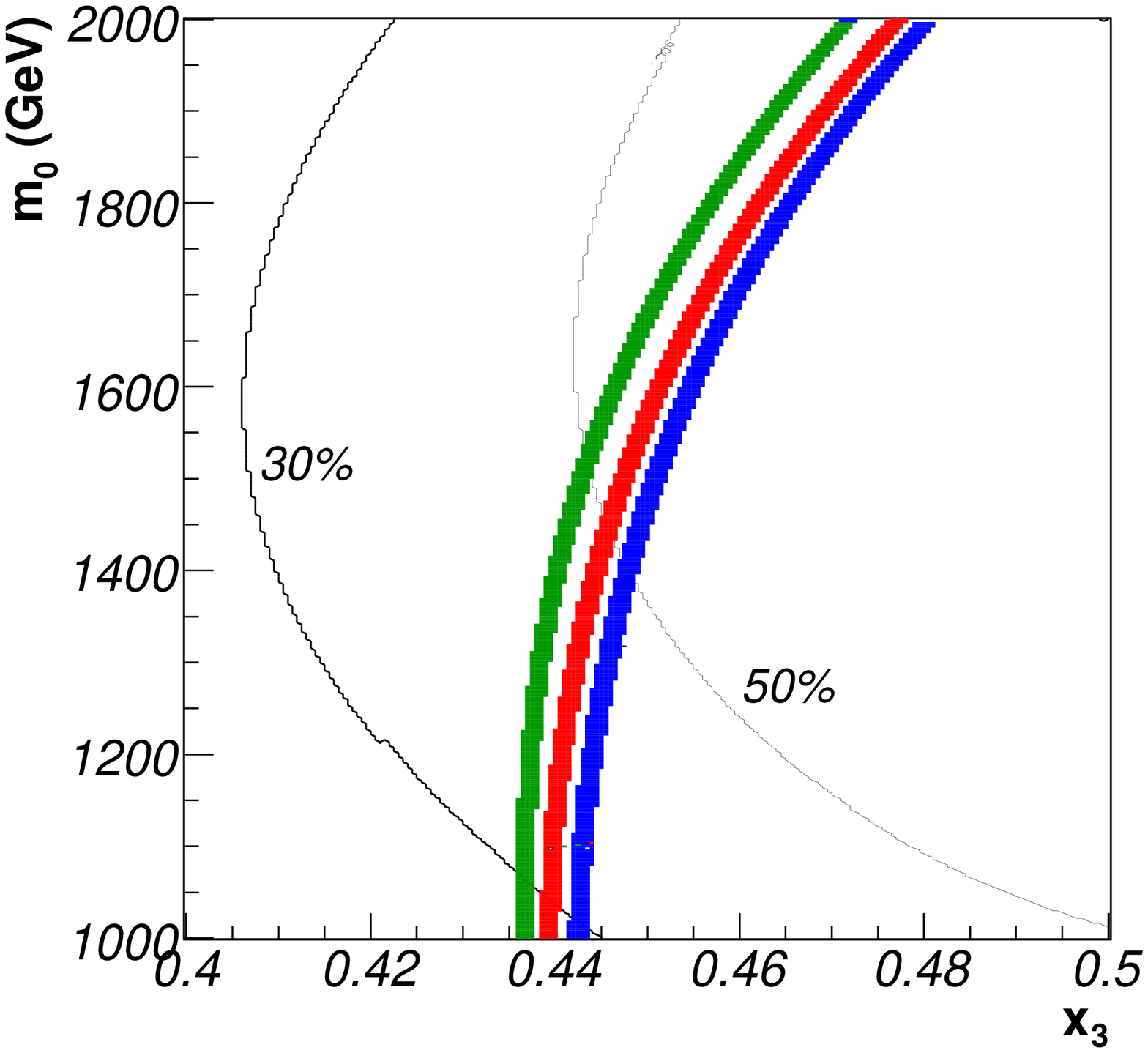}
    \includegraphics[scale=0.4]{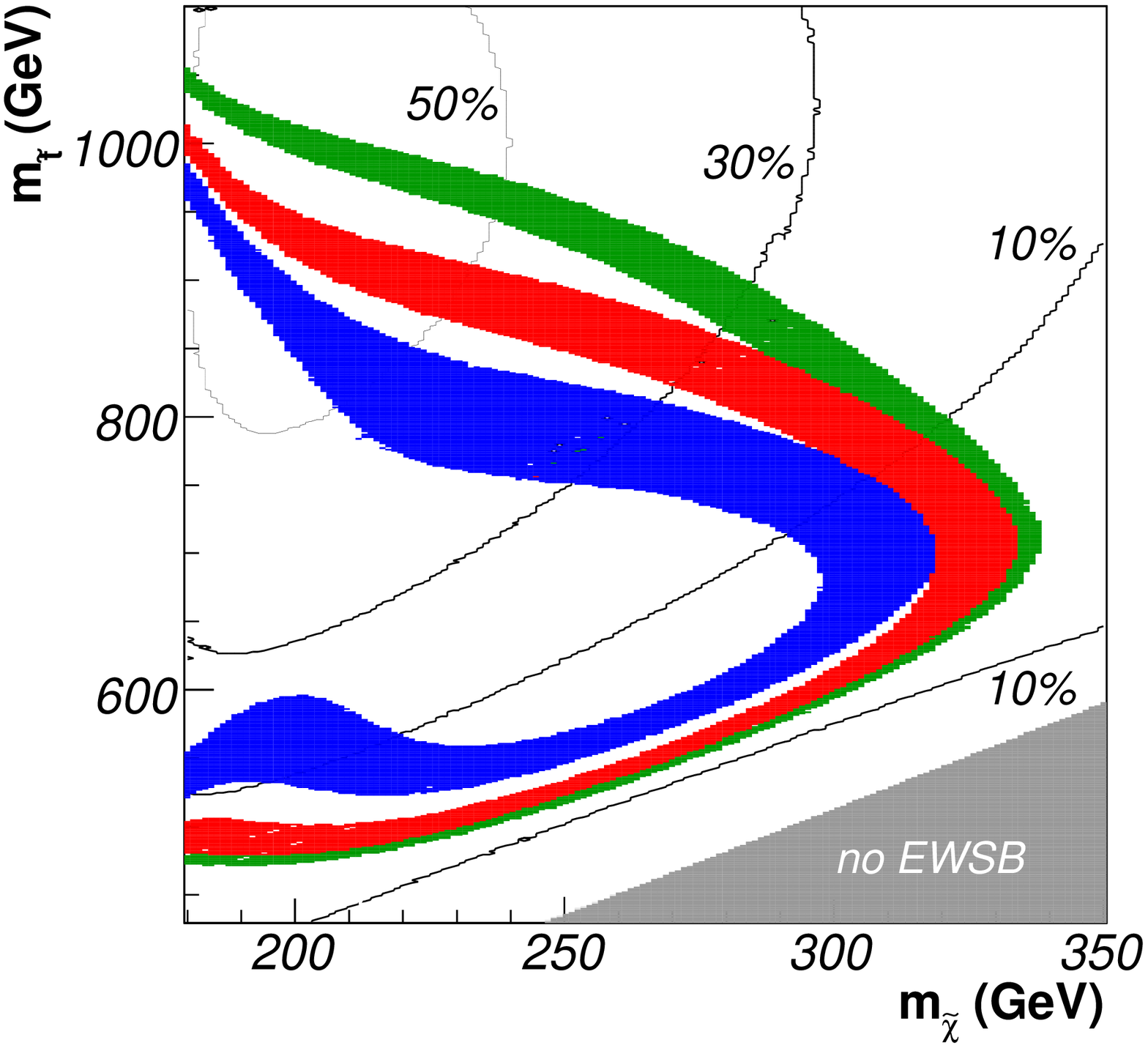}
    \includegraphics[scale=0.4]{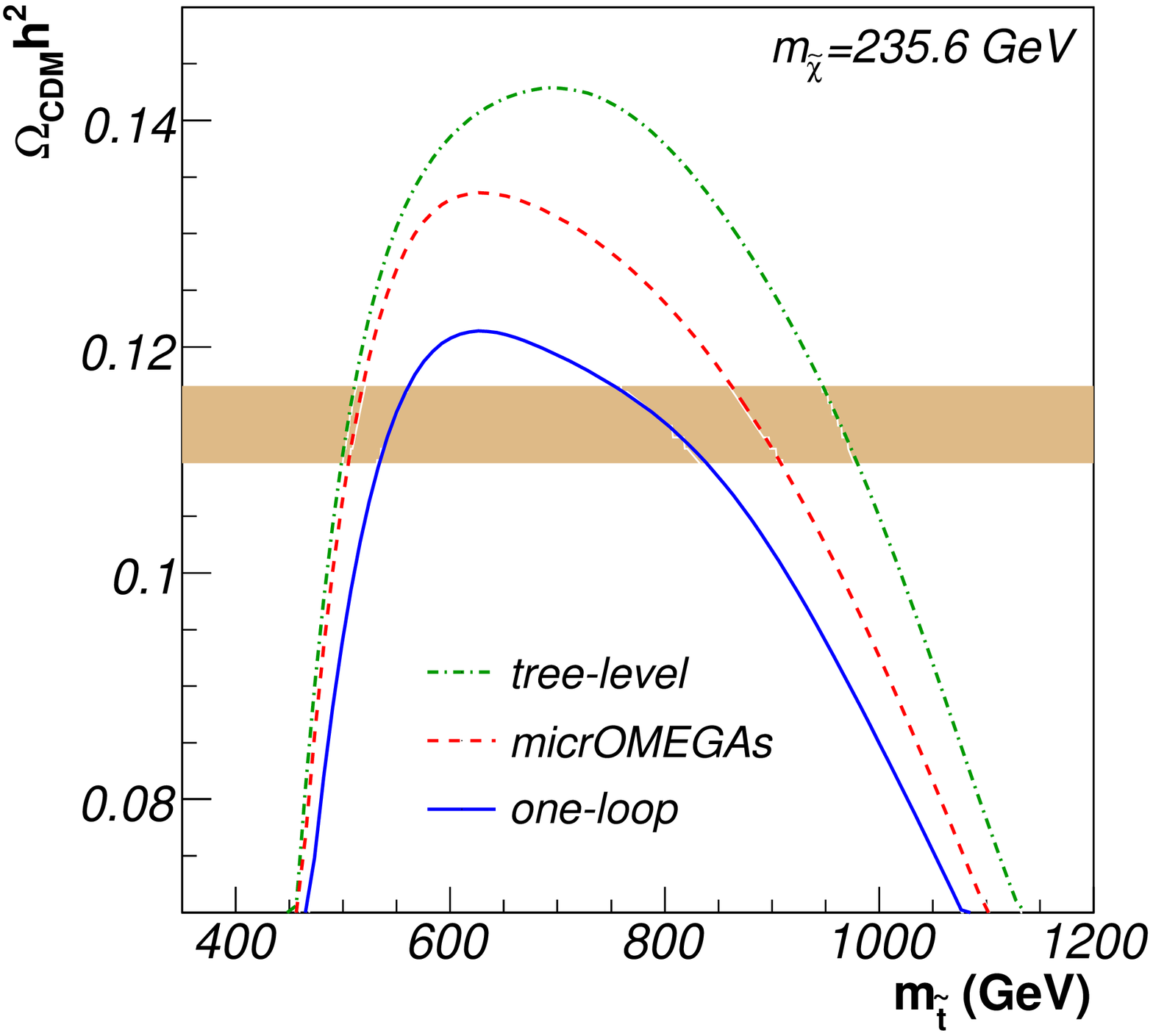}
  \end{center}
  \caption{Top and Bottom left: Scans around our parameter point V in the $m_0$-$M_2$-plane (top left), $m_{\tilde{\chi}}$-$m_{\tilde{t}}$-plane (bottom left), and $m_0$-$x_3$-plane (top right). Bottom right: The prediction of the neutralino relic density $\Omega_{\rm CDM}h^2$ as a function of the stop mass $m_{\tilde{t}_1}$ for a fixed neutralino mass $m_{\tilde{\chi}}=235.6$ GeV.  The labels are the same as in Fig.\ \ref{figscan4}.}
\label{figscan5}
\end{figure}
Let us now turn to the impact of the corrections in this ``compressed SUSY'' scenario. The corresponding favored regions in the $m_0$-$M_2$  and the $m_0$-$x_3$ planes, the projection on the $m_{\tilde{t}_1}$-$m_{\tilde{\chi}}$ plane as well as the prediction of the neutralino relic density as a function of the gaugino mass parameter $M_2$ around our parameter point IV are shown in Fig.~\ref{figscan4}. Again, in the regions where annihilation into a top quark-antiquark pair is kinematically allowed, the effect of the SUSY-QCD corrections is sizable, resulting in an important shift of the favored regions in the parameter space. The preferred region of parameter space is shifted to higher values of the scalar mass parameter $m_0$ and consequently to higher stop masses $m_{\tilde{t}_1}$. As already discussed above, a heavier stop mass compensates the increasing effect of the additional loop diagrams on the annihilation cross-section. In the right bottom panel of Fig.~\ref{figscan4} one sees again that the prediction of the relic density is decreased by the order of 15\% with respect to the tree-level calculation and by about 10\% with respect to the approximation implemented in {\tt micrOMEGAs}.

For our point V with a dominant $Z^0$-boson exchange, the correction accounts for about 20\% of the {\tt micrOMEGAs} cross-section. Although {\tt micrOMGEAs} does not include any correction for the exchange of a $Z^0$-boson, the corresponding curve is approximately 20\% over the tree-level prediction, which can be explained by the presence of effective couplings for the sub-leading squark-exchange. The difference between the approximation included in {\tt micrOMEGAs} and our full one-loop calculation originates from the supplementary corrections, especially for the exchange of the $Z^0$-boson, shown in Figs.\ \ref{FDloop} and \ref{FDbrem} and discussed in Sec.~\ref{sec2}.
We study again the influence of our corrections on the regions of parameter space that are favored with respect to the WMAP limits of Eq.\ (\ref{cWMAP}). The upper left panel of Fig.~\ref{figscan5} shows these regions in the $m_0$-$M_2$ plane for our scenario V and fixed values of $x_1$ and $x_3$ as given in Tab.\ \ref{tab2}. In the lower left panel, we show the same contours projected on the corresponding plane of the physical neutralino and stop masses $m_{\tilde{\chi}}$ and $m_{\tilde{t}_1}$. In each plot, we also indicate the isocontours corresponding to a contribution from top quark-antiquark final states of 50\%, 30\%, and 10\% to the total neutralino annihilation cross-section. The points lying in the grey shaded areas do not allow for physical solutions of the renormalization group equations.
The correction to the relic density is reduced from the 20\% it was at the cross section level (as compared with the {\tt micrOMEGAs} cross section), because the quark-antiquark final state constitutes only about 50\% of the annihilation cross section. The remaining contributions include mostly $W^+W^-$ and $Z^0Z^0$ final states. Nevertheless, as it was also the case for the mSUGRA scenarios analyzed in Ref.~\cite{Letter2}, the impact of the one-loop SUSY-QCD corrections is larger than the experimental uncertainty. Therefore distinct bands are observed in wide regions of both the $m_0$-$M_2$ and in the $m_{\tilde{\chi}}$-$m_{\tilde{t}_1}$ planes. Note that for $M_2 \lesssim 450$ GeV, the annihilation of a neutralino pair into top quarks is kinematically forbidden. The dominating channels for that region are mainly the annihilation into combinations of the gauge bosons and the light Higgs boson.
In the $m_0$-$x_3$ plane, shown in the right top panel of Fig.~\ref{figscan5}, the WMAP-favored points are confined to rather narrow bands corresponding to the different levels of included corrections. This underlines that $M_3$ is one of the key parameters to which phenomenology is very sensitive. It is interesting that for lower values of $x_3$ their positions are almost independent of $m_0$, while for higher $x_3$ the dependence becomes stronger. The shift from the tree-level prediction to our full one-loop result is in the direction of higher gluino masses $M_3 = x_3 M_2$. This is explained by the fact that increasing the gluino mass implies an increase in the squark masses, which in turn implies a decrease of the annihilation cross-section. This effect is then combined with the increase of the cross-section due to the one-loop corrections discussed above, so that the relic density remains in the narrow range which is favored by cosmological data.
Finally, in the right bottom panel of Fig.~\ref{figscan5}, we show the prediction for the neutralino relic density $\Omega_{\tilde{\chi}}h^2$ as a function of the lightest stop mass $m_{\tilde{t}_1}$. The neutralino mass has been fixed to the value $m_{\tilde{\chi}}=235.6$ GeV of our point V. The graph corresponds to a cut through the $m_{\tilde{\chi}}$-$m_{\tilde{t}_1}$ plane shown in the lower left panel of Fig.~\ref{figscan5}. The favored region of Eq.~(\ref{cWMAP}) is indicated by a shaded area. Since the SUSY-QCD corrections increase the annihilation cross section by about 50\%, the prediction for the neutralino relic density is reduced by about the same amount.

\section{Conclusions \label{sec5}}

The theoretical calculation of the dark matter relic density is an interesting
tool to obtain rather stringent constraints on the MSSM parameter space, both at
the electroweak and at the grand unification scale. Cosmological precision
measurements therefore play an important role in the extraction of SUSY mass
parameters from experimental data. In times of increasing experimental
sensitivity, in particular for the cosmological parameters such as the relic
density of cold dark matter, it is essential to increase the accuracy of the
theoretical calculation. Consequently, higher-order corrections to the dark matter
annihilation cross section become important, since they enter directly into the
calculation of the relic density.

We have presented here the full analytic details of our calculation of the
${\cal O}(\alpha_s)$ corrections to the annihilation of a neutralino pair into a
massive quark-antiquark pair, i.e.\ of the one-loop and real gluon emission
corrections. Annihilation processes into heavy quarks have been shown to be
important in large cosmologically allowed regions of SUSY GUT models without
scalar or gaugino mass unification, and the effects of our SUSY-QCD corrections
have been investigated numerically for these two classes of models. We have identified
regions of parameter space, which correspond to current cosmological limits on
the dark matter relic density and which feature important annihilations of the
neutralino pair into a top quark-antiquark pair through the exchange of a
$Z^0$-boson or a scalar quark. Higgs boson resonances were shown to be suppressed
in those regions. For five selected parameter points, the effects of the
corrections were shown to be sizable, since they enhanced the annihilation
cross-section by typically 20 and up to 50\% with respect to the tree-level
calculation. 

As a consequence, the theoretical prediction of the neutralino relic density is
also affected by the contributions at the one-loop level. Since the cross section
is increased by typically 20 and up to 50\%, the relic density is reduced by about
the same amount. We have shown that the impact of our corrections is more
important than the uncertainty on the observational limits at the $2\sigma$
confidence level. It is therefore essential to take these corrections into account
when analyzing cosmological data with the goal of extracting SUSY mass parameters,
which may be shifted by typically 5 GeV and up to 50 GeV, and of determining
the favored regions of the SUSY parameter space.

\begin{acknowledgments}
The authors would like to thank A.~Pukhov for his help in implementing the results
into the {\tt micrOMEGAs} code and W.~Porod for useful discussions. The work of
K.K.\ is supported by the ANR projects ANR-06-JCJC-0038-01 and ToolsDMColl
BLAN07-2-194882. The work of B.H.\ is supported by the DFG project PO 1337/1-1.
\end{acknowledgments}

\appendix
\section{Notation and couplings \label{asec1}}
We follow closely the notation of the couplings defined in Ref.\
\cite{GunionHaber} and the conventions used in Ref.\ \cite{Kovarik2005,PhDs} and
begin by listing all necessary couplings of the $Z^0$-boson.
The couplings of fermions to the $Z^0$-boson in the MSSM are identical to those
in the SM, and the Lagrangian is
\begin{eqnarray}
 \mathcal L & = & -
 \frac{g}{c_W}Z^0_{\mu}\,\bar f \g^{\mu}(C^f_L P_L + C^f_R P_R) f\,,
\end{eqnarray}
where $g$ is the weak coupling constant, $C^f_L = I_f^{3L} - e_f
s_{\scriptscriptstyle W}^2$, $C^f_R = - e_f s_{\scriptscriptstyle W}^2$,
$s_{\scriptscriptstyle W}$ ($c_{\scriptscriptstyle W}$) is the (co-)sine of the
weak mixing angle $\theta_{\scriptscriptstyle W}$, and $I^{3L}_f$ and $e_f$ are
the weak isospin and electric charge of the fermion $f$.
The Lagrangian for the interaction of the $Z^0$-boson with two neutralinos is
given by
\begin{eqnarray}
 \mathcal{L} &=& \frac{g}{2 c_W} Z^0_{\mu}\, \bar{{\tilde
 \x}}^0_{i}\g^{\mu}\left(O_{ij}^{''L}P_L + O_{ij}^{''R}P_R\right)\nt_j\,,
\end{eqnarray}
where
\begin{eqnarray}
 & O_{ij}^{''L} ~=~ - \frac{1}{2} Z_{i3} Z_{j3} + \frac{1}{2}
 Z_{i4} Z_{j4} ~=~ - O_{ij}^{''R} \, &
\end{eqnarray}
depend bilinearly on the neutralino mixing matrix $Z$.
The Lagrangian of the $Z^0$-boson coupling to two sfermions
\begin{eqnarray}
 \mathcal L = -i\, {g\over c_W}\, 
 Z_\mu^0\, \sf_i^\ast z^\sf_{ij}\,
 \!\stackrel{\leftrightarrow}{\partial^\mu}\!\sf_j
\end{eqnarray}
is proportional to
\begin{eqnarray}
 z^\sf_{ij} &=& C^f_L \, R^\sf_{i1} R^\sf_{j1} + C^f_R \,
 R^\sf_{i2} R^\sf_{j2},
\end{eqnarray}
which depends bilinearly on the sfermion mixing matrix
\begin{eqnarray}
 R = (R_{iL},R_{iR}) = \left(\begin{array}{cc}\cos\theta_{\ti f} &
 \sin\theta_{\ti f}\\ -\sin\theta_{\ti f} & \cos\theta_{\ti f}\end{array}\right).
\end{eqnarray}

We continue with the couplings containing neutral Higgs bosons, where we use the
notation $H_k^0 = \{h^0, H^0, A^0, G^0\}$. $t/\st$ stands for an up-type
(s)fermion and $b/\sb$ for a down-type one.
For the neutral Higgs-boson-fermion-fermion couplings, the interaction Lagrangian
reads
\begin{equation}
 \mathcal{L} = \sum_{k=1}^2 s_k^f\, H_k^0 \bar f f + \sum_{k=3}^4
 s_k^f\, H_k^0 \bar f \gamma^5 f
\end{equation}
with the couplings
\begin{equation}
    \begin{array}{ll}
        s_1^t ~=~ - g \frac{m_t \cos\alpha}{2m_{\scriptscriptstyle W}\sin\beta} ~=~
        -\frac{h_t}{\sqrt{2}}\cos\alpha,  \qquad\qquad & s_1^b ~=~ g\frac{m_b
        \sin\alpha}{2m_{\scriptscriptstyle W}\cos\beta} ~=~
        \frac{h_b}{\sqrt{2}}\sin\alpha \,, \non
        \\[5mm]
        s_2^t ~=~ - g \frac{m_t \sin\alpha}{2m_{\scriptscriptstyle W}\sin\beta} ~=~
        -\frac{h_t}{\sqrt{2}}\sin\alpha, \qquad\qquad & s_2^b ~=~ -g\frac{m_b
        \cos\alpha}{2m_{\scriptscriptstyle W}\cos\beta} ~=~
        -\frac{h_b}{\sqrt{2}}\cos\alpha \,, \non
        \\[5mm]
        s_3^t ~=~ ig\frac{m_t\cot\beta}{2m_{\scriptscriptstyle W}} ~=~
        i\frac{h_t}{\sqrt{2}}\cos\beta,  \qquad\qquad & s_3^b ~=~
        ig\frac{m_b\tan\beta}{2m_{\scriptscriptstyle W}} ~=~
        i\frac{h_b}{\sqrt{2}}\sin\beta \,, \non
        \\[5mm]
        s_4^t ~=~ ig\frac{m_t}{2m_{\scriptscriptstyle W}} ~=~
        i\frac{h_t}{\sqrt{2}}\sin\beta,  \qquad\qquad & s_4^b ~=~
        -ig\frac{m_b}{2m_{\scriptscriptstyle W}} ~=~
        -i\frac{h_b}{\sqrt{2}}\cos\beta \,. \non
        \\[5mm]
    \end{array}
\end{equation}
Here, $h_t$ and $h_b$ are the Yukawa couplings
\begin{eqnarray}
h_t ~=~ \frac{g\, m_t}{\sqrt 2 m_{\scriptscriptstyle W} \sin\b}\,,
\qquad h_b ~=~ \frac{g\, m_b}{\sqrt 2 m_{\scriptscriptstyle W}
\cos\b} \,.
\end{eqnarray}
The interaction Lagrangian for neutral Higgs bosons and neutralinos is given by 
\begin{eqnarray}\non
\mathcal L &=& - \frac{g}{2} \sum_{k=1}^2 H_k^0\, 
{\bar{\tilde\chi}}_l^0 F_{lmk}^0 \, \nt_m - i \frac{g}{2} \sum_{k=3}^4 
H_k^0\, {\bar{\tilde\chi}}_l^0\, F_{lmk}^0 \g_5\, \nt_m 
\end{eqnarray}
with 
\begin{eqnarray} \non
F_{lmk}^0 &=& \hphantom{+}\frac{e_k}{2} \Big[ Z_{l3} Z_{m2} + Z_{m3} 
Z_{l2} - \tan\theta_{W} \left( Z_{l3} Z_{m1} + Z_{m3} Z_{l1} \right) 
\Big] 
\\[2mm]
&& + \frac{d_k}{2} \Big[ Z_{l4} Z_{m2} + Z_{m4} Z_{l2} - \tan\theta_{W} 
\left( Z_{l4} Z_{m1} + Z_{m4} Z_{l1} \right) \Big] \ = \ F_{mlk}^0 \,, 
\end{eqnarray}
where $d_k$ and $e_k$ take the values 
\begin{eqnarray}\non
d_k = \{ -\cos\a, -\sin\a, \cos\b, \sin\b \}\,, \qquad e_k = \{ 
-\sin\a, \cos\a, -\sin\b, \cos\b \} \,. 
\end{eqnarray}
Following Ref.\ \cite{GunionHaber}, the neutral Higgs-boson-sfermion-sfermion
couplings can be written as
\begin{eqnarray}
G_{ijk}^\sf &\equiv& G\Big(H_k^0 \sf_i^\ast \sf_j\Big) = \Big[
R^\sf G_{LR, k}^\sf (R^\sf)^T \Big]_{ij} \,,
\end{eqnarray}
where the left-right couplings $G_{LR, k}^\sf$ for third-generation up- and
down-type sfermions are
\begin{eqnarray}\non
G_{LR, 1}^\st &=& \left(\!
\begin{array}{cc}
-\sqrt 2 h_t m_t c_\a + g_{\scriptscriptstyle Z}
m_{\scriptscriptstyle Z} (I_t^{3L}\!-\!e_t s_{\scriptscriptstyle
W}^2) s_{\a+\b} & -\frac{h_t}{\sqrt 2} (A_t\, c_\a + \mu s_\a) \\
-\frac{h_t}{\sqrt 2} (A_t\, c_\a + \mu s_\a) & -\sqrt 2 h_t m_t
c_\a + g_{\scriptscriptstyle Z} m_{\scriptscriptstyle Z} e_t
s_{\scriptscriptstyle W}^2 s_{\a+\b}
\end{array} \!\right) \,,
\\[2mm] \non
G_{LR, 1}^\sb &=& \left(\!
\begin{array}{cc}
\sqrt 2 h_b m_b s_\a + g_{\scriptscriptstyle Z}
m_{\scriptscriptstyle Z} (I_b^{3L}\!-\!e_b s_{\scriptscriptstyle
W}^2) s_{\a+\b} & \frac{h_b}{\sqrt 2} (A_b\, s_\a + \mu c_\a) \\
\frac{h_b}{\sqrt 2} (A_b\, s_\a + \mu c_\a) & \sqrt 2 h_b m_b s_\a
+ g_{\scriptscriptstyle Z} m_{\scriptscriptstyle Z} e_b
s_{\scriptscriptstyle W}^2 s_{\a+\b}
\end{array} \!\right) \,,
\\[2mm] \non
G_{LR, 2}^\sf &=& G_{LR, 1}^\sf \qquad {\rm with}\ \a
\rightarrow \a - \pi/2 \,,
\\[2mm] \non
G_{LR, 3}^\st &=& -\sqrt 2 h_t \left(\!
\begin{array}{cc}
0 & -\frac{i}{2} ( A_t\, c_\b + \mu\, s_\b ) \\
\frac{i}{2} ( A_t\, c_\b + \mu\, s_\b ) & 0
\end{array} \!\right) \,,
\\[2mm] \non
G_{LR, 3}^\sb &=& -\sqrt 2 h_b \left(\!
\begin{array}{cc}
0 & -\frac{i}{2} ( A_b\, s_\b + \mu\, c_\b ) \\
\frac{i}{2} ( A_b\, s_\b + \mu\, c_\b ) & 0
\end{array} \!\right) \,,
\\[2mm] \non
G_{LR, 4}^\sf &=& G_{LR, 3}^\sf \qquad {\rm with}\ \b
\rightarrow \b - \pi/2 \,.
\end{eqnarray}
Here, we have used the abbreviations $s_x \equiv \sin x$ and $c_x \equiv \cos x$
and $\a$ denotes the mixing angle of the $\{h^0, H^0\}$-system.

For the neutralino-sfermion-fermion couplings, the Lagrangian reads
\begin{eqnarray}
 \mathcal L & = & \bar f \left(a_{ik}^\sf P_R + b_{ik}^\sf
 P_L\right) \nt_k\,\sf_i + \bar{\ti \x}^0_k \left(a_{ik}^\sf P_L +
 b_{ik}^\sf P_R\right) f \sf_i^\ast
\end{eqnarray}
with the coupling matrices
\begin{eqnarray}
 a_{ik}^\sf &=& h_f Z_{kx} R_{i2}^\sf + g f_{Lk}^f R_{i1}^\sf \,,
 \hspace{18mm} b_{ik}^\sf ~=~ h_f Z_{kx} R_{i1}^\sf + g f_{Rk}^f
 R_{i2}^\sf
\end{eqnarray}
and
\begin{eqnarray}
 f_{Lk}^f = \sqrt2 \left(
 (e_f-I_f^{3L})\tan\theta_{\scriptscriptstyle W} Z_{k1} + I_f^{3L}
 Z_{k2} \right)\,,\quad f_{Rk}^f = -\sqrt2
 e_f\tan\theta_{\scriptscriptstyle W} Z_{k1} \,.
\end{eqnarray}
The coupling of gluinos to sfermions and fermions can be derived from the
Lagrangian
\begin{eqnarray}
	\mathcal L & = & -\sqrt{2}g_sT^{a}_{st}\left[\bar f_s \left(R_{iL} P_R - R_{iR} P_L\right) \tilde{g}^a\,\sf_{i,t} + \bar{\ti g}^a \left(R_{iL} P_L - R_{iR} P_R\right) f_s \sf_{i,t}^\ast\right]\,,	
\end{eqnarray}
where $g_s = \sqrt{4\pi\alpha_s}$ is the strong coupling constant and $R$
is the sfermion mixing matrix defined above. Finally, the Lagrangian
corresponding to the couplings of the gluon to fermions and sfermions is
\begin{eqnarray}
	\mathcal L & = & - g_sT^{a}_{st}\, g^a_{\mu}\,\bar f_{s} \g^{\mu} f_t - i\, g_sT^{a}_{st}\,g_\mu^a\, \sf_{i,s}^\ast
	\!\stackrel{\leftrightarrow}{\partial^\mu}\!\sf_{j,t}\,,
\end{eqnarray}
where the $T^a$ represent the usual SU(3) color matrices.

\section{Virtual corrections \label{asec2}}
Here, we give explicit forms of the next-to-leading order amplitudes
$\mathcal{M}_{\rm 1-loop}$ mentioned in Sec.~\ref{sec2}. We use generic amplitudes
with generic couplings, which we then specify using the couplings given in
App.~\ref{asec1}. Note that all kinds of indices are to be summed over even if not
explicitly stated. We leave out the results for self-energies, which can be found
in \cite{PhDs} using the same conventions and couplings.
\subsection{Vertex corrections}
\begin{figure}
	\begin{picture}(450,100)
       \put(0,0){\includegraphics[scale=1.2]{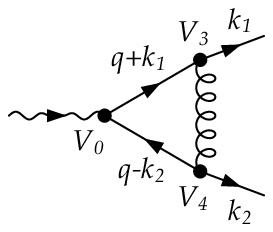}}
       \put(100,62){$V_0: \gamma^{\nu} (g_0^L P_L + g_0^R P_R)$}
       \put(100,47){$V_3: \gamma^{\rho} (g_3^L P_L + g_3^R P_R)$}
       \put(100,32){$V_4: \gamma^{\rho} (g_4^L P_L + g_4^R P_R)$}
       \put(230,0){\includegraphics[scale=1.2]{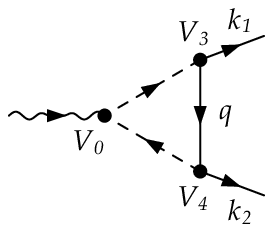}}
       \put(330,62){$V_0: g_0 (2q+k_1-k_2)^{\nu}$}
       \put(330,47){$V_3: (g_3^L P_L + g_3^R P_R)$}
       \put(330,32){$V_4: (g_4^L P_L + g_4^R P_R)$}
	\end{picture}
	\caption{Vertex corrections to the $Z^0$-boson coupling to quarks, where the arrows indicate the flow of the four-momenta and $q$ is the independent loop momentum.}\label{gendiagZ} 
\end{figure}

The contribution of the vertex corrections depicted in Fig.~\ref{FDloop} to the
matrix element $\mathcal{M}_{\rm 1-loop}$ can be written in terms of six loop
diagrams. Denoting the momenta of the incoming neutralinos $p_1$ and
$p_2$ and those of the outgoing quark and anti-quark as $k_1$ and $k_2$, 
the next-to-leading corrections to the $s$-channel exchange of the $Z^0$-boson can be parameterized as
\begin{eqnarray}
	\mathcal{M}_{\rm 1-loop} &=& \Big[\overline{v}(p_2)\gamma^\mu (A_L P_L + A_R P_R) u(p_1)\Big] \frac{-ig_{\mu\nu}}{s-m_Z^2}\\ \nonumber &&\times \Big[\overline{u}(k_1) \Big\{ \gamma^\nu (B^1_L P_L + B^1_R P_R) + k_1^{\nu}(B^2_L P_L + B^2_R P_R) + k_2^{\nu}(B^3_L P_L + B^3_R P_R) \Big\} v(k_2)\Big]
\end{eqnarray}
where $A_L$, $A_R$ are the chiral couplings of two neutralinos to the $Z^0$-boson
\begin{equation}
	A_L = -A_R = - \frac{g}{2c_W}\left( Z_{i3} Z_{j3} + 
	Z_{i4} Z_{j4}\right)
\end{equation} 
and $B^i_L$, $B^i_R$ are general form factors, which will now be given for the loops with a gluon and a gluino exchange.
The form factors for the loop diagram containing a gluon (on the left side in Fig.~\ref{gendiagZ}) are
\begin{eqnarray}
	 B^1_L &=& \frac{1}{8\pi^2}\, g^L_3 g^L_4 \Big[ g^L_0 \big(B_0 - 2 C_{00} - s (C_0 + C_1 + C_2) + m_f^2\, (2 C_0 + 3 C_1 + 3 C_2)\big) + g^R_0 m_f^2 \,(C_1 + C_2) \Big]\,,	\\
	 B^2_L &=& -\frac{1}{4\pi^2}\, g_3^L g_4^L m_f \Big[ g^L_0 C_{11} + g^R_0 ( C_{12} + C_2 )\Big]\,,\\
	 B^3_L &=& \frac{1}{4\pi^2}\, g_3^L g_4^L m_f \Big[g^L_0 (C_1 + C_{12}) + g^R_0 C_{22}\Big]\,,\\[2mm]
	B^i_R &=& B^i_L (g_0^L\leftrightarrow g_0^R)\,,\qquad\quad i=1,2,3\,.
\end{eqnarray}
Using the notation from App.\ \ref{asec1}, the generic couplings in this case are
\begin{eqnarray}
	g^L_0 = \frac{g}{c_W}\, C^f_L\,,\qquad g^R_0 = \frac{g}{c_W}\, C^f_R\,,\qquad\quad g^L_3 = g^R_3 = g^L_4 = g^R_4 = -g_s T^{a}_{st}\,.
\end{eqnarray}
The scalar loop integrals follow the definition given in Ref.\ \cite{Denner1991} and their arguments are
\begin{equation}\label{eqloopint1}
	C_i\big(k_1,-k_2;0,m_f,m_f\big)\,,\qquad\quad B_0\big(-k_1-k_2;m_f,m_f\big)\,,
\end{equation}
where $m_f$ is the mass of the final-state fermion. 

The same form factors for the diagram with a gluino exchange (see Fig.~\ref{gendiagZ} right) are given as
\begin{eqnarray} 
	 B^1_L &=& \frac{1}{8\pi^2}\, g_0\, g_3^R g_4^L C_{00}\,,\\
	 B^2_L &=&  \frac{1}{16\pi^2}\, g_0 \Big[ g_4^L \big(- g_3^L m_{\tilde g} (C_0 + 2 C_1) + g_3^R m_f (C_1 + 2 C_{11})\big) -
            g_4^R g_3^L m_f (C_2 + 2 C_{12})\Big]\,,\\
	 B^3_L &=& -\frac{1}{16\pi^2}\, g_0 \Big[ g_4^L \big(- g_3^L m_{\tilde g} (C_0 + 2 C_2) + g_3^R m_f (C_1 + 2 C_{12})\big)+ 
            g_4^R g_3^L m_f (C_2 + 2 C_{22})\Big]\,,\\
			B^i_R &=& B^i_L (g_3^L\leftrightarrow g_3^R, g_4^L\leftrightarrow g_4^R)\,,\qquad\quad i=1,2,3\,,
\end{eqnarray}
where the couplings are
\begin{eqnarray}
	g_0 = -\frac{g}{c_W}\, z^{\sf}_{ij}\,,\qquad && g^L_3 = \phantom{-}\sqrt{2}g_sT^{a}_{st}R_{i2} \,,\qquad g^R_3 = -\sqrt{2}g_sT^{a}_{st}R_{i1} \,,\\ \nonumber && g^L_4 = -\sqrt{2}g_sT^{a}_{st}R_{j1} \,,\qquad g^R_4 = \phantom{-}\sqrt{2}g_sT^{a}_{st}R_{j2}\,.
\end{eqnarray}
Here, the arguments of the loop integrals are
\begin{equation}\label{eqloopint2}
	C_i\big(k_1,-k_2;m_{\ti g},m_{\ti f_i},m_{\ti f_j}\big)\,,\qquad\quad B_0\big(-\!k_1-\!k_2;m_{\ti f_i},m_{\ti f_j}\big)\,.	
\end{equation}

\begin{figure}
	\begin{picture}(450,100)
       \put(0,0){\includegraphics[scale=1.2]{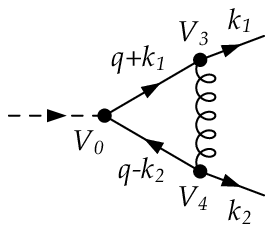}}
       \put(100,62){$V_0: (g_0^L P_L + g_0^R P_R)$}
       \put(100,47){$V_3: \gamma^{\rho} (g_3^L P_L + g_3^R P_R)$}
       \put(100,32){$V_4: \gamma^{\rho} (g_4^L P_L + g_4^R P_R)$}
       \put(230,0){\includegraphics[scale=1.2]{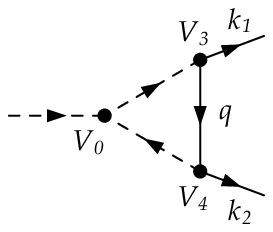}}
       \put(330,62){$V_0: g_0$}
       \put(330,47){$V_3: (g_3^L P_L + g_3^R P_R)$}
       \put(330,32){$V_4: (g_4^L P_L + g_4^R P_R)$}
	\end{picture}\caption{Vertex corrections to the $H_k^0$-boson coupling to fermions, where the arrows indicate the flow of the four-momenta and $q$ is the independent loop momentum.}\label{gendiagH}
\end{figure}

The generic structure of the amplitude, which parameterizes the corrections to the $s$-channel Higgs-boson exchanges, is much simpler and can be written as
\begin{eqnarray}
	\mathcal{M}_{\rm 1-loop} = \sum_{k=1}^4 \Big[ \overline{v}(p_2) \big( C_{L,k} P_L + C_{R,k} P_R \big) u(p_1)\Big] \frac{i}{s-m_{H_k}^2}\Big[ \overline{u}(k_1) \big( D^1_{L,k} P_L + D^1_{R,k} P_R \big) v(k_2) \Big],
\end{eqnarray}
where $C_{L,k}$ and $C_{R,k}$ are the couplings of the two neutralinos to the Higgs boson
\begin{eqnarray}
  C_{L,k} &=& C_{R,k} = -\frac{g}{2}\,F^0_{11k}	\qquad\quad\mbox{for}\ k=1,2\,,\\
  C_{L,k} &=& -C_{R,k} = \frac{g}{2}\,F^0_{11k}	\qquad\quad\mbox{for}\ k=3,4\,.
\end{eqnarray}
There are only two form factors $D^1_L$ and $D^1_R$, which receive contributions from two diagrams (see Fig.~\ref{gendiagH}). The contribution in the case of the gluon exchange is
\begin{eqnarray}
	D_L^1 &=& \frac{1}{8\pi^2}\,g_3^L g_4^L \Big[ g_0^L \big(2 B_0 - m_f^2 C_0 + ( - s + 3 m_f^2) (C_0 + C_1 + C_2)\big) - g_0^R m_f^2(C_1 + C_2)\Big]\,,\\
	D^1_R &=& D^1_L (g_0^L\leftrightarrow g_0^R)\,,
\end{eqnarray}
where the couplings are
\begin{eqnarray}
	g^L_0 = g^R_0 = s^f_k \quad (k=1,2)\,,\qquad g^L_0 = - g^R_0 = - s^f_k\quad k=(3,4)\,,\qquad g^L_3 = g^R_3 = g^L_4 = g^R_4 = -g_s T^{a}_{st}\,.
\end{eqnarray}
The loop integrals are identical to those defined in Eq.~(\ref{eqloopint1}). 

The same form factors for the gluino contribution are
\begin{eqnarray}
	D_L^1 &=& \frac{1}{16\pi^2}\, g_0 \Big[ -g_3^L g_4^L m_{\tilde g} C_0 + g_3^R g_4^L m_f C_1 + g_3^L g_4^R m_f C_2\Big]\,,\\
	D^1_R &=& D^1_L (g_3^L\leftrightarrow g_3^R, g_4^L\leftrightarrow g_4^R)\,,	
\end{eqnarray}
where the couplings are
\begin{eqnarray}
	g_0 = G_{ijk}^{\tilde{f}}\,,\qquad && g^L_3 = \phantom{-}\sqrt{2}g_sT^{a}_{st}R_{i2} \,,\qquad g^R_3 = -\sqrt{2}g_sT^{a}_{st}R_{i1} \,,\\ \nonumber && g^L_4 = -\sqrt{2}g_sT^{a}_{st}R_{j1} \,,\qquad g^R_4 = \phantom{-}\sqrt{2}g_sT^{a}_{st}R_{j2}\,,
\end{eqnarray}
and the loop integrals are defined in the same manner as in Eq.~(\ref{eqloopint2}).

\begin{figure}
	\begin{picture}(450,100)
       \put(0,0){\includegraphics[scale=1.2]{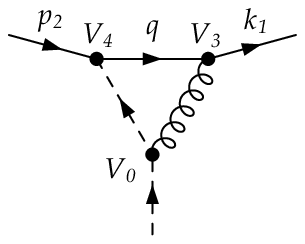}}
       \put(100,62){$V_0: g_0 (q+k_1-2p_2)^{\nu}$}
       \put(100,47){$V_3: \gamma^{\nu} (g_3^L P_L + g_3^R P_R)$}
       \put(100,32){$V_4: (g_4^L P_L + g_4^R P_R)$}
       \put(230,0){\includegraphics[scale=1.2]{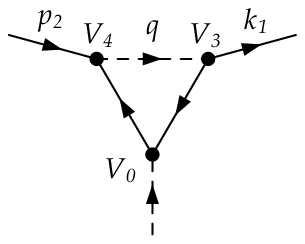}}
       \put(330,62){$V_0: (g_0^L P_L + g_0^R P_R)$}
       \put(330,47){$V_3: (g_3^L P_L + g_3^R P_R)$}
       \put(330,32){$V_4: (g_4^L P_L + g_4^R P_R)$}
	\end{picture}\caption{Vertex corrections to the neutralino coupling to a fermion and a sfermion where the arrows indicate the flow of the four-momenta and $q$ is the independent loop momentum.}\label{gendiagTU}
\end{figure}

The remaining vertex corrections are those connected with the $t$- and $u$-channel
squark exchanges, where we have defined $t=(p_1-k_2)^2$ and $u=(p_1-k_1)^2$. We
give here just the results for the $t$-channel, since the $u$-channel can obtained
by using the crossing symmetry. The generic amplitude is
\begin{eqnarray}
	\mathcal{M}_{\rm 1-loop} &=& \sum_{i=1}^2 \Big[ \overline{u}(k_1) \big( E^1_{L,i} P_L + E^1_{R,i} P_R \big) u(p_2) \Big] \frac{i}{t-m_{\tilde{f}_{i}}^2}\Big[ \overline{v}(p_1) \big(F_{L,i} P_L + F_{R,i} P_R \big) v(k_2) \Big]\\ \nonumber &+& \sum_{i=1}^2 \Big[ \overline{u}(k_1) \big( E_{L,i} P_L + E_{R,i} P_R \big) u(p_2) \Big] \frac{i}{t-m_{\tilde{f}_{i}}^2} \Big[ \overline{v}(p_1) \big( F^1_{L,i} P_L + F^1_{R,i} P_R \big) v(k_2)\Big] ,
\end{eqnarray}
where $E_{L/R,i}$ and $F_{L/R,i}$ are tree-level couplings given as
\begin{eqnarray}
  E_{L,i} = b^{\ti f}_{i1}\,,\qquad
  E_{R,i} = a^{\ti f}_{i1}\,,\qquad
  F_{L,i} = E_{R,i}\,,\qquad F_{R,i} = E_{L,i} \,.
\end{eqnarray} 
The contribution of the gluon exchange diagram (see Fig.~\ref{gendiagTU} left) is
\begin{eqnarray}
	 E_L^1 &=& \frac{1}{16\pi^2}\, g_0 g_3^L \Big[ g_4^L \big(B_0 + 2 m_f^2 
	(C_0-C_1) + (s + t) C_1 + m_{\ti \x}^2 (C_2-C_1)\big) 
	- g_4^R m_{\ti \x} m_f (C_0 + C_2 + C_1) \Big]\,,\\
	E_R^1 &=& E_L^1 (g_4^L\leftrightarrow g_4^R)\,,\\
	F_L^1 &=& E_R^1\,,\\
	F_R^1 &=& E_L^1\,,	
\end{eqnarray}
where the couplings are
\begin{eqnarray}
g_0 = -g_s T^{a}_{st}\delta_{ij}\,,\qquad g^L_3 = g^R_3 = -g_s T^{a}_{st}\qquad g^L_4 = b^{\ti f}_{j1}\,,\qquad g^R_4 = a^{\ti f}_{j1}
\end{eqnarray}
and the scalar loop coefficients have the arguments
\begin{equation}
	C_i(-k_1,-p_2;m_f,0,m_{\ti f_j})\,,\qquad\quad B_0(k_1-p_2;0,m_{\ti f_j})\,.
\end{equation} 

The gluino contribution gives
\begin{eqnarray}
	  E_L^1 &=& -\frac{1}{16\pi^2} \Big[ g_0^L g_3^L g_4^L m_{\tilde g} m_f C_0 + g_0^L g_3^L g_4^R m_{\tilde g} m_{\ti \x} (C_0 + C_2) + 
	     g_0^L g_3^R g_4^L m_f^2 (C_0 + C_1) + g_0^R g_3^R g_4^L m_{\tilde g} m_f C_1\\ \nonumber && + g_0^L g_3^R g_4^R m_{\ti \x} m_f (C_0 + C_1 + C_2) + 
	 g_0^R g_3^L g_4^L (B_0 + m_{\ti f_j}^2 C_0 + m_f^2 C_1 + m_{\ti \x}^2 C_2) + g_0^R g_3^L g_4^R m_f m_{\ti \x} C_2\Big],\\
	E_R^1 &=& E_L^1 (g_0^L\leftrightarrow g_0^R, g_3^L\leftrightarrow g_3^R, g_4^L\leftrightarrow g_4^R)\,,\\
	F_L^1 &=& E_R^1\,,\\
	F_R^1 &=& E_L^1\,,	
\end{eqnarray}
where the couplings are given by
\begin{eqnarray}
&& g^L_0 = \phantom{-}\sqrt{2}g_sT^{a}_{st}R_{i2} \,,\qquad g^L_3 = \phantom{-}\sqrt{2}g_sT^{a}_{st}R_{j2} \,,\qquad g^L_4 = a^{\ti f}_{j1} \,,\\ \nonumber && g^R_0 = -\sqrt{2}g_sT^{a}_{st}R_{i1} \,,\qquad g^R_3 = - \sqrt{2}g_sT^{a}_{st}R_{j1}\,,\qquad g^R_4 = b^{\ti f}_{j1} \,,
\end{eqnarray}
and the scalar loop functions are
\begin{equation}
	C_i(-k_1,-p_2;m_{\ti f_j},m_{\ti g},m_f)\,,\qquad\quad B_0(k_1-p_2;m_{\ti g},m_f)\,.
\end{equation}
\subsection{Box corrections}
\begin{figure}
  \begin{picture}(450,100)
    \put(0,0){\includegraphics[scale=0.8]{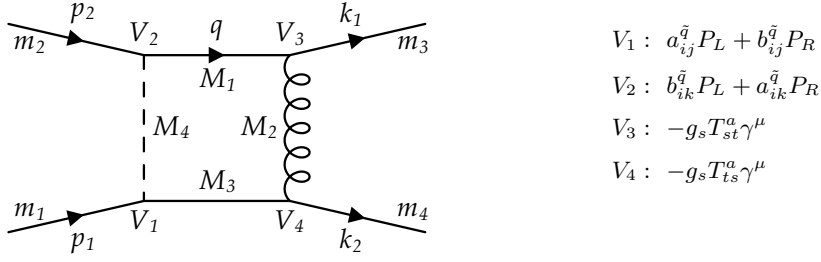}}
    \put(230,60){      
      \begin{math}\begin{array}{ll}
          V_1: & a_{ij}^{\tilde{q}} P_L + b_{ij}^{\tilde{q}} P_R \\[2mm]
          V_2: & b_{ik}^{\tilde{q}} P_L + a_{ik}^{\tilde{q}} P_R \\[2mm]
          V_3: & -g_s T^a_{st} \gamma^{\mu} \\[2mm]
          V_4: & -g_s T^a_{ts} \gamma^{\mu} 
        \end{array}\end{math}
    }
  \end{picture}
  \caption{Conventions and notations used for momenta, masses, and couplings in the calculation of the box diagram arising from the exchange of a gluon between the two final-state quarks in neutralino-pair annihilation. The arrows indicate the direction of defined four-momenta. The coupling strengths are defined in Sec.\ \ref{asec1}.}
  \label{notbox1}
\end{figure}

\begin{figure}
  \begin{picture}(450,100)
    \put(0,0){\includegraphics[scale=0.8]{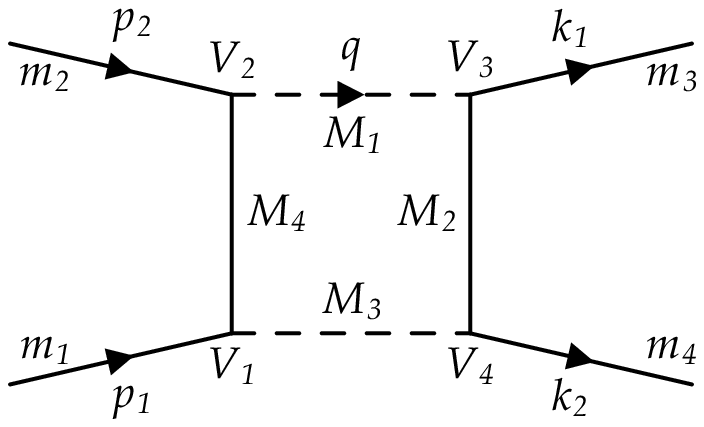}}
    \put(230,60){      
      \begin{math}\begin{array}{ll}
          V_1: & b_{ik}^{\tilde{q}} P_L + a_{ik}^{\tilde{q}} P_R \\[2mm]
          V_2: & a_{jk}^{\tilde{q}} P_L + b_{jk}^{\tilde{q}} P_R \\[2mm]
          V_3: & -\sqrt{2} g_s T^a_{st} \big( R_{jL} P_L - R_{jR} P_R \big) \\[2mm]
          V_4: & -\sqrt{2} g_s T^a_{st} \big( R_{iR} P_L - R_{iL} P_R \big)
        \end{array}\end{math}
    }
  \end{picture}
  \caption{Same as Fig.\ \ref{notbox1} for the exchange of a gluino between the two final-state quarks.}
  \label{notbox2}
\end{figure}

At the one-loop level, the neutralino annihilation into quark-antiquark pairs
receives box contributions arising from the exchange of a gluon or a gluino
between the final state quarks (see Fig.\ \ref{FDloop}). In order to express the
corresponding amplitudes in a rather generic way, we introduce the notations shown
in Figs.\ \ref{notbox1} and \ref{notbox2} for the gluon and gluino box,
respectively. The momenta of the incoming neutralinos are again labeled $p_1$ and
$p_2$, those of the outgoing quarks are $k_1$ and $k_2$. The corresponding
external masses are $m_i$ ($i=1,\dots,4$), while the masses of the internal
particles are denoted $M_i$ ($i=1,\dots,4$). The left- and right-handed coupling
strengths, which correspond to the vertices appearing in the schematic diagram,
are generically denoted by $g_i^{L,R}$ ($i=1,\dots,4$). For the sake of compact
expressions, we use in the following the abbreviations
\begin{equation}
  {\cal V}_i ~=~ g_i^L P_L + g_i^R P_R , \qquad\qquad
  \overline{{\cal V}}_i ~=~ g_i^L P_R + g_i^R P_L ,
\end{equation}
where $P_L$ and $P_R$ are the left- and right-handed chirality projectors, respectively. The independent loop-momentum $q$ is defined to be the momentum of the particle having the mass $M_1$ as shown in Figs.\ \ref{notbox1} and \ref{notbox2}. The arising tensor integrals have been reduced to scalar integrals 
\begin{equation}
  D_{\{i,ij\}} ~\equiv~ D_{\{i,ij\}}\big(-\!k_1,-(p_1+p_2),-p_2;M_1,M_2,M_3,M_4\big) 
\end{equation}
for $i,j=0,\dots,3$. For their definition and the applied convention we refer the reader to Ref.\ \cite{Denner1991}.

In case of the gluon box, the amplitude can be written as 
\begin{eqnarray}
  i{\cal M}_{\rm box}^{(g)} &=& -\bar{u}(k_1) {\cal F}^1_0 u(p_2) \bar{v}(p_1) {\cal F}^2_0 v(k_2) D_0 -
      \sum_{i=1}^3 \sum_{k=1}^2 \bar{u}(k_1) {\cal F}^{2k-1}_i u(p_2) \bar{v}(p_1) {\cal F}^{2k}_i v(k_2) D_i \nonumber \\
  & & -\bar{u}(k_1) {\cal F}^1_{00} u(p_2) \bar{v}(p_1) {\cal F}^2_{00} v(k_2) D_{00}  
      - \sum_{i,j=1}^3 \bar{u}(k_1) {\cal F}^1_{ij} u(p_2) \bar{v}(p_1) {\cal F}^2_{ij} v(k_2) D_{ij} ,
\end{eqnarray}
where the form factors are given by
\begin{equation}
  \begin{array}{rclp{1.2cm}rcl}
    {\cal F}^1_0 &=& m_1 \gamma^{\mu} {\cal V}_3 {\cal V}_2  &\qquad\qquad\qquad&
    {\cal F}^2_0 &=& m_3 {\cal V}_1 \gamma_{\mu} {\cal V}_4 + M_1 \overline{\cal V}_1 \gamma_{\mu} {\cal V}_4 - \pslash_2 \overline{\cal V}_1 \gamma_{\mu} {\cal V}_4 \\[2mm]
    {\cal F}^1_1 &=& M_3 \gamma^{\mu} \overline{\cal V}_3 {\cal V}_2 - 2 k_1^{\mu} \overline{\cal V}_3 {\cal V}_2 & &
    {\cal F}^2_1 &=& m_3 {\cal V}_1 \gamma_{\mu} {\cal V}_4 + M_1 \overline{\cal V}_1 \gamma_{\mu} {\cal V}_4 - \pslash_2 \overline{\cal V}_1 \gamma_{\mu} {\cal V}_4 \\[1mm]
    {\cal F}^3_1 &=& - m_1 \gamma^{\mu} {\cal V}_3 {\cal V}_2 & &
    {\cal F}^4_1 &=& \kslash_1 \overline{\cal V}_1 \gamma_{\mu} {\cal V}_4 \\[1mm]
    {\cal F}^1_2 &=& - \gamma^{\mu} \pslash_1 \overline{\cal V}_3 {\cal V}_2 - M_2 \gamma^{\mu} {\cal V}_3 \overline{\cal V}_2 & &
    {\cal F}^2_2 &=& m_3 {\cal V}_1 \gamma_{\mu} {\cal V}_4 + M_1 \overline{\cal V}_1 \gamma_{\mu} {\cal V}_4 - \pslash_2 \overline{\cal V}_1 \gamma_{\mu} {\cal V}_4 \\[2mm]
    {\cal F}^3_2 &=& m_1 \gamma^{\mu} {\cal V}_3 {\cal V}_2   & &
    {\cal F}^4_2 &=& M_1 \overline{\cal V}_1 \gamma_{\mu} {\cal V}_4 - \pslash_2 \overline{\cal V}_1 \gamma_{\mu} {\cal V}_4 \\[1mm]
    {\cal F}^1_3 &=& - M_2 \gamma^{\mu} {\cal V}_3 \overline{\cal V}_2   & &
    {\cal F}^2_3 &=& m_3 {\cal V}_1 \gamma_{\mu} {\cal V}_4 + M_1 \overline{\cal V}_1 \gamma_{\mu} {\cal V}_4 - \pslash_2 \overline{\cal V}_1 \gamma_{\mu} {\cal V}_4 \\[1mm]
    {\cal F}^3_3 &=& m_1 \gamma^{\mu} {\cal V}_3 {\cal V}_2  & &
    {\cal F}^4_3 &=& \pslash_2 \overline{\cal V}_1 \gamma_{\mu} {\cal V}_4\\[2mm]
    {\cal F}^1_{00} &=& \gamma^{\mu} {\cal V}_3 \gamma^{\nu} {\cal V}_2 & &
    {\cal F}^2_{00} &=& {\cal V}_1 \gamma_{\nu} \gamma_{\mu} {\cal V}_4 \\[2mm]
    {\cal F}^1_{11} &=& 2 k_1^{\mu} \overline{\cal V}_3 {\cal V}_2 - M_3 \gamma^{\mu} \overline{\cal V}_3 {\cal V}_2  & &
    {\cal F}^2_{11} &=& \kslash_1 \overline{\cal V}_1 \gamma_{\mu} {\cal V}_4 \\[1mm]
    {\cal F}^1_{12} &=& 2 k_1^{\mu} \overline{\cal V}_3 {\cal V}_2 - M_3 \gamma^{\mu} \overline{\cal V}_3 {\cal V}_2  & &
    {\cal F}^2_{12} &=& - M_1 \overline{\cal V}_1 \gamma_{\mu} {\cal V}_4 + \pslash_2 \overline{\cal V}_1 \gamma_{\mu} {\cal V}_4 \phantom{\pslash_2 \overline{\cal V}_1 \gamma_{\mu} {\cal V}_4}\\[1mm]
    {\cal F}^1_{13} &=& 2 k_1^{\mu} \overline{\cal V}_3 {\cal V}_2 - M_3 \gamma^{\mu} \overline{\cal V}_3 {\cal V}_2  & &
    {\cal F}^2_{13} &=& \pslash_2 \overline{\cal V}_1 \gamma_{\mu} {\cal V}_4 \\[2mm]
  \end{array}
\end{equation}
\begin{equation}
  \begin{array}{rclp{1.5cm}rcl}
    {\cal F}^1_{21} &=& \gamma^{\mu} \pslash_1 \overline{\cal V}_3 {\cal V}_2 + M_2 \gamma^{\mu} {\cal V}_3 \overline{\cal V}_2  & &
    {\cal F}^2_{21} &=& \kslash_1 \overline{\cal V}_1 \gamma_{\mu} {\cal V}_4 \\[1mm]
    {\cal F}^1_{22} &=& \gamma^{\mu} \pslash_1 \overline{\cal V}_3 {\cal V}_2 + M_2 \gamma^{\mu} {\cal V}_3 \overline{\cal V}_2  & &
    {\cal F}^2_{22} &=& - M_1 \overline{\cal V}_1 \gamma_{\mu} {\cal V}_4 + \pslash_2 \overline{\cal V}_1 \gamma_{\mu} {\cal V}_4 \\[1mm]
    {\cal F}^1_{23} &=& \gamma^{\mu} \pslash_1 \overline{\cal V}_3 {\cal V}_2 + M_2 \gamma^{\mu} {\cal V}_3 \overline{\cal V}_2  & &
    {\cal F}^2_{23} &=& \pslash_2 \overline{\cal V}_1 \gamma_{\mu} {\cal V}_4 \\[2mm]
    {\cal F}^1_{31} &=& M_2 \gamma^{\mu} {\cal V}_3 \overline{\cal V}_2  & &
    {\cal F}^2_{31} &=& \kslash_1 \overline{\cal V}_1 \gamma_{\mu} {\cal V}_4  \\[1mm]
    {\cal F}^1_{32} &=& M_2 \gamma^{\mu} {\cal V}_3 \overline{\cal V}_2  & &
    {\cal F}^2_{32} &=& - M_1 \overline{\cal V}_1 \gamma_{\mu} {\cal V}_4 + \pslash_2 \overline{\cal V}_1 \gamma_{\mu} {\cal V}_4 \\[1mm]
    {\cal F}^1_{33} &=& M_2 \gamma^{\mu} {\cal V}_3 \overline{\cal V}_2  & &
    {\cal F}^2_{33} &=& \pslash_2 \overline{\cal V}_1 \gamma_{\mu} {\cal V}_4 .
  \end{array}
\end{equation}

Using the same notation, the amplitude of the gluino box is given by
\begin{eqnarray}
  i{\cal M}_{\rm box}^{(\tilde{g})} &=& \bar{v}(p_2) {\cal F}^1_0 u(p_1) \bar{u}(k_1) {\cal F}^2_0 v(k_2) D_0 +
      \sum_{i=1}^3 \sum_{k=1}^2 \bar{v}(p_2) {\cal F}^{2k-1}_i u(p_1) \bar{u}(k_1) {\cal F}^{2k}_i v(k_2) D_i \nonumber \\
  & & +~\bar{v}(p_2) {\cal F}^1_{00} u(p_1) \bar{u}(k_1) {\cal F}^2_{00} v(k_2) D_{00}  
      + \sum_{i,j=1}^3 \bar{v}(p_2) {\cal F}^1_{ij} u(p_1) \bar{u}(k_1) {\cal F}^2_{ij} v(k_2) D_{ij}
\end{eqnarray}
with the form factors
\begin{equation}
  \begin{array}{rclcrcl}
    {\cal F}^1_0 &=& m_4 {\cal V}_2 {\cal V}_1 + M_2 \overline{\cal V}_2 {\cal V}_1 &\qquad\qquad\qquad&
    {\cal F}^2_0 &=& m_2 {\cal V}_3 {\cal V}_4 + M_3 \overline{\cal V}_3 {\cal V}_4 \\[2mm]
    {\cal F}^1_1 &=& {\cal V}_2 \kslash_1 {\cal V}_1 & &
    {\cal F}^2_1 &=& M_3 \overline{\cal V}_3 {\cal V}_4 + m_2 {\cal V}_3 {\cal V}_4 \\[1mm]
    {\cal F}^3_1 &=& M_2 \overline{\cal V}_2 {\cal V}_1 + m_4 {\cal V}_2 {\cal V}_1 & &
    {\cal F}^4_1 &=& M_3 \overline{\cal V}_3 {\cal V}_4 \\[1mm]
    {\cal F}^1_2 &=& M_2 \overline{\cal V}_2 {\cal V}_1 - M_1 {\cal V}_2 \overline{\cal V}_1  & &
    {\cal F}^2_2 &=& M_3 \overline{\cal V}_3 {\cal V}_4 + m_2 {\cal V}_3 {\cal V}_4 \\[2mm]
    {\cal F}^3_2 &=& M_2 \overline{\cal V}_2 {\cal V}_1 + m_4 {\cal V}_2 {\cal V}_1& &
    {\cal F}^4_2 &=& M_3 \overline{\cal V}_3 {\cal V}_4 - M_4 {\cal V}_3 \overline{\cal V}_4 \\[1mm]
    {\cal F}^1_3 &=& M_2 \overline{\cal V}_2 {\cal V}_1  & &
    {\cal F}^2_3 &=& M_3 \overline{\cal V}_3 {\cal V}_4 + m_2 {\cal V}_3 {\cal V}_4 \\[1mm]
    {\cal F}^3_3 &=& M_2 \overline{\cal V}_2 {\cal V}_1 + m_4 {\cal V}_2 {\cal V}_1 & &
    {\cal F}^4_3 &=& -{\cal V}_3 \pslash_2 {\cal V}_4\\[2mm]
    {\cal F}^1_{00} &=& - {\cal V}_2 \gamma^{\mu} {\cal V}_1 & &
    {\cal F}^2_{00} &=& {\cal V}_3 \gamma_{\mu} {\cal V}_4 \\[2mm]
    {\cal F}^1_{11} &=& - {\cal V}_2 \kslash_1 {\cal V}_1  & &
    {\cal F}^2_{11} &=& M_3 \overline{\cal V}_3 {\cal V}_4 \\[1mm]
    {\cal F}^1_{12} &=& - {\cal V}_2 \kslash_1 {\cal V}_1  & &
    {\cal F}^2_{12} &=& M_3 \overline{\cal V}_3 {\cal V}_4 - M_4 {\cal V}_3 \overline{\cal V}_4 \\[1mm]
    {\cal F}^1_{13} &=& - {\cal V}_2 \kslash_1 {\cal V}_1  & &
    {\cal F}^2_{13} &=& {\cal V}_3 \pslash_2 {\cal V}_4  \\[2mm]
    {\cal F}^1_{21} &=& M_2 \overline{\cal V}_2 {\cal V}_1 - M_1 {\cal V}_2 \overline{\cal V}_1  & &
    {\cal F}^2_{21} &=& M_3 \overline{\cal V}_3 {\cal V}_4  \\[1mm]
    {\cal F}^1_{22} &=& M_2 \overline{\cal V}_2 {\cal V}_1 - M_1 {\cal V}_2 \overline{\cal V}_1  & &
    {\cal F}^2_{22} &=& M_3 \overline{\cal V}_3 {\cal V}_4 - M_4 {\cal V}_3 \overline{\cal V}_4  \\[1mm]
    {\cal F}^1_{23} &=& M_2 \overline{\cal V}_2 {\cal V}_1 - M_1 {\cal V}_2 \overline{\cal V}_1  & &
    {\cal F}^2_{23} &=& {\cal V}_3 \pslash_2 {\cal V}_4  \\[2mm]
    {\cal F}^1_{31} &=& M_2 \overline{\cal V}_2 {\cal V}_1  & &
    {\cal F}^2_{31} &=& M_3 \overline{\cal V}_3 {\cal V}_4  \\[1mm]
    {\cal F}^1_{32} &=& M_2 \overline{\cal V}_2 {\cal V}_1  & &
    {\cal F}^2_{32} &=& M_3 \overline{\cal V}_3 {\cal V}_4 - M_4 {\cal V}_3 {\cal V}_4 \\[1mm]
    {\cal F}^1_{33} &=& M_2 \overline{\cal V}_2 {\cal V}_1  & &
    {\cal F}^2_{33} &=& {\cal V}_3 \pslash_2 {\cal V}_4 .
  \end{array}
\end{equation}


\end{document}